\documentclass[]{aastex631}

\usepackage{amsmath}
\usepackage{float}
\usepackage{xcolor}

\newcommand{\beten}{\mbox{${}^{10}{\rm Be}$}}
\newcommand{\beseven}{\mbox{${}^{7}{\rm Be}$}}
\newcommand{\beratio}{\mbox{${}^{10}{\rm Be}/{}^{9}{\rm Be}$}}
\newcommand{\clthreesix}{\mbox{${}^{36}{\rm Cl}$}}
\newcommand{\clratio}{\mbox{${}^{36}{\rm Cl}/{}^{35}{\rm Cl}$}}
\newcommand{\altwosix}{\mbox{${}^{26}{\rm Al}$}}
\newcommand{\alratio}{\mbox{${}^{26}{\rm Al}/{}^{27}{\rm Al}$}}
\newcommand{\neratio}{\mbox{${}^{21}{\rm Ne}/{}^{22}{\rm Ne}$}}
\newcommand{\liratio}{\mbox{${}^{7}{\rm Li}/{}^{6}{\rm Li}$}}
\newcommand{\permil}{\mbox{\text{\textperthousand}}}
\newenvironment{Steve}{\color{black}}

\begin{document}

\title{The Extent of Solar Energetic Particle Irradiation in the Sun's Protoplanetary Disk}


\author[0000-0002-1571-0836]{Steven J. Desch}
\affiliation{School of Earth and Space Exploration \\ Arizona State University \\ PO Box 876004, Tempe AZ 85287-6004}

\author
[0000-0002-6051-9002]{Ashley K. Herbst}
\affiliation{School of Earth and Space Exploration \\ Arizona State University \\ PO Box 876004, Tempe AZ 85287-6004}

\author[0000-0001-7892-5423]{Richard L. Hervig}
\affiliation{School of Earth and Space Exploration \\ Arizona State University \\ PO Box 876004, Tempe AZ 85287-6004}

\author[0000-0003-4149-517X]
{Benjamin Jacobsen}
\affiliation{Nuclear and Chemical Sciences Division \\ Lawrence Livermore National Laboratory \\ 7000 East Avenue, Livermore CA 94550}
 
\begin{abstract}

Solar flares emit X rays and high-energy (MeV-GeV) ions (Solar Energetic Particles, or SEPs). 
Astronomical observations show solar mass-protostellar fluxes are a factor {\Steve $\Phi \approx 3 \times 10^2 - 3 \times 10^3$} times higher than the present-day Sun. 
Constraining $\Phi$ in the early solar system is important for modeling ionization in the Sun's protoplanetary disk, the extent of magnetorotational instability or magnetocentrifugal outflows, or even production of short-lived radionuclides.
Recent interpretations of meteoritic data---cosmogenic Ne in hibonite grains, initial $(\beratio)_0$ ratios in Ca-rich, Al-rich inclusions (CAIs), or even inferences of live $\beseven$ in CAIs---have suggested values $\Phi > 10^5$, even as large as $\Phi \approx 6 \times 10^6$, which would make the young Sun extraordinarily active, even for a protostar. 
We constrain $\Phi$ by re-examining these data.
We conclude: cosmogenic Ne was produced in hibonite grains as they resided in the disk;
${}^{36}{\rm Cl}$ was created in Cl-poor grains after the disk dissipated; 
$\beten$ was inherited from the molecular cloud, with almost no ($< 1\%$) $\beten$ created in the disk;
and there is no evidence whatsoever for any live $\beseven$ in CAIs.
We show these data are consistent with a value {\Steve $\Phi \approx 3 \times 10^3$} for the first $> 5$ Myr of the solar nebula.
The early Sun evidently emitted {\Steve a flux of X rays and SEPs not atypical} for a protostar.

\end{abstract}


\section{Introduction} \label{sec:intro}

Many questions surround the evolution of protoplanetary disks, and one of the most important unknowns is how effectively they are irradiated by {\Steve flares and} energetic particles from their central stars.
During solar flares, the present-day Sun emits X rays and copious energetic ($\sim$ MeV to GeV) protons and $\alpha$ particles called Solar Energetic Particles (SEPs; also called Solar Cosmic Rays, or SCRs).
The long-term ($\sim 10^6$ yr) time-averaged flux of energetic particles (protons, $\alpha$ particles, etc.) at Earth (at 1 AU, near the Sun's equator) today is $\approx 100 \, {\rm cm}^{-2} \, {\rm s}^{-1}$. 
\citep[e.g.,][\S 2.1]{TrappitschLeya2013}, {which corresponds roughly} to an SEP luminosity $\approx 5 \times 10^{-9} \, L_{\odot}$.
{\Steve
The long-term time-averaged X-ray luminosity of the Sun today is on the order of $6 \times 10^{-7} \, L_{\odot}$ (see \S 2.1).
}
Flares from protostars can ionize the H$_2$ gas in their surrounding disks by Extreme Ultraviolet (EUV) radiation, by X rays \citep{IgeaGlassgold1999}, and by emission of SEPs \citep{TurnerDrake2009,FraschettiEtal2018}.
The resultant ionization structure is a key factor in determining whether disks evolve via magnetorotational instability or magnetocentrifugal outflows \citep{LesurEtal2023}, and stellar ionization will drive significant chemistry in the protoplanetary disk \citep{BerginEtal2007,LongEtal2024}.
SEPs can also drive nuclear reactions, creating short-lived radionuclides, although it is unclear how much this occurred in the early Solar System \citep{DeschEtal2023pp7}.
An important goal of protoplanetary disk studies is to constrain the SEP flux of the early Sun and other protostars.
We define $\Phi$ to be the flux (at 1 AU) of SEPs with energies $> 10$ MeV, relative to the flux today of $100 \, {\rm cm}^{-2} \, {\rm s}^{-1}$.
Our goal is to determine $\Phi$ for the early Sun.

Observations of protostars provide one way of determining $\Phi$.  
SEPs are emitted in flares, which emit X rays.
The typical X-ray luminosities of {\Steve solar-mass} protostars are much higher than the present-day Sun but lie in a restricted range: $2 \times 10^{-4} - 2 \times 10^{-3} \, L_{\odot}$ \citep{GetmanFeigelson2021}, corresponding to {\Steve $\Phi = 3 \times 10^{2} - 3 \times 10^{3}$} (see \S 2.1).
Assuming the same {\Steve ratio of SEP flux to X-ray flux}, $\Phi$ is unlikely to exceed {\Steve $\approx 4 \times 10^{4}$} in most protostellar systems.

We can also use meteoritic data to determine $\Phi$ in the early Solar System and compare this to protostellar data.
If a high value $\Phi > 4 \times 10^4$ were demanded to explain meteoritic data, this would imply the Sun's origin was unusual, or that our observations of protostars were incomplete.
It might also imply production of short-lived radionuclides (SLRs) like $\altwosix$ could be achieved by SEP irradiation in the disk, undermining or even contradicting the evidence that the Sun acquired SLRs from its molecular cloud in a high-mass star-forming region \citep{DeschEtal2023pp7}.
Conversely, if meteoritic data were adequately explained by $\Phi \leq 4 \times 10^4$, this would imply the Sun's origin was not unusual, and that our Solar System can be used to understand planet formation in other systems.

%
%
Unfortunately, the models built to explain certain meteoritic data yield very different estimates of $\Phi$.  
\citet{LeeEtal1998} and \citet{GounelleEtal2001}, in their modeling of SLR production, assumed typical X-ray luminosities $4.5 \times 10^{29} \, {\rm erg} \, {\rm s}^{-1}$, corresponding to $\Phi \approx 2 \times 10^3$, close to the typical value and well within the range inferred for protostars.
In more recent investigations, much higher values have been assumed.
\citet{SossiEtal2017}, presuming $\beten$ and observed variations in V isotopes in some inclusions were attributable to SEP irradiation, inferred a fluence of particles $6 \times 10^{18} \, {\rm cm}^{-2}$ obtained within 3 yr at 0.01 AU, corresponding to $\Phi = 6 \times 10^4$. 
\citet{LiuEtal2024}, presuming live $\beseven$ produced by SEP irradiation was incorporated into some meteoritic inclusions, inferred a particle flux $(0.5 - 1) \times 10^{10} \, {\rm cm}^{-2} \, {\rm s}^{-1}$ at 0.03 AU, corresponding to $\Phi \approx 9 \times 10^{4}$.
\citet{Jacquet2019}, in modeling production of live $\beten$ in the disk, assumed an SEP luminosity $4.7 \times 10^{-3} \, L_{\odot}$, corresponding to $\Phi = 9 \times 10^5$. 
\citet{YangEtal2024}, assuming cosmogenic Ne in hibonite grains was produced in only a few years while the grains were presumably launched above the disk, also inferred an SEP flux corresponding to $\Phi = 10^{6.8} = 6 \times 10^6$.
It may appear that the meteoritics community has found multiple lines of evidence for a particle flux $\sim 10^5 - 10^7$ times the present-day flux in the Solar System, orders of magnitude higher than the large $\Phi \sim 10^3$ enhancements previously considered {\Steve and} supported by observations of protostars. 
{(Again, these refer to averages over time ranging from years to millions of years.)}

While an enhancement in SEP flux of {\Steve $\Phi \sim 10^{3}$} would be typical for protostars
{(based on the discussion above)}, values $\Phi > 10^5$ would not be, and time-averaged values $\Phi \sim 10^7$ are unreasonably high.
If the SEP and X-ray luminosities of the early Sun were a factor $\Phi = 6 \times 10^{6}$ times the present-day value, this would mean the early Sun's X-ray luminosity {\Steve would be $4 \, L_{\odot}$, i.e., on average more of the proto-Sun's energy would be emitted as X rays than as blackbody radiation}.
These numbers simply are not supported by observations, which calls into question whether the meteoritic data above can be attributed to SEP irradiation at all.
Or, if they can, what limits do they place on the SEP flux?  

In this paper, we examine several types of meteoritic data that could be used to constrain the SEP flux in the solar nebula.
We start with a discussion in \S~\ref{sec:astrophysics} of the astronomical measurements pertinent to SEP fluxes, which constrain the SEP enhancement factor to be {\Steve $\Phi < 3 \times 10^3$} if the Sun is typical.
We also provide astrophysical models of the protoplanetary disk and how SEPs irradiate it, and how small solid particles are distributed within the disk or possibly launched above the disk, where they potentially could be exposed directly to SEPs from the early Sun.

In \S~\ref{sec:chlorine} we examine the meteoritic evidence for live $\clthreesix$ in the early Solar System, which comes from Cl-S and Cl-Ar systematics in CAIs (Ca-rich, Al-rich inclusions) that have been altered on their parent bodies.
We consider the possibility that $\clthreesix$ was inherited from the molecular cloud, one model considered by \citep{DeschEtal2023pp7}, but we argue that it was not inherited, that it arose from SEP irradiation, and its one-time abundance fixes the SEP flux enhancement to {\Steve $\Phi < 10^6$, perhaps as low as $\Phi \approx 10^3$,} late in disk evolution.

In \S~\ref{sec:neon} we discuss cosmogenic Ne found in hibonite grains in meteorites, which is widely accepted to have formed from SEPs irradiating free-floating grains during the lifetime of the solar nebula.
We demonstrate that the amounts and $\neratio$ ratios of the cosmogenic Ne are consistent with irradiation of grains within the disk, near the disk surfaces, {\Steve assuming an SEP flux enhancement of $\Phi \approx 3 \times 10^3$ to $2 \times 10^4$.}
\citet{YangEtal2024} inferred $\Phi = 6 \times 10^6$ assuming hibonite grains acquired cosmogenic Ne only while directly exposed to SEPs, in the few years they would be above the disk after being launched in a disk wind.
We demonstrate that it is not possible to launch particles{ \Steve by disk winds} above the disk to be directly exposed, and that even if they were, they would acquire a $\neratio$ ratio lower than is observed in meteoritic hibonite.

In \S~\ref{sec:beten} we discuss the possibility that live $\beten$ was produced by SEP irradiation in the solar nebula.
Of all the SLRs, $\beten$ is the one most effectively produced by SEP irradiation, as the target nuclei (e.g., ${}^{16}{\rm O}$) are plentiful and ${}^{9}{\rm Be}$ is rare.
It has been calculated that an SEP flux enhancement of $\Phi \approx 8 \times 10^5$ could lead to production of all the $\beten$ observed in CAIs \citep{Jacquet2019}, but that in such a scenario, CAIs formed at different times and places in the disk would acquire very different initial $(\beratio)_0$ ratios.
We demonstrate that in fact the $(\beratio)_0$ ratios of CAIs are remarkably uniform, clustered around a single value $\approx 7 \times 10^{-4}$, with spread equal to that expected from measurement uncertainties, with {\bf zero} CAIs showing credible evidence for larger values.
We also show that CAIs with smaller values are consistent with later formation or resetting after decay of $\beten$ for $\approx 1$ Myr. 
We rule out SEP irradiation as a significant (i.e., $>$ few percent) source of $\beten$ in CAIs, and we constrain the SEP flux enhancement factor to be $\Phi < 10^5$ or so.

Finally, in \S~\ref{sec:beseven} we discuss the evidence for ${}^{7}{\rm Be}$, for which evidence in CAIs has been claimed multiple times. 
With a half-life of only 53 days, this SLR, if verified, would demand production within the solar nebula and SEP fluxes enhanced by factors $\Phi > 10^4 - 10^5$ \citep{LiuEtal2024}.
Inferences of ${}^{7}{\rm Li}/{}^{6}{\rm Li}$ excesses come only after subtracting a model-dependent amount of cosmogenic Li, and we demonstrate that this has been overestimated and the ${}^{7}{\rm Li}/{}^{6}{\rm Li}$ data overcorrected. 
We also demonstrate that the uncertainties have not been propagated correctly; proper accounting for this results in the correlations of ${}^{7}{\rm Li}/{}^{6}{\rm Li}$ vs.\ ${}^{9}{\rm Be}/{}^{6}{\rm Li}$ not resolved from zero slope.
There is no evidence whatsoever for live $\beseven$ in the early solar system. 

We draw conclusions in \S~\ref{sec:conclusions}.
We conclude that an SEP flux {\Steve $\Phi < 1 \times 10^4$} times the present-day flux persisted from the beginning of the Solar System to at least the time the gas dissipated after $> 5$ Myr.
This SEP irradiation created cosmogenic Ne in hibonite grains in the disk, and after the disk gas dissipated, continued to make $\clthreesix$ in solid materials. 
There is no evidence this irradiation created significant live $\beten$, or any live $\beseven$, incorporated into meteoritic inclusions.
We discuss implications.

\section{Irradiation of the Protoplanetary Disk}
\label{sec:astrophysics}

There are essentially two hypotheses for how materials in the solar nebula could experience irradiation by SEPs.
One is that all materials resided in the disk itself.
Gas at the disk surfaces was irradiated, as well as particles that did not settle to the midplane where they would be shielded from irradiation.
The other hypothesis is that dust grains were irradiated by being temporarily lofted above (or below) the disk, presumably by launching in a magnetocentrifugal outflow (disk wind).
Here we provide physical models of each scenario to allow later tests of each hypothesis.

\subsection{SEP fluxes from protostars and the early Sun}

Because it is an important input, we first consider the SEP flux from the Sun.
{SEPs are ions (and electrons) accelerated to relativistic speeds by a variety of processes, including shocks associated with coronal mass ejections (CMEs), and magnetic reconnection events. \citep{Reames2013}.
The former are associated with ``gradual" events, and the latter with ``impulsive" flares.
Ions accelerated in gradual flares tend to be roughly 96\% protons, 4\% $\alpha$ particles, and $< 0.1\%$ heavier ions.
Ions accelerated in impulsive flares are only roughly 65\% protons, 3\% $\alpha$ particles, and the remainder heavier ions
\citep{Reames2013}.
}

There are more low-energy particles than high-energy particles.
This is often modeled as a power law, with the flux $F$ (number per area per time) of SEPs with energies (per nucleon) between $E$ and $E + dE$ being $(dF/dE) \, dE \propto E^{-p} \, dE$, where the spectral index $p$ may vary from 2.7 for gradual flares, to 4.0 for impulsive flares \citep[e.g.,][]{GounelleEtal2001}.
Alternatively, the energy spectrum may be better described using the following expression
\citep{McGuireVonRosenvinge1984,TrappitschLeya2013}:
\begin{equation}
\frac{dF}{dE} = F(E > 10 \, {\rm MeV}) \, \frac{ \exp(+R_{10} / R_0) }{ R_0 } \, \frac{ \left( E + E_0 \right) }{ Z e \, \left( E^2 + 2 E E_0 \right)^{1/2} } \, \exp \left( -\frac{ \left( E^2 + 2 E E_0 \right)^{1/2} }{ Z e R_0 } \right),
\label{eq:SEPflux}
\end{equation}
where $R_{10} = (E^2 + 2 E E_0)^{1/2} / (Z e)$ when $E = 10 \, {\rm MeV}$, and $R_0$ is the rigidity.
{
The rigidity can be different for each ion and depends on what type of event accelerated the ions.
For protons in impulsive flares we might typically take $R_0 \approx 70 - 130$ MV, with higher values favored \citep{TrappitschLeya2013}.
}
Here $E_0 = 938 \, {\rm MeV}$ is the rest-mass energy of the proton.
This function is a quasi power law, with $p \approx 2.7$ to $4.0$ at energies of a few hundred MeV, but shallower at lower $E$ and steeper at higher $E$.
{
For $R_0 = 130$ MV,}
the average energy per SEP ({among those} with $E > 10$ MeV) is about 45 MeV.
{\Steve
The long-term ($\sim 10^6$ yr) time-averaged SEP flux is inferred from the measured abundances of cosmogenic radionuclides like ${}^{26}{\rm Al}$ in samples of lunar regolith, and can be compared with production rates of ${}^{14}{\rm C}$ on Earth and with satellite measurements
\citep{PoluianovEtal2018,UsoskinEtal2023}.
Extrapolating from the estimates by \citet{PoluianovEtal2018} of the SEP fluxes at 20-80 MeV (per nucleon), we estimate the omnidirectional flux of SEPs with kinetic energies $> 10$ MeV (per nucleon) is $F(E > 10 \, {\rm MeV}) \approx 190 \, {\rm cm}^{-2} \, {\rm s}^{-1}$.
However, there is a tendency for more SEPs during the Sun's quiet phases to end up in the ecliptic plane, by a factor comparable to 1.6, whereas for the Sun's more active phases (and presumably also protostars), SEPs appear distributed more isotropically \citep{WaszewskiEtal2025}. 
The long-term averaged SEP flux at 1 AU today is probably within a factor of two or less of the commonly adopted value, $F(E > 10 \, {\rm MeV}) = 100 \, {\rm cm}^{-2} \, {\rm s}^{-1}$.
}

Unfortunately, the production rate of SEPs is only directly observable for the present-day Sun, and not for the early Sun or even other protostars. 
However, the same flares that would generate SEPs also generate hard (keV) X rays, and it is known that protostars have much greater X-ray luminosities than the Sun, and therefore presumably higher SEP fluxes.
{\Steve
The keV X-ray luminosity of the Sun today varies between $7 \times 10^{-8}$ and 
$1.2 \times 10^{-6} \, L_{\odot}$
\citep{JohnstoneGudel2015}.
The average X-ray luminosity of Sun-like stars---presumably the same as the Sun's long-term average X-ray luminosity---is $6.3 \times 10^{-7} \, L_{\odot}$
\citep{ZhuPreibisch2025}.
This suggests a long-term average ratio $L_{\rm SEP} / L_{\rm X} \approx 0.01$.}
It is presumed that the $L_{\rm SEP} / L_{\rm X}$ ratio is not significantly different for protostars, although \citet{FeigelsonEtal2002} speculated the efficiency could be one order of magnitude higher for intense protostellar flares.

The X-ray luminosities of solar-mass protostars can be thought of as the sum of a quiescent, or `characteristic,' component, plus a superposition of strong flares, including mega-flares releasing $> 1.6 \times 10^{36} \, {\rm erg}$ of X-ray energy \cite{GetmanFeigelson2021}.
The `characteristic' (quiescent) X-ray luminosities of $0.4 - 1.0 \, M_{\odot}$ protostars in the Orion Nebula (ages 0.5 - 10 Myr, median age 2 Myr) fall in a restricted range $2 \times 10^{-4}$ to $2 \times 10^{-3} \, L_{\odot}$ \citep{WolkEtal2005,GetmanFeigelson2021}.
Mega-flares, lasting on average 50 ks (no more than 100 ks) each, occur on average 1.7 (1.1 to 2.7) times per year.
During the day or so a mega-flare is active, it generates an X-ray luminosity as high as $\sim 10^{-2} \, L_{\odot}$, but this flux does not last long; averaged over {timescales of at least years}, such flares contribute only 10-20\% of the overall X-ray flux of young protostars \citep{GetmanFeigelson2021}.
{\Steve
We therefore consider $\Phi = (2 \times 10^{-3} \, L_{\odot}) / (6.3 \times 10^{-7} \, L_{\odot}) = 3 \times 10^{3}$ to be a maximum typical value in protostars when averaging over timescales of years or more.}
Even in the most active solar-mass protostars observed, the time-averaged X-ray luminosities never exceed $1 \times 10^{32} \, {\rm erg} \, {\rm s}^{-1}$ $=0.026 \, L_{\odot}$
\citep{GetmanFeigelson2021}, corresponding to {\Steve $\Phi = 4 \times 10^{4}$} for unusually active stars. 
If solar-mass protostars exist with $\Phi = 6 \times 10^{6}$, they are rare enough not to have been observed, i.e., represent $\ll 1\%$ of all solar-mass protostars.

\subsection{Irradiation by SEPs in protoplanetary disks}

{Whether accelerated near the Sun in an impulsive flare, or accelerated by a CME far from the Sun, ions reach 1 AU by traveling along magnetic field lines associated with the Parker spiral. 
Even a 1 GeV proton will gyrate around a field line with typical interplanetary magnetic field strength (tens of nT, or a fraction of $1 {\rm mG}$) with Larmor radius $< 0.001$ AU, and will not travel to 1 AU in a straight line.
To reach Earth itself, the SEPs must diffuse off of the interplanetary magnetic field and onto the Earth's magnetic field. 
They do so at magnetic cusps and then spiral along Earth's field lines to its north and south magnetic poles. 

Protostars are also expected to have flares, launch CMEs, and generate SEPs that will propagate along the ``interplanetary" magnetic field lines carried by stellar winds.
Mass loss from protostars is dominated by bipolar outflows, with mass loss rates $\sim 10^{-11}$ to $10^{-8} \, M_{\odot} \, {\rm yr}^{-1}$ for Class II protostars, making it difficult to observationally determine the mass loss from the relatively weaker stellar wind. 
For comparison, the solar wind is associated with a mass loss of only a few $\times 10^{-14} \, M_{\odot} \, {\rm yr}^{-1}$, although protostellar winds might be orders of magnitude stronger.
As discussed below (\S 2.3), we assume that disk winds are launched from beyond about 1 AU, but that interior to that, the stellar wind extends to the disk surface.
This allows SEPs to travel along ``interplanetary" magnetic field lines and reach the disk, at least interior to roughly 1 AU.

Models of this interaction hardly exist, so we make the following simplifying assumptions.
We assume the magnetopause (where the ram pressure of the stellar wind is balanced by thermal and magnetic pressure of the disk) lies essentially at the surface of the disk (e.g., at $z = \pm 5 H$), in analogy to the magnetopause at the top of Venus's atmosphere.
As SEPs spiral around the ``interplanetary" magnetic field, they can enter the disk and then propagate from that point on in straight lines.
As such, SEPs are directed into the disk; we assume their propagation directions are isotropic with respect to the vertical direction.
If a protoplanetary disk were to be placed at 1 AU in the present-day Solar System, we assume the top and bottom surfaces would each see an SEP flux of $50 \, {\rm cm}^{-2} \, {\rm s}^{-1}$, and we assume these fluxes are multipled by $\Phi$ in a protostellar system.}

{
We model the disk as follows,}
with inputs from the solar nebula models of \citet{DeschEtal2018}.
At any heliocentric distance $r$, the variation of density with height $z$ above the midplane is 
\begin{equation}
\rho(r,z) = \frac{ \Sigma(r) }{ \sqrt{2\pi} H } \, 
\exp \left( -\frac{ z^2 }{ 2 H^2 } \right),
\end{equation}
where $\Sigma(r)$ is the column density in the disk, $H = C / \Omega$ is the scale height, $C$ the sound speed, and $\Omega$ the Keplerian orbital frequency.
We assume the disk temperature at $t = 1$ Myr at $r = 1-3$ AU from the Sun was about {$T(r) = 370 \, (r / 1 \, {\rm AU})^{-0.56} \, {\rm K}$, which yields a scale height of $H = 5.8 \times 10^{11} \, {\rm cm}$ at 1 AU.
We assume a surface density at $r = 1$ AU and $t = 1 \, {\rm Myr}$ of about $1200 \, {\rm g} \, {\rm cm}^{-2}$, yielding a midplane density at $1 \, {\rm AU}$ of $8.2 \times 10^{-10} \, {\rm g} \, {\rm cm}^{-3}$. 
At $z = 5 \, H$, the thermal pressure would be $\approx 4000 \, {\rm nPa}$, balancing the stellar wind ram pressure. 
For the Sun this is $\approx 1 - 6 \, {\rm nPa}$ and would presumably be orders of magnitude stronger in a protostellar wind.}

Once in the disk, SEPs lose their kinetic energy, overwhelmingly due to ionization of gas. 
Each encounter with an H$_2$ molecule that it ionizes leads to loss of about 20 eV of SEP energy \citep{UmebayashiNakano1981}.
Protons with kinetic energies $E = 10$ MeV will be stopped when they pass through a column density $\Sigma_{\rm stop} = 0.075 \, {\rm g} \, {\rm cm}^{-2}$, 
100 MeV protons by $\Sigma_{\rm stop} = 5.6 \, {\rm g} \, {\rm cm}^{-2}$,
and 500 MeV protons by $\Sigma_{\rm stop} = 140 \, {\rm g} \, {\rm cm}^{-2}$
\citep{CravensDalgarno1978,PadovaniEtal2009}.
The stopping length scales roughly as $\Sigma_{\rm stop}(E) \propto E^2$, so high-energy SEPs penetrate deeper into the disk, changing the SEP energy spectrum at depth.

{We have written a numerical code to calculate the propagation of SEPs in the disk at 1 AU.
We create particles with initial energies $E_0$ in 250 energy bins logarithmically spaced between 0.1 MeV and 10 GeV, and divided into 64 angular bins spaced in $\mu = \cos\theta$ ($\theta$ is the deviation from vertical), using standard Gauss-Legendre weighting \citep{PressEtal1992}.
We divide the disk gas into 500 uniformly spaced vertical layers between $z = -5 H$ and $z = +5 H$, with the mass per area $d\Sigma$ in each layer derived from Equation 2. 
For each particle type ($\mu$, $E_0$), we calculate the loss of energy propagating from $z = +5 H$ down to the new layer (or up from $z = -5 H$ propagating up to the next layer) as $E_{\rm loss} = E \, (d\Sigma / \mu) / \Sigma_{\rm stop}$, and calculate its energy in the next layer as $E = E - E_{\rm loss}$, until the particle has energy $< 1$ MeV and is considered stopped.
Repeating for all $\mu$ and $E_0$, we then consider each particle with initial energy $E_0$ to represent a number of particles given by Equation 1, and we calculate the number of SEPs in each layer at height $z$ within each energy bin, integrating over angle.
}

\subsection{Irradiation by SEPs in a disk wind}

Magnetocentrifugal outflows and their accompanying disk winds are increasingly recognized to play an important role in the evolution of protoplanetary disks \citep[e.g.,][]{PascucciEtal2023}.
Magnetocentrifugal outflows involve magnetic fields vertically threading the protoplanetary disk, normal to the midplane. 
Above and below the disk, the inertia of gas frozen on the field lines causes them to bend outward.
If field lines bend by more than an angle of $30^{\circ}$ from vertical, the dynamical equilibrium is unstable, and gas may be accelerated upward and outward along field lines \citep{BlandfordPayne1982,BaiStone2013}, eventually to be launched in a bipolar outflow.

{Disk winds probably dominate the mass loss in bipolar outflows, but are launched only from distances of $\approx 0.5 - 3$ AU from the star \cite{PascucciEtal2023}.
Mass accretion rates of $10^{-8} \, M_{\odot} \, {\rm yr}^{-1}$ demand the disk be threaded by vertical fields on the order of 30 mG \cite{PascucciEtal2023}, corresponding to a magnetic pressure of about 3600 nPa.
The thermal pressure at 1 AU and $z = 5 H$ in our model is about 4000 nPa.
In what follows, we assume the ram pressure of the stellar wind is of this order at 1 AU, allowing the stellar wind to penetrate to the disk atmosphere inside 2 AU; the disk wind described below can be launched only outside this radius.
For comparison, the solar wind has ram pressure 1 to 6 nPa.
These values are plausible but not demanded.}

In all models of magnetocentrifugal outflows, the disk is threaded by a magnetic field that is vertical at the midplane, by symmetry.
Because of this general feature, winds are only launched far from the midplane.
In models of disks dominated by ambipolar diffusion, with low Ohmic diffusion, in which the magnetic flux is frozen in the ionized gas, gas velocities begin to approach the local sound speed at heights $\left|z\right| \approx 1-2 \, H$ from the midplane \citep{WardleKonigl1993,SalmeronEtal2011}. 
The most favorable scenario for launching particles would be at the inner edge of the disk at $r \sim 0.1 \, {\rm AU}$, where gas would be fully ionized, and winds could be launched from $\left|z\right| \approx 1-2 \, H$.
Beyond $\sim 0.1 \, {\rm AU}$, disks are less ionized, chararacterized by ``dead zones" with strong Ohmic diffusion that prevents magnetic field lines in the disk midplane region from being advected and bent.
Field lines remain vertical within about $\left|z\right| < 4-5 \, H$ of the midplane \citep{SalmeronEtal2011,Teitler2011,BaiStone2013}.
In the detailed models of \citet{BaiStone2013}, the density structure of the disk is affected by the outflow only at $\left|z\right| > 4.5 \, H$, more than four scale heights from the midplane, and the gas velocity becomes supersonic only at $\left|z\right| > 6 \, H$.

In general, disk winds are launched far from the midplane, where gas densities are lower, and this greatly hinders the launching of particles, in several ways.

First, very large particles will not even exist at $\left|z\right| > 4.5 \, H$, instead settling to the disk midplane.
If the gas has a scale height $H$ but particles have a scale height $H_{\rm d} < H$, then at the base of the disk wind at $z = 4.5 \, H$, the dust-to-gas ratio is reduced by a factor $f_{\rm dep} = 
\exp \left[ -(z/H_{\rm d})^2 / 2 +(z/H)^2 / 2 \right]$
$=\exp \left[ -(z/H)^2 \, \left( (H/H_{\rm d})^2 - 1 \right) / 2 \right]$ $=\exp \left[ -10.1 \left( (H/H_{\rm d})^2 - 1 \right) \right]$.
If the dust scale height is reduced by even a factor of 2 compared to the gas scale height ($H_{\rm d} = 0.5 H$), then the dust-to-gas ratio in the disk wind is reduced by a factor $\exp(-30.3)$, i.e., by 13 orders of magnitude. 
The scale height $H_{\rm d}$ of particles is related to the scale height of the gas, $H$, through the relation
\begin{equation}
\frac{H_{\rm d}}{H} = 
\frac{ \alpha^{1/2} / {\rm St}_0 }
{ \left[ 1 + \alpha / {\rm St}_0^2 \right]^{1/2} },
\end{equation}
where $\alpha$ is the turbulence parameter and the Stokes number ${\rm St}_0 = \Omega \, t_{\rm stop}$ is the product of the Keplerian orbital frequency $\Omega$ and the aerodynamic stopping time $t_{\rm stop} = \rho_{\rm s} a / (\rho_{\rm g} C)$, evaluated here at the midplane
\citep{DubrulleEtal1995}.
Effectively, particles with $\alpha^{1/2} / {\rm St}_{0} < 0.6$, i.e., those larger than about centimeter-sized (assuming $\Sigma \approx 1000 \, {\rm g} \, {\rm cm}^{-2}$ and $\alpha \approx 10^{-4}$), have settled out of the layers launched by disk winds.

Second, it is well established but underappreciated that even if particles exist in the surface layers launched in disk winds, that does not mean they will be launched along with the gas.
Particles are accelerated upward only if the acceleration due to gas drag, $\Delta V / t_{\rm stop}$, exceeds the downward gravitational acceleration, $\Omega^2 \, z$.
Here $\Delta V$ is the relative velocity between particle and gas, which cannot exceed the gas velocity.
This means that only particles smaller than a maximum radius $a_{\rm max}$ can be launched from a height $z$ above the midplane, where
\begin{equation}
a_{\rm max} = \frac{ \Sigma }{ \sqrt{2\pi} \rho_{\rm s} \, (\left|z\right| / H) } \, \frac{ \rho_{\rm g}(z) }{ \rho_{\rm g}(0) } \, \frac{ V_{\rm g} }{ C }.
\end{equation}
In the simulations of \citet{BaiStone2013}, at a height $\left|z\right| = 4.6 \, H$, just above the base of the disk wind, the gas velocity is $\approx 0.1 \, C$, and $\rho_{\rm g}(z) / \rho_{\rm g}(0) \approx 3 \times 10^{-5}$ [$\approx \exp \left( -4.6^2 / 2 \right)$], yielding $a_{\rm max} = 0.9 \, \mu{\rm m}$ for $\Sigma \approx 1000 \, {\rm g} \, {\rm cm}^{-2}$.
Within the wind, the product of gas density and velocity is constant (reflecting a uniform mass flux), so it is only more difficult to launch particles from higher in the disk wind. In general, only sub-micron particles can be launched in disk winds.

Very similar conclusions were reached by \citet{BansKonigl2012} and  \citet{GiacaloneEtal2019}.
The latter provided a formula for $a_{\rm max}$ based on self-similar models they constructed for the disk, presumed to apply between inner and outer disk radii $r_{\rm in} \sim 0.1 \, {\rm AU}$ and $r_{\rm out} \sim 10^2 \, {\rm AU}$:
\begin{equation}
a_{\rm max} = 0.35 \, \left( \frac{ \dot{M}_{\rm out} }{ 10^{-8} \, M_{\odot} \, {\rm yr}^{-1} } \right) \, \left( \frac{ M_{\star} }{ M_{\odot} } \right)^{-1} \, \left( \frac{ T }{ 200 \, {\rm K} } \right)^{1/2} \, \left( \frac{ z/r }{ 0.06 } \right)^{-1} \, \left( \frac{ \ln ( r_{\rm out} / r_{\rm in} ) }{ \ln 10^3 } \right)^{-1} \, \left( \frac{ r }{ 1 \, {\rm AU} } \right)^{-1/4} \, \mu{\rm m}.\label{eq:launch}
\end{equation}
Here $\dot{M}_{\rm out}$ is the mass flux carried by the disk wind, $M_{\star}$ is the stellar mass, and $z/r$ and $T$ are the disk flaring and gas temperature at the base of the wind.
Again, for typical conditions, particles larger than micron-sized are not launched by disk winds.

Third, even if atypical conditions do allow launching of particles, the particles generally will not acquire very large velocities.
Particles remain dynamically coupled to the gas, and roughly share its velocity, only until the gas density drops to the point that the aerodynamic stopping time $t_{\rm stop} = \rho_{\rm g} C / (\rho_{\rm s} a)$ [which is the timescale to exchange momentum with the gas], exceeds the timescale over which the gas is changing velocity, which is roughly the orbital timescale, $\Omega^{-1}$.
That is, particles of radius $a$ remain coupled only until the gas density drops below a value $\rho_{\rm g} < \rho_{\rm s} a / H$.
If the disk wind carries a mass flux $\dot{M}_{\rm out}$ from an area $A$, the velocity of the gas is $V = \dot{M}_{\rm out} / (\rho_{\rm g} A)$, and the velocity particles will acquire before decoupling from the gas is no greater than 
\begin{equation}
V_{\rm max} \approx \frac{ \dot{M}_{\rm out} }{ A } \, \frac{ H }{ \rho_{\rm s} a}.
\end{equation}
Assuming $\dot{M}_{\rm out} \sim 10^{-8} \, M_{\odot} \, {\rm yr}^{-1}$, $A \sim \pi r^2$, and $H/r \approx 0.05$, we find $V_{\rm max} \sim 60 \, (a / 1 \, \mu{\rm m})^{-1} \, {\rm km} \, {\rm s}^{-1}$ at $r \sim 0.1$ AU, and is an order of magnitude smaller at 1 AU. 
Micron-sized grains remain coupled to the gas long enough to achieve escape velocity, but particles $\sim 100 \, \mu{\rm m}$ in radius decouple from the gas before they are accelerated to even $\sim 1 \, {\rm km} \, {\rm s}^{-1}$.
They drop out of the gas after a very short trip in the wind.

We note that changes in velocity on the order of $1 \, {\rm km} \, {\rm s}^{-1}$ or less will not change the orbital radius of particles significantly.
To acquire enough orbital energy to reach 3 AU (where chondrites might form), a particle starting at $r_0$ would have to see its orbital velocity increase by 
\begin{equation}
\Delta V = \left( \frac{G M_{\odot}}{ r_0 } \right)^{1/2} \, \left[ \left( \frac{ 6 \, {\rm AU} }{ r_0 + 3 \, {\rm AU} } \right)^{1/2} - 1 \right].
\end{equation}
To travel from $r_0 = 1 \, {\rm AU}$ to 3 AU requires $\Delta V = 7 \, {\rm km} \, {\rm s}^{-1}$, and to travel from $r_0 = 0.1 \, {\rm AU}$ to 3 AU requires $\Delta V = 37 \, {\rm km} \, {\rm s}^{-1}$. 
A change in orbital velocity $\sim 1 \, {\rm km} \, {\rm s}^{-1}$ alters the orbital radius by only a few percent.


In general, disk winds will contain particles no larger than micron-sized \citep{GiacaloneEtal2019}.
For particles larger than about $1 \, \mu{\rm m}$ in radius, gravitational acceleration will either immediately exceed the acceleration due to gas drag, preventing the particles from launching; or it will exceed the gas drag before particles can be accelerated by more than about $1 \, {\rm km} \, {\rm s}^{-1}$, which will not allow the particles to reach significantly greater heliocentric distances.
Previous models have been more optimistic about the ability of disk winds to launch particles because they arbitrarily start the particles with large vertical velocities $>$ tens of km/s \citep{Shang1998thesis,Hu2010,LiffmanEtal2016,YangEtal2024}.
In simulations that do not make this unjustified assumption, particles either are barely launched, returning to the same heliocentric distance in the disk \citep{SalmeronEtal2011}, or are not launched at all \citep{BansKonigl2012,GiacaloneEtal2019}.

To illustrate these points, we have carried out numerical simulations of the trajectories of particles participating in a disk wind.
We initialize particles at the base of the disk wind at $z \approx 4.5 H$, with the local gas velocity, then compute the positions of the particles over time, $t$, integrating the accelerations caused by gravity and the drag force from the gas in the disk wind to find the instantaneous velocities and positions.
We use the results of \citet{BaiStone2013} to approximate the gas velocities at each height $z$ as follows:
\begin{eqnarray}
V_{{\rm g}r} & = & 0 - 3 \, C \, \zeta \\
V_{{\rm g}\phi} & = & V_{\rm K} \, \left( 1 - \eta \right) - 2 \, C \, \zeta \\ 
V_{{\rm g}z} & = & 0 + 4 \, C \, \zeta,
\end{eqnarray}
where $\eta \approx 10^{-3}$, $V_{\rm K}$ is the Keplerian orbital velocity, and $\zeta = (z/H - 4.5) / (8 - 4.5)$ for $z \geq 4.5 \, H$, and $\zeta = 0$ for $z \leq 4.5 \, H$.
We approximate the gas density using Equation 2 for $z \leq 4.5 \, H$, and assuming 
\begin{equation}
\rho_{\rm g}(z) = \rho_{\rm g}(z = 4.5 H) \, \frac{ V_{{\rm g}z}(z = 4.5 H) }{ V_{{\rm g}z}(z) }
\end{equation}
for $z \geq 4.5 \, H$, signifying uniform mass flux.
We then calculate the vector components for the acceleration due to the drag force and gravity,
\begin{eqnarray}
a_{r} & = & +\frac{ V_{{\rm g}r} - V_{{\rm p}r} }{ t_{\rm stop} }
-\frac{ G M_{\odot} }{ \left( r^2 + z^2 \right)^{3/2} } \, r 
\\
a_{\phi} & = & 
+\frac{ V_{{\rm g}\phi} - V_{{\rm p}\phi} }{ t_{\rm stop} }
\\
a_{z} & = & +\frac{ V_{{\rm g}z} - V_{{\rm p}z} }{ t_{\rm stop} }
-\frac{ G M_{\odot} }{ (r^2 + z^2)^{3/2} } \, z, 
\end{eqnarray}
where again $t_{\rm stop} = \rho_{\rm s} a / (\rho_{\rm g} \, C)$.
We integrate over time $t$, updating position and velocity using a fourth-order Runge-Kutta integrator with variable timestep to integrate $\vec{r}(t + \Delta t) = \vec{r}(t) + \vec{v} \times \Delta t$ and $\vec{v}(t + \Delta t) = \vec{v}(t) + \vec{a} \times \Delta t$.
We integrate for a maximum of 50 years.

\begin{figure}[ht]
\centering
\includegraphics[width=\linewidth]{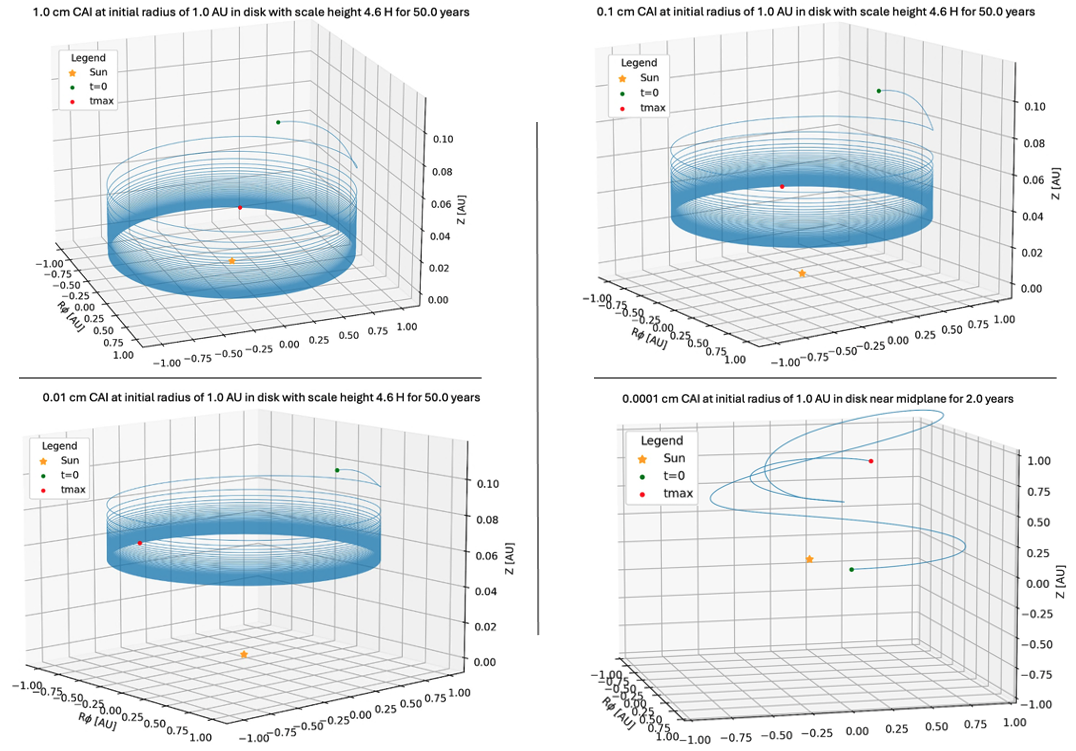}
\caption{
Trajectories (in a fixed frame) of particles with various radii (1 cm, 1 mm, $100 \, \mu{\rm m}$, and $1 \, \mu{\rm m}$, in disk winds at 1 AU as modeled by \citet{BaiStone2013}.
The three larger particles are started at a height $z = 4.6 H$ and orbit the star but eventually settle below the $z = 4.5 H$ height from which winds are launched.
The $1 \, \mu{\rm m}$ particle is placed in a disk with surface density $10\times$ larger than the others and is started at the midplane, and is marginally able to escape.
These simulations confirm expectations \citep[e.g.,][]{GiacaloneEtal2019} that only micron-sized particles can be launched in disk winds, and $100 \, \mu{\rm m}$-sized particles cannot.
}
\label{fig:fling}
\end{figure}

Representative trajectories are shown in {\bf Figure~\ref{fig:fling}}.
The first three plots show particles with radii of 1 cm, 1 mm, and $100 \, \mu{\rm m}$, initialized at $r = 1$ AU and $z = 4.6 \, H$ in a minimum-mass solar nebula disk with $\Sigma = 1700 \, {\rm g} \, {\rm cm}^{-2}$ at 1 AU.
All three particles orbit the Sun but settle vertically within decades to below the $z = 4.5 \, H$ height from which winds are launched.
In the last panel, a grain with radius $a = 1 \, \mu{\rm m}$ is placed near the midplane of a disk with $\Sigma = 1.7 \times 10^4 \, {\rm g} \, {\rm cm}^{-2}$.
This smaller particle, in a denser disk, is likely to be launched in the wind and escape. 
These results confirm the calculations above and the expectations from \citet{BansKonigl2012} and \citet{GiacaloneEtal2019}.

In {\bf Table~\ref{table:fling}}, we consider grains of various radii $a$ placed at two distance from the Sun, $r_0 = 0.1$ AU or $r_0 = 1$ AU.
We list the factor $f_{\rm dep}$ by which particles are depleted at the base of the disk wind (at $z = 4.5 \, H$), relative to the uniformly-mixed case.
Particles microns-sized and smaller are well-mixed, while centimeter-sized particles have settled and are very depleted from these heights.
We list the height $z$ (in scale heights $H$) above the midplane at which the particles dynamically decouple from the gas.
Where this exceeds $4.5 H$, we also list the velocity relative to the star, in units of the Keplerian velocity $V_{\rm K}$ (30 km/s at 1 AU, 95 km/s at 0.1 AU).
Although some of these particles are formally `launched' by the disk wind, they all land back on the disk at a radius $r_{\rm max}$ that we list. 
None of them travels more than $\approx 0.1$ AU from where they started. 

It is possible to launch particles in disk winds if they are small enough (micron-sized), and if the mass-loss rates are high enough, but for realistic disk winds, particles do not travel more than about 0.1 AU, or are not launched at all. 
{Another important consideration is that if particles can be launched in a disk wind, they are unlikely to be subjected to any SEP flux at all, as SEPs will naturally follow the ``interplanetary" magnetic field lines connected to the star and escaping to infinity.
In what follows, we ignore this consideration and assume particles in disks winds will be subject to the same SEP flux $F(> 10 \, {\rm MeV}) = \Phi \times 100 \, {\rm cm}^{-2} \, {\rm s}^{-1}$.}

\begin{deluxetable*}{cccccc}
\tablecaption{CAI Trajectories in Disk Winds}
\label{table:fling}
\tablehead{
\colhead{$r_{0}$} &
\colhead{\textit{a}} &
\colhead{$f_{\rm dep}$ } &
\colhead{Launch $z/H$} &
\colhead{Launch $V / V_{\rm K}$} &
\colhead{$r_{\rm max}$ (AU)}
}
\startdata
\phantom{r = 0.1 AU}& 1 $\mu{\rm m}$   & $9.98 \times 10^{-1}$  & 20.5 & 1.31  & 0.1896 \\
\phantom{r = 0.1 AU}& 10 $\mu{\rm m}$  & $9.80 \times 10^{-1}$  & 8.0  & 0.83  & 0.1008 \\
0.1 AU & 100 $\mu{\rm m}$ & $8.15 \times 10^{-1}$  & 6.8  & 0.98  & 0.1013 \\
\phantom{r = 0.1 AU}& 1 mm  & $1.29 \times 10^{-1}$  & (4.5)  & - 
& 0.1006 \\
\phantom{r = 0.1 AU}& 1 cm  & $8.53 \times 10^{-10}$ & (3.8)  & - 
& 0.1001 \\
\hline
& 1 $\mu{\rm m}$   & $9.38 \times 10^{-1}$  & 10.3 & 0.75 & 1.041 \\
& 10 $\mu{\rm m}$  & $5.27 \times 10^{-1}$  & 8.0 & 0.90  & 1.052 \\
1.0 AU & 100 $\mu{\rm m}$ & $1.55 \times 10^{-3}$  & 4.6 & 1.00 & 1.014 \\
& 1 mm  & $1.76 \times 10^{-29}$  & (4.0) & 
- 
& 1.007 \\
& 1 cm  & $\approx 0$  & (3.3) & 
- 
& 1.000 \\
\enddata
\end{deluxetable*}

\bigskip

\section{Live $\clthreesix$ in the early Solar System}
\label{sec:chlorine}

\subsection{Initial $(\clratio)_0$ ratios}

The SLR $\clthreesix$ is a very short-lived isotope with half-life $0.301 \pm 0.002$ Myr (National Nuclear Data Center, Brookhaven National Laboratory).
It decays either to ${}^{36}{\rm Ar}$ (branching ratio 98\%) or ${}^{36}{\rm S}$ (branching ratio 2\%).
The first evidence for its one-time existence was from ${}^{36}{\rm S}$ excesses in sodalite in a CAI from the Ningqiang (C-ung) carbonaceous chondrite \citep{LinEtal2005}.
Four analyses of sodalite showed initial $(\clratio)_0 \approx 5 \times 10^{-6}$ at the time of the sodalite formation, but with significant scatter, $(\clratio)_0$ ranging from $(4.6 \pm 0.6)$ to $(11 \pm 2) \times 10^{-6}$.
This analysis assumed a relative sensitivity factor (RSF) of sulfur vs.\ chlorine of 0.58. 
Subsequent analyses assume RSF $= 0.83$, and to compare to those, the inferred average value should be $(\clratio)_0 \approx 3.7 \times 10^{-6}$ \citep{HsuEtal2006}.

Sodalite is a product of aqueous alteration on the parent body, so the value of $(\clratio)_0$ pertains to the time of sodalite formation, which might be expected to match the time of formation inferred from Al-Mg systematics.
An upper limit $(\alratio)_0 < 7 \times 10^{-6}$ was found \citep{LinEtal2005}, which (assuming $\alratio = 5.23 \times 10^{-5}$ at $t\!=\!0$, the time of most CAI formation; see Equation~\ref{eq:AlMg}) implies the alteration took place $t > 2.1$ Myr later.
If it is hypothesized that none of the $\clthreesix$ was created within the solar nebula, then it would have to predate the solar nebula, and the initial value in the Solar System at $t\!=\!0$ would have to have been $(\clratio)_{\rm SS} > (3.7 \times 10^{-6}) \, \exp \left( + (2.1 \, {\rm Myr} / 0.434 \, {\rm Myr}) \right)$ $= 5 \times 10^{-4}$.

Similar analyses in other inclusions lead to similar results, as compiled in {\bf Table~\ref{table:chlorine}}.
From ${}^{36}{\rm S}$ excesses in the {\it Pink Angel} CAI and a chondrule in Allende (CV3), values $(\clratio)_0 = (3.7 \pm 0.8) \times 10^{-6}$ and $(4.2 \pm 1.0) \times 10^{-6}$ were inferred \citep{HsuEtal2006}.
Subsequently $\clthreesix$ was also inferred from excesses of ${}^{36}{\rm Ar}$, but at levels implying an initial $(\clratio)_0$ about 200 times lower \citep{TurnerEtal2013}.
This implies that much of the $\clthreesix$ decayed in the fluid and the Ar was not retained \citep{TurnerEtal2013}.
{\it Pink Angel} is characterized by  $(\alratio)_0 < 1.7 \times 10^{-6}$,
implying formation after 3.5 Myr.

From ${}^{36}{\rm S}$ excesses in a CAI ({\it CAI\#2}) in Allende (CV3), a value $(\clratio)_0 = (1.4 \pm 0.3) \times 10^{-6}$ was inferred \citep{UshikuboEtal2007}.
Upper limits $(\alratio)_0 < 4.4 \times 10^{-7}$ and $< 2.1 \times 10^{-7}$ in two sodalite domains \citep{UshikuboEtal2007} imply formation after 5.7 Myr.
From ${}^{36}{\rm S}$ excesses in  wadalite in CAIs in Allende (CV3),  a value $(\clratio)_0 = (18.1 \pm 1.3) \times 10^{-6}$ was inferred \citep{JacobsenEtal2011}.
In co-existing grossular, $(\alratio)_0 < 3.9 \times 10^{-6}$ \citep{JacobsenEtal2011}, implying formation after 2.7 Myr.

From measurements of ${}^{36}{\rm S}$ excesses in sodalite in {\it Curious Marie}, a CAI in Allende (CV3), a value $(\clratio)_0 = (23 \pm 6) \times 10^{-6}$ was inferred \citep{TangEtal2017}.
In the same CAI, a $(\alratio)_0 = (6.2 \pm 0.9) \times 10^{-5}$ {\it model} (not internal) isochron was inferred \citep{TangEtal2017}.
This ratio was inferred by assuming $\delta^{26}{\rm Mg}^{*} = 0\permil$ at ${}^{27}{\rm Al}/{}^{24}{\rm Mg} = 0$, and $\delta^{26}{\rm Mg}^{*} = +46\permil$ at an assumed bulk composition ${}^{27}{\rm Al}/{}^{24}{\rm Mg} = 95$.
Repeating their analysis, we find that two outliers spots (sodalite \#1 and sodalite + nepheline \#2) are not consistent with a uniform $\delta^{26}{\rm Mg}^{*}$; the averages of the remaining points are ${}^{27}{\rm Al}/{}^{24}{\rm Mg} = 128$ and $\delta^{26}{\rm Mg}^{*} = +44.4\permil$, yielding $(\alratio)_0 = 4.8 \times 10^{-5}$, with an uncertainty likely to be tens of percent.
Moreover, although \citet{TangEtal2017} interpreted this to be the $({}^{26}{\rm Al}/{}^{27}{\rm Al})_0$ value in the CAI when the alteration took place, we do not.
A very straightforward interpretation is that {\it Curious Marie} formed as a fine-grained CAI in the nebula at $t=0$ with a canonical value like other CAIs, comprising mostly melilite and grossular, but possibly incorporating hibonite, corundum, or other Al-rich, Mg-poor phases, yielding a bulk abundance ${}^{27}{\rm Al}/{}^{24}{\rm Mg} \approx 120$.
It then was incorporated into Allende at $\approx 3$ Myr, and experienced continued ${}^{26}{\rm Al}$ decay. 
At $\approx 6$ Myr, after all ${}^{26}{\rm Al}$ had decayed, aqueous alteration on the Allende parent body dissolved all the Mg in the minerals, which would then have an average value $\delta^{26}{\rm Mg}^{*} = +45\permil$, yielding a homogeneous fluid with $\delta^{26}{\rm Mg}^{*} = +45\permil$.
The fluid in the CAI effectively did not exchange with any other fluid, and a fraction of the Mg in it re-precipitated as sodalite and nepheline before the fluid and the rest of the Mg in it was lost. 
Precipitation (or condensation in the nebula) could yield $\delta^{25}{\rm Mg} < 0$.
This scenario is consistent with the homogeneous values of $\delta^{26}{\rm Mg}^{*}$ throughout {\it Curious Marie}, which only signify when it formed, sometime soon after $t = 0$.

In addition to these published data, there is one more analysis of note, in a conference abstract by \citet{FengEtal2012lpsc}.
From ${}^{36}{\rm S}$ excesses in lawrencite (${\rm FeCl}_2$) in the MS17 clast (classified as originating in an EL3 chondrite) in Almahata Sitta, they inferred clustering of data into three isochrons with different 
$(\clratio)_0$: $(2.5 \pm 0.2) \times 10^{-6}$, $(11 \pm 0.8) \times 10^{-6}$, and $(\clratio)_0 = (94 \pm 8) \times 10^{-6}$.
These appear to be the ratios in reservoirs pertinent to the enstatite chondrite formation region.

Unlike other short-lived radionuclides like ${}^{26}{\rm Al}$ or ${}^{10}{\rm Be}$, evidence for $\clthreesix$ is found only in minerals that are the products of fluid alteration on the parent body.
Sodalite and wadalite are products of aqueous alteration on the CV parent body. 
Lawrencite may be a primary phase, formed from condensation from nebula gas, but it may also be the product of metasomatism on enstatite chondrite parent bodies, like the mineral djerfisherite \citep[e.g.,][]{WilburEtal2023}.
As such, the $\clthreesix$ does not need to exist at the birth of the solar system, but could be created shortly before the fluid alteration, and in relatively small quantities.

Indeed, the data are not consistent with {zero} $\clthreesix$ being created in the solar nebula. 
At face value, the CAI {\it Curious Marie} implies $(\clratio)_0 = 2 \times 10^{-5}$ at $t\!=\!0$, whereas the values implied by the other CAIs are much higher. 
Even discounting {\it Curious Marie}, the wide range of implied $(\clratio)_{\rm SS}$, from $> 5 \times 10^{-4}$ to $> 0.7$, seems to rule out a single uniform value at $t\!=\!0$. 
Aside from this, ratios $> 10^{-2}$ are inconsistent with the values expected in the Sun's molecular cloud contaminated by supernovae \citep{JacobsenEtal2011,DeschEtal2023pp7}.
We therefore assign no chronological significance to $\clthreesix$, and instead interpret the data to mean $\clthreesix$ was created at late times within the solar nebula, e.g., by particle irradiation of nebular materials, as suggested by \citet{JacobsenEtal2011}.
At the various times in the nebula, it appears that $\clthreesix$ was heterogeneous, with $\clratio \approx (1 - 20) \times 10^{-6}$.

\begin{deluxetable}{lcrrrr}
\tabletypesize{\scriptsize}
\tablewidth{0pt}
\tablecaption{Chlorine 36 Data}
\label{table:chlorine}
\tablehead{
\colhead{Sample} & \colhead{Ref.} & \colhead{$(\clratio)_0$} & \colhead{$(\alratio)_0$} & \colhead{Formation} & \colhead{Hypothesized} \\
$\,$ & $\,$ & $\,$ & $\,$ & time (Myr) & $(\clratio)_{\rm SS}$ 
}
\startdata
Ningqiang sodalite & 1 & $\approx 3.7 \times 10^{-6}$ & $< 7 \times 10^{-6}$ & $> 2.1$ & $> 5 \times 10^{-4}$ \\
Allende (Pink Angel)  & 2 & $(4.2 \pm 1.0) \times 10^{-6}$ & $< 1.7 \times 10^{-6}$ & $> 3.5$ & $> 1 \times 10^{-2}$ \\
Allende sodalite & 3 & $(1.4 \pm 0.3) \times 10^{-6}$ & $< 4.4 \times 10^{-7}$ & $> 5.7$ & $> 0.7$ \\
Allende wadalite & 4 & $(18.1 \pm 1.3) \times 10^{-6}$ & $< 3.9 \times 10^{-6}$ & $> 2.7$ & $> 9 \times 10^{-3}$ \\
Allende (Curious Marie) & 5 & $(23 \pm 6) \times 10^{-6}$ & $(6.2 \pm 0.9) \times 10^{-5}$ & $\approx 0$ & $2 \times 10^{-5}$ \\
EL3 clast lawrencite & 6 & $(94 \pm 8) \times 10^{-6}$ & - & $> 1.8$? & $> 9 \times 10^{-3}$ \\
\enddata
\tablecomments{References: 1. \citet{LinEtal2005}; 2. \citet{HsuEtal2006}; 3. \citet{UshikuboEtal2007}; 4. \citet{JacobsenEtal2011}; 5. \citet{TangEtal2017}; 6. \citet{FengEtal2012lpsc}.}
\end{deluxetable}

\subsection{Production of $\clthreesix$ in the solar nebula}

\citet{JacobsenEtal2011} suggested production of $\clthreesix$ by irradiation of HCl-bearing ices in the solar nebula, primarily through the reaction ${}^{37}{\rm Cl}(p,pn){}^{36}{\rm Cl}$.
\citet{DeschEtal2011lpsc} suggested an additional channel, production by irradiation of solar nebula gas by the reaction ${}^{38}{\rm Ar}(p,ppn){}^{36}{\rm Cl}$.
\citet{BowersEtal2013} considered a variety of reactions and SEP energy spectra, and concluded that reactions on S such as ${}^{33}{\rm S}(\alpha,p){}^{36}{\rm Cl}$,
${}^{34}{\rm S}(\alpha,pn){}^{36}{\rm Cl}$ and
${}^{34}{\rm S}({}^{3}{\rm He},p){}^{36}{\rm Cl}$ should dominate production of $\clthreesix$, for near-chondritic abundances of the target nuclei.
In particular, we focus on the reaction ${}^{34}{\rm S}(\alpha,pn){}^{36}{\rm Cl}$, which for a gradual flare irradiation (which dominates over long periods of time), provides 45\% of the $\clthreesix$ \citep{BowersEtal2013}.
This reaction is driven mostly by $\alpha$ particles with energies $5-10 \, {\rm MeV}$ per nucleon, for which the cross section is $\sigma \approx 100-500 \, {\rm mbarn}$.

It is straightforward to show that irradiation of solar nebula gas cannot lead to sufficient $\clthreesix$.
For the SEP energy spectrum in Equation~\ref{eq:SEPflux}, the flux of $\alpha$ particles with energies between 5 and 10 MeV per nucleon is a fraction $0.03 \times 0.36 \times F(E > 10 \, {\rm MeV})$, assuming $\alpha$ particles comprise 3\% of the flux {(as for impulsive flares)}.
The number of $\alpha$ particles with the right energy, intercepted by the disk in an area $A$ in a time $\Delta t$, is 
{\begin{equation}
\#{\rm SEPs} = (0.03) \times (0.36) \, F_{\rm SEP} \times A \times \left( \Delta t \right).
\end{equation} }
The probability that a relevant SEP will undergo the reaction ${}^{34}{\rm S}(\alpha,pn){}^{36}{\rm Cl}$ is \begin{equation}
{\cal P}_{{\rm SEP -> 36Cl}} = n_{\rm 34S} \, \sigma \, \frac{ \Sigma_{\rm stop} }{ 1.4 \, m_{\rm H} \, n_{\rm H} },
\end{equation}
where we take $\sigma = 300$ mbarn \citep{BowersEtal2013}.
The product of these two gives the number of $\clthreesix$ nuclei produced in this part of the disk, in a time $\Delta t$.
These nuclei would be mixed with Cl nuclei throughout the disk {over an annulus with area $A$ and surface density $\Sigma$.:}
\begin{equation}
\# {}^{35}{\rm Cl} = \frac{ (\Sigma) \, A }{ 1.4 m_{\rm H} } \, \frac{ n_{\rm 35Cl} }{ n_{\rm H} }.
\end{equation}
Taking the ratio, and setting $\Delta t = \tau = 0.43$ Myr, the mean-life of $\clthreesix$, we find the steady-state ratio in the disk: 
\begin{equation}
\frac{ {}^{36}{\rm Cl} }{ {}^{35}{\rm Cl} } = \frac{ \# {\rm SEPs} \times {\cal P}_{\rm SEP -> 36Cl} }{ \# {}^{35}{\rm Cl} } = 
\left( 0.03 \right) \, \left( 0.36 \right)
F_{\rm SEP} \,
\frac{ n_{\rm 34S} }{ n_{\rm 35Cl} } \,
\frac{\Sigma_{\rm stop}}{ \Sigma } \, \sigma  \, \tau.
\label{eq:slr}
\end{equation}
Using the present-day flux from the Sun at 1 AU, $F_{\rm SEP} = 100 \, {\rm cm}^{-2} \, {\rm s}^{-1}$, $A = 4.9 \times 10^{27} \, {\rm cm}^{2}$, $\sigma = 300 \, {\rm mbarn}$, $\Sigma_{\rm stop} \approx 0.075 \, {\rm g} \, {\rm cm}^{-2}$ for this energy range, $\Sigma \approx 1000 \, {\rm g} \, {\rm cm}^{-2}$, and taking $n_{\rm 34S} / n_{\rm 35Cl} = 4.7$ \citep{Lodders2003}, we find {$\clratio \approx 1.5 \times 10^{-15}$.} 
The flux of SEPs from the early Sun was perhaps $10^4$ times greater than the present-day flux, but this would still produce a $\clratio$ ratio {six} orders of magnitude smaller than the observed $\clratio \sim 10^{-5}$.

Production of short-lived radionuclides in the solar nebula is limited by the fact that almost all of the energy of SEPs is consumed ionizing gas molecules, and not in inducing nuclear reactions; and by the fact that the $\clthreesix$ so produced is diluted by solar abundances of Cl. 
Similar arguments apply even to the production of ${}^{10}{\rm Be}$ (\S~\ref{sec:beten}).

A more promising avenue for production of $\clthreesix$ is direct irradiation of small (radius $R \approx 1$ mm) solid material (especially if it is depleted in Cl) after gas has dissipated from the disk.
{Solids could be directly exposed to the stellar wind and SEPs in it.}
We note that there is little observed decrease in X-ray fluxes of protostars over their first few Myr \citep{WolkEtal2005}.
In a time $\tau$, a particle intercepts a number $0.05 \, F_{\rm SEP} \, \pi R^2 \, \tau$ of relevant SEPs ($\alpha$ particles with energies 5-10 MeV/nucleon).
The probability of a relevant SEP causing a reaction ${}^{34}{\rm S}(\alpha,pn){}^{36}{\rm Cl}$ is $n_{\rm 34S} \, \sigma \, (4 R /3)$.
The total number of ${}^{35}{\rm Cl}$ nuclei in the particle is $n_{\rm 35Cl} \, (4\pi R^3 / 3)$.
Therefore the steady-state abundance of $\clthreesix$ in the solid material is
\begin{equation}
\frac{ {}^{36}{\rm Cl} }{ {}^{35}{\rm Cl} } = (0.03) \times (0.36) \, F_{\rm SEP} \, \sigma \, \tau \, \frac{ n_{\rm 34S} }{ n_{\rm 35Cl} },
\end{equation}
which equals {\Steve $2 \times 10^{-8}$}, assuming an SEP flux {\Steve $\Phi \approx 1 \times 10^3$} times more active than the present-day Sun, and assuming a CI-like composition for the solid material.
However, we expect the S/Cl to greatly exceed chondritic values in the solid.
While S can condense into many phases, including the common mineral troilite (FeS), essentially all of the Cl should be in the gas phase, as HCl.
Even though refractory Cl-bearing minerals of Cl could exist in the nebula (e.g., sodalite is stable below 948 K; \citet{Lodders2003}), 
the sodalite and wadalite in the CAIs in Table~\ref{table:chlorine} appear to be exclusively secondary minerals.
Instead, Cl is likely to remain in the gas as HCl, condensing only as ${\rm HCl}\cdot3{\rm H}_2{\rm O}$, at temperatures $\approx 160$ K (at $10^{-4}$ bar total pressure) to $\approx 140$ K (at $10^{-6}$ bar)
\citep{ZolotovMironenko2007lpsc}.

During the disk phase, the temperatures in the enstatite chondrite-forming region are predicted to be $> 180$ K during the disk stage \citep{DeschEtal2018}, and solids at 1 AU will assume even higher blackbody temperatures $\approx 240$ K after the gas dissipates, assuming the Sun's luminosity is $0.5 \, L_{\odot}$ at 6 Myr \citep{BaraffeEtal2002}.
The S/Cl ratio in solids would remain high.
In the carbonaceous chondrite-forming region at 3-4 AU, pressures are expected to be $< 10^{-6}$ bar and temperatures are expected to be $\approx 93-118$ K during the disk phase \citep{DeschEtal2018}, so HCl hydrate would condense.
However, after gas dissipates and pressures drop to $\ll 10^{-6}$ bar, allowing irradiation of solid materials, the blackbody temperatures of solids will be 
$\approx 166$ K, exceeding the condensation temperatures $< 140$ K.
These temperatures allow the condensation of H$_2$O ice but not HCl hydrates.
The S/Cl ratios in the solids would remain high during irradiation in the carbonaceous chondrite-forming region as well.
We assume that the S/Cl ratio in the irradiated solids is several orders of magnitude higher than chondritic, allowing the $\clratio$ ratio in them to far exceed the $(\clratio)_0$ values in Table~\ref{table:chlorine}.

These irradiated S-bearing but Cl-free materials could be accreted by asteroids, forming a reservoir on the surface with $\clratio$ ratios consistent with the values observed in CV and EL chondrites.
Melting of accreted ice (e.g., by impacts) would allow Cl to be mobilized via water and carried to depth on the CV parent body, where it could react with CAI minerals.
Perhaps a similar process could occur on the EL parent body, allowing metasomatism to producing lawrencite.

\subsubsection{Summary}

The one-time existence of $\clthreesix$ at high abundances in some meteorites is a strong indication of SEP irradiation during the early Solar System.
The high abundance, and great variability of $\clratio$ across samples, argues against an origin in the Sun's molecular cloud, and in favor of later production. 
Production of sufficient quantities of $\clthreesix$ is not likely until after the solar nebula gas dissipates, as otherwise almost all of the energy is lost ionizing ${\rm H}_{2}$ molecules, and the resultant $\clthreesix$ is diluted with abundant ${}^{35}{\rm Cl}$.
Irradiation of small solid particles bearing S (e.g., as troilite) but depleted in Cl (e.g., {\Steve 99.9\%} of Cl remains in the gas as HCl) would lead to $\clratio$ ratios exceeding the values observed in alteration products in CV and EL chondrite parent bodies, even assuming SEP fluxes {\Steve $\Phi \approx 10^3$} times the present-day values.
These fluxes would apply soon after the disk gas dissipated, which in the inner solar nebula occurred at $\approx 2.2$ Myr  and in the outer solar nebula occurred after 5 Myr \citep{DeschEtal2018}.
These suggest that the Sun's SEP flux may have remained constant for several Myr after the disk dissipated. 
This is consistent with observations, which show little decrease in X-ray fluxes of protostars over their first few Myr \citep{WolkEtal2005}.

\section{Hibonite grains with cosmogenic Neon}
\label{sec:neon}

As found by \citet{KoopEtal2018}, about 78\% of hibonite grains and 6\% of spinel grains they analyzed contain cosmogenic Ne, with distinctive $\neratio$ ratios.
These were argued to have been pre-accretionary signatures of irradiation by solar cosmic rays (i.e., SEPs) based on a few lines of evidence.
First, the irradiation could not have been on the parent body in a regolith, because of the variation in exposure ages experienced by different grains in the same meteorite.
The pre-accretionary irradiation could not have been from Galactic Cosmic Rays (GCRs) because the concentrations of cosmogenic Ne exceed the amounts that could be produced by GCR irradiation within the lifetime of the nebula (a few Myr), before the parent body (Murchison) accreted.
For example, one of the best-studied inclusions was the $\approx 120 \, \mu{\rm m}$-diameter hibonite inclusion {\it PLAC-21}.
After correcting for GCR exposure in the meteoroid (which was negligible), this grain was found to have an amount of cosmogenic ${}^{21}{\rm Ne}_{\rm cos} = (63.87 \pm 0.66) \times 10^{-15} \, {\rm cm}^{3} \, {\rm STP}$.
For its composition, the GCR production rate of ${}^{21}{\rm Ne}_{\rm cos}$ was calculated to be $30.8 \times 10^{-10} \, {\rm cm}^{3} \, {\rm g}^{-1} \, {\rm STP} \, {\rm Myr}^{-1}$.
Combined with an estimated mass $9.4 \times 10^{-7} \, {\rm g}$, the inferred exposure age would be $22$ Myr, whereas Murchison probably {accreted} within $< 4$ Myr \citep{DeschEtal2018}.
Finally, the $\neratio$ ratios were typically well below the value expected for GCR irradiation, $\approx 0.8$.
For example, the ratio was 0.69 for {\it PLAC-21}, but the production rates from GCR irradiation would have predicted 0.81.
In contrast, \citet{LeyaEtal1998} have calculated the production rates of cosmogenic Ne by SEPs under different scenarios.  
Irradiation of hibonite produces Ne almost exclusively from Al, resulting in Ne with a ratio, $\neratio \approx 0.5$, {distinctly} lower than that produced by GCRs.

SEP irradiation is accepted, but at question is whether this irradiation by SEPs happened within the disk, or requires launching of grains above the disk.
\citet{YangEtal2024} recently concluded, based on a model with questionable assumptions {about the stopping lengths of SEPs}, that irradiation of hibonite grains could not have occurred within the disk.
They then argued for irradiation by grains during 1-10 year excursions above the disk, based on an incorrect model of how particles could be launched.
This still required SEP fluxes orders of magnitude greater than expected for the early Sun {(assuming SEPs can even penetrate the magnetic field lines associated with disk winds)}, and would result in a $\neratio$ different from what was observed in the hibonite grains.
Here we show that hibonite grains receive the required fluence of SEPs during the fraction of the time they spend near disk surfaces, provided $\Phi \approx 1 \times 10^{4}$. 
We demonstrate that the $\neratio$ ratio of this cosmogenic neon is also consistent with such values.

\subsection{Hypothesis \#1: Grains were irradiated in the disk}

The concentrations of cosmogenic Ne in hibonite grains require fluences consistent with exposure to an active early Sun---at least $\sim 10^4$ times as active as the present-day, if typical for protostars---for at least hundreds of years.
This fluence does not need to be acquired all at once, though.
Hibonite grains presumably formed in the first few $\times 10^5$ years of the disk, but were not accreted into CM chondrites like Murchison {until about} 3 Myr \citep{DeschEtal2018}.
This means they could have acquired their cosmogenic Ne intermittently, during the short periods of times they spent near the disk surface and exposed to SEPs.
The relevant questions are: What fluence did hibonite grains experience?  What fluence would they acquire while diffusing vertically through the disk? 
And would such irradiation reproduce the $\neratio$ ratio of their cosmogenic Ne?

The production rates of cosmogenic Ne by proton bombardment of Al, which will dominate production of Ne in hibonite [${\rm CaAl}_{12}{\rm O}_{19}$], have been determined by \citet{TrappitschLeya2013}.
Assuming an SEP flux like that given in Equation~\ref{eq:SEPflux}, with $F(E > 10 \, {\rm MeV}) = 100 \, {\rm cm}^{-2} \, {\rm s}^{-1}$ and $R_0 = 125 \, {\rm MV}$, these production rates are $2.2 \times 10^{-8} \, {\rm cm}^{3} \, {\rm STP} \, {\rm g}^{-1} \, {\rm Myr}^{-1}$ for ${}^{21}{\rm Ne}$, and $4.5 \times 10^{-8} \, {\rm cm}^{3} \, {\rm STP} \, {\rm g}^{-1} \, {\rm Myr}^{-1}$ for ${}^{22T}{\rm Ne}$, where ${}^{22T}{\rm Ne}$ denotes both direct production of ${}^{22}{\rm Ne}$ and contributions from production of ${}^{22}{\rm Na}$, which decays to ${}^{22}{\rm Ne}$ with a half-life of 2.6 yr.
Aluminum comprises a fraction 0.48 of the mass in hibonite.
The {\it PLAC-21} grain has a typical concentration, $6.8 \times 10^{-8} \, {\rm cm}^{3} \, {\rm STP} \, {\rm g}^{-1}$.
Therefore {\it PLAC-21} must have spent a cumulative time 6.4 Myr exposed to a flux like the present-day flux at 1 AU, or 640 yr if the flux were $10^4$ times larger.

Whether hibonite grains experience such fluences depends on the amount of time they spend near the disk surfaces.
We consider hibonite grains that have diffused radially outward from regions $< 1$ AU where temperatures are high enough ($> 1400$ K) for hibonite to be a common condensate, out to beyond 1 AU, where even midplane temperatures are low enough ($< 500$ K after 0.5 Myr; \citet{DeschEtal2018}) that cosmogenic Ne would not be lost as the hibonite grains diffuse in and out of the midplane region.
According to the model of \citet{DeschEtal2018}, the average surface density of the gas in the 1-3 AU region decreases from $\approx 1.6 \times 10^{3} \, {\rm g} \, {\rm cm}^{-2}$ at $t = 0.5$ Myr, to $\approx 0.6 \times 10^{3} \, {\rm g} \, {\rm cm}^{-2}$ at $t = 3.0$ Myr; we take as an average value $\Sigma \approx 1000 \, {\rm g} \, {\rm cm}^{-2}$.
Even particles as large as {\it PLAC-21} (radius $a = 60 \, \mu{\rm m}$) would remain well-mixed with gas throughout this vertical column.
They would have  Stokes numbers ${\rm St} = \sqrt{2\pi} \rho_{\rm s} \, a / \Sigma$ $\approx 5 \times 10^{-5}$, and even with a low turbulence parameter $\alpha \sim 10^{-4}$, they would have scale heights $H_{\rm d}$ indistinguishable from the gas scale height $H$ \citep[\S 2.3;][]{DubrulleEtal1995}.
The time spent at any height $z$ above the midplane therefore is proportional to the density of gas, and the fraction of time they spend within a column density $\Delta \Sigma$ is therefore $(\Delta \Sigma) / \Sigma$.

{Having calculated the fluxes of particles in each energy bin, $dF(E)/dE \, dE$, at each vertical location $z$ in the disk, we use the analog of Equation 19 to calculate the abundances of ${}^{21}{\rm Ne}$ and ${}^{22}{\rm Ne}$ produced by proton spallation of Al atoms within a hibonite grain:
\begin{eqnarray}
\left.
\frac{{}^{21}{\rm Ne}}{{}^{27}{\rm Al}} \right|_{z} = & (0.65) \, \int_{0}^{\infty} \, \frac{d F(E,z)}{dE} \, \sigma_{21}(E) \, dE \, ( \Delta t ) \\
\left.
\frac{{}^{22}{\rm Ne}}{{}^{27}{\rm Al}} \right|_{z} = & (0.65) \, \int_{0}^{\infty} \, \frac{d F(E,z)}{dE} \, \sigma_{22}(E) \, dE \, ( \Delta t ) 
\end{eqnarray}  }
{
These ratios can be multiplied by 404 to convert them into ${\rm cm}^{3}-{\rm STP} \, {\rm g}^{-1}$.}
In spinel (${\rm MgAl}_{2}{\rm O}_{4}$), and overwhelmingly in hibonite (${\rm CaAl}_{12}{\rm O}_{19}$), ${}^{21}{\rm Ne}$ and ${}^{22}{\rm Ne}$ are created by the reactions ${}^{27}{\rm Al}(p,4p3n){}^{21}{\rm Ne}$ and ${}^{27}{\rm Al}(p,4p2n){}^{22}{\rm Ne}$ plus ${}^{27}{\rm Al}(p,3p3n){}^{22}{\rm Na}$.
These cross sections are known \citep{Reedy1992lpsc,SistersonEtal1999lpsc,LeyaEtal1998}.
We use the results of \citet{SistersonEtal1999lpsc} to approximate the cross sections for ${}^{27}{\rm Al}(p,x){}^{21}{\rm Ne}$ and ${}^{27}{\rm Al}(p,x){}^{22T}{\rm Ne}$ as follows:
\begin{equation}
\sigma_{21}(E) = 
\begin{cases}
0 \, {\rm mbarn}, & 
E < 30 \, {\rm MeV} \\
35 \, {\rm mbarn} \, \left( E - 30 \, {\rm MeV} \right) / \left( 25 \, {\rm MeV} \right) & 
30 \, {\rm MeV} \leq E < 55 \, {\rm MeV} \\
35 \, {\rm mbarn} - 13 \, {\rm mbarn} \, \left( E - 55 \, {\rm MeV} \right) / \left( 45 \, {\rm MeV} \right) & 
55 \, {\rm MeV} \leq E < 100 \, {\rm MeV} \\
22 \, {\rm mbarn} & 
100 \, {\rm MeV} \leq E
\end{cases}
\end{equation}
and
\begin{equation}
\sigma_{22T}(E) = 
\begin{cases}
0 \, {\rm mbarn}, & 
E < 30 \, {\rm MeV} \\
80 \, {\rm mbarn} \, \left( E - 30 \, {\rm MeV} \right) / \left( 15 \, {\rm MeV} \right) & 
30 \, {\rm MeV} \leq E < 45 \, {\rm MeV} \\
80 \, {\rm mbarn} - 52 \, {\rm mbarn} \, \left( E - 45 \, {\rm MeV} \right) / \left( 25 \, {\rm MeV} \right) & 
45 \, {\rm MeV} \leq E < 70 \, {\rm MeV} \\
28 \, {\rm mbarn} & 
70 \, {\rm MeV} \leq E.
\end{cases}
\end{equation}
By integrating these cross sections over energy, weighted by an unattenuated SEP flux given by Equation~\ref{eq:SEPflux}, we find the average cross section for production of ${}^{21}{\rm Ne}$ is 21 mbarn, and that of ${}^{22}{\rm Ne}$ is 43 mbarn.
The ratio produced by such irradiation is $\neratio = 0.52$, in excellent agreement with previous findings \citep{KoopEtal2018}, including those of 
\citet{TrappitschLeya2013}, who found the production rates of ${}^{21}{\rm Ne}$ and ${}^{22T}{\rm Ne}$ from Al were $2.2$ and $4.5$ $\times 10^{-8} \, {\rm cm}^{3} \, {\rm STP} \, {\rm g}^{-1} \, {\rm Myr}^{-1}$, respectively, yielding $\neratio = 0.49$, for the SEP spectrum above.

\begin{figure}[ht]
\centering
\includegraphics[width=0.54\linewidth]{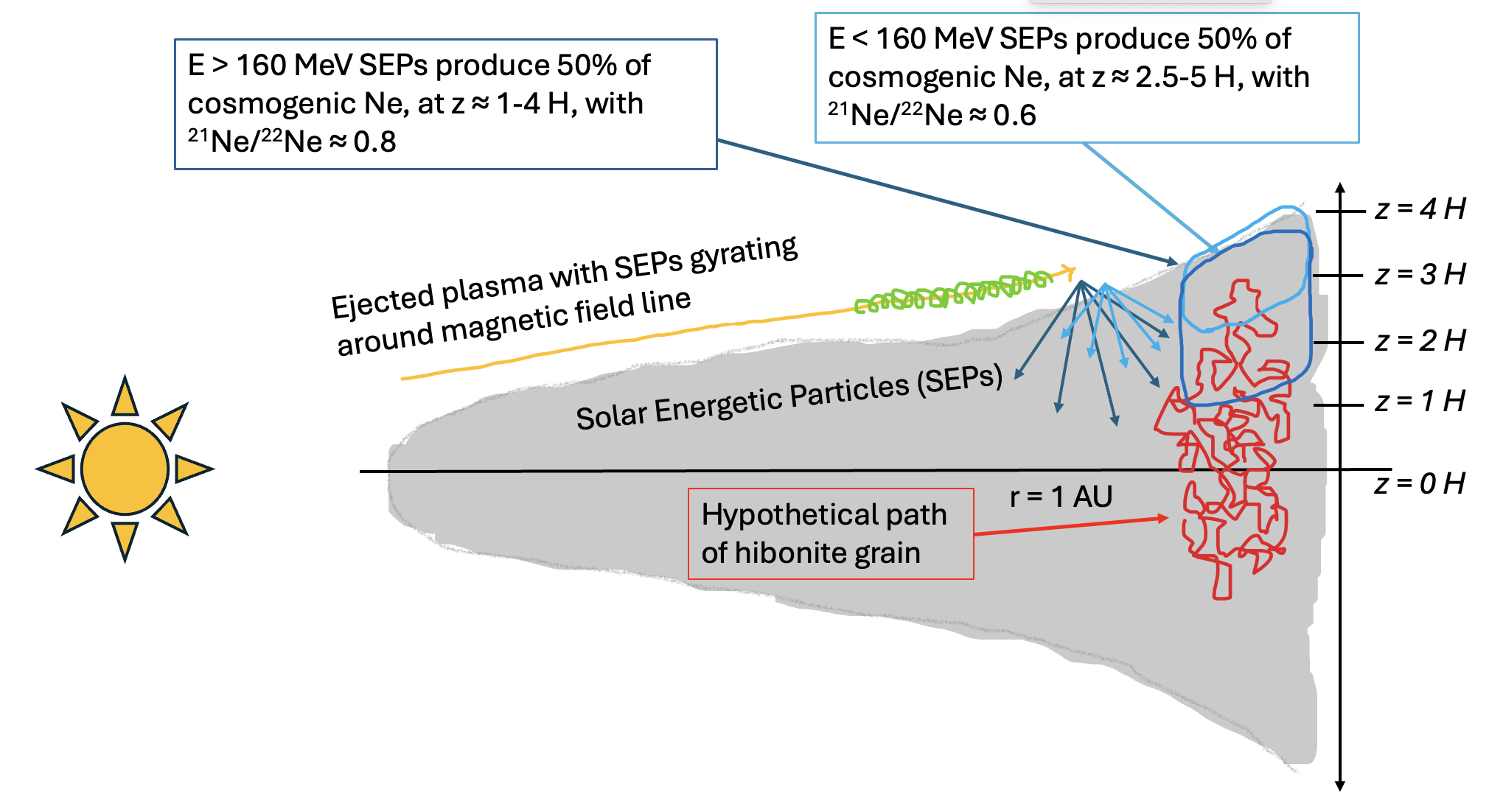}
\includegraphics[width=0.44\linewidth]{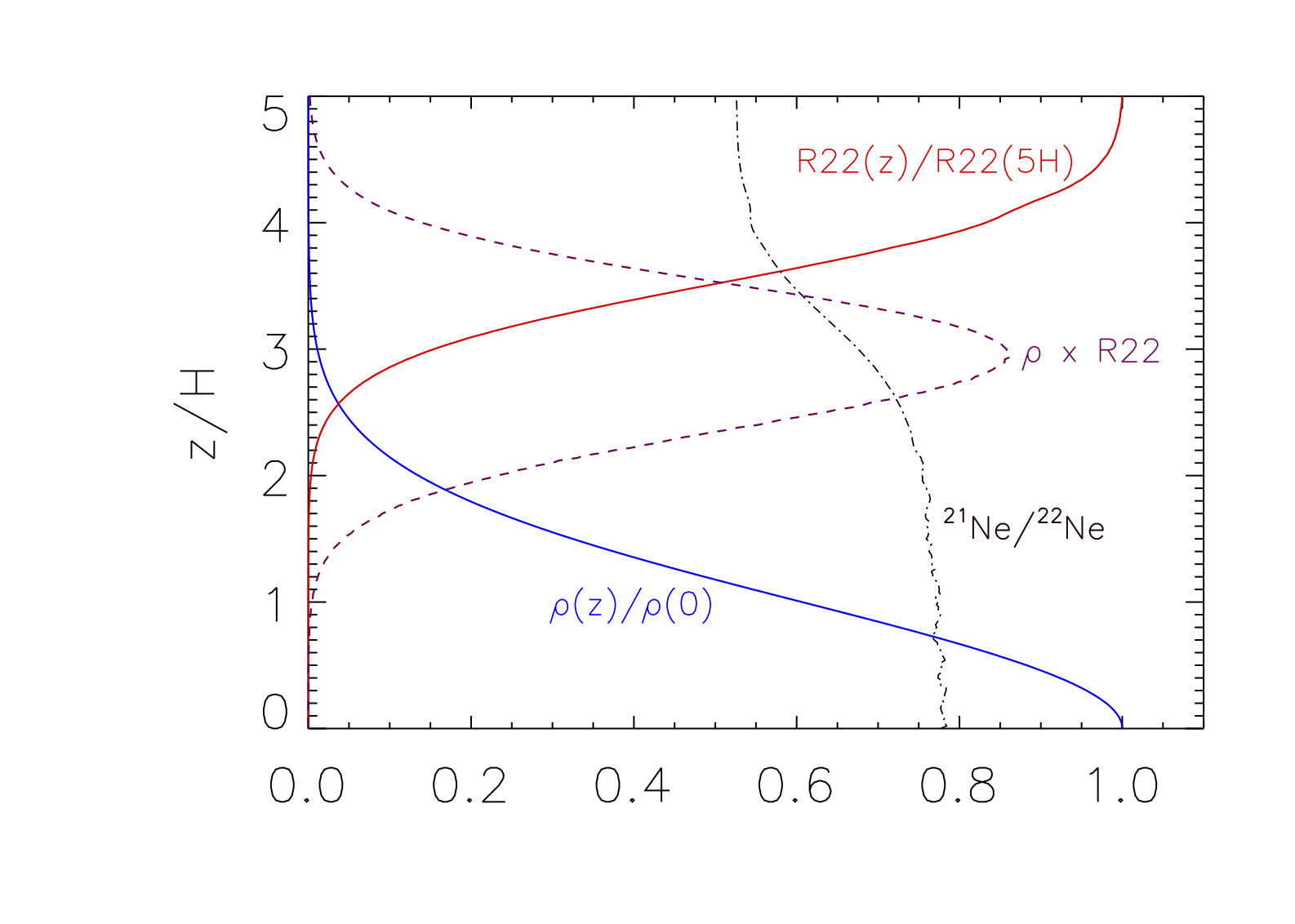}
\caption{{
{\it Left}): Cartoon depicting how hibonite grains acquire cosmogenic Ne.
SEPs in a stellar wind gyrate around the magnetic field lines and can enter the disk in an isotropic way where the wind impinges on the disk.
Lower-energy ($E < 160$ MeV) protons (light blue) are more readily absorbed, and generate cosmogenic Ne with ${}^{21}{\rm Ne}/{}^{22}{\rm Ne} \approx 0.6$, only from $2.5$ to $5 H$; but higher-energy protons (dark blue) penetrate farther and generate Ne ${}^{21}{\rm Ne}/{}^{22}{\rm Ne} \approx 0.8$, from $1$ to $4 H$.
Hibonite grains diffuse vertically (red trajectories) and spend time at various heights $z$ above the midplane.
({\it Right}): Variation of ${}^{22}{\rm Ne}$ production rate from SEPs of all energies (red solid curve), and the relative time spent by dust grains (blue solid curve) as a function of height $z$ above the midplane.
The product (purple dashed curve, here multiplied by 500) illustrates that most cosmogenic Ne is produced between 2 and 4 scale heights above the midplane. 
Also plotted (dotted curve) is the ${}^{21}{\rm Ne}/{}^{22}{\rm Ne}$ ratio of cosmogenic Ne, which is $\approx 0.52$ near the surface (appropriate for unattenuated SEPs), but $\approx 0.79$ near the midplane, where only high-energy SEPs can penetrate. 
On average, ${}^{21}{\rm Ne}/{}^{22}{\rm Ne} \approx 0.67$.}}
\label{fig:neon}
\end{figure}

{Figure~\ref{fig:neon} depicts this process for the case where the disk surface density is $\Sigma = 1000 \, {\rm g} \, {\rm cm}^{-2}$ and the rigidity is $R_0 = 130$ MV.
A radius $r \approx 1$ AU is assumed.
SEPs enter the disk at approximately $5 H$ and propagate in all directions.
For the purposes of the calculation, they can be assumed to move in straight lines even if they spiral around a weak, presumably vertical, magnetic field within the disk. 
In practice, initially lower-energy ($E_0 < 160$ MeV) protons are absorbed between 2.5 and $5 H$, and produce Ne with ${}^{21}{\rm Ne}/{}^{22}{\rm Ne} \approx 0.6$.
Protons with $E_0 > 160$ MeV are absorbed across a range of heights, 1 to $4 H$, and yield Ne with ${}^{21}{\rm Ne}/{}^{22}{\rm Ne} \approx 0.8$.
Hibonite grains diffusing vertically through the disk sample all of these environments.
Most of the cosmogenic Ne in grains will be produced when they are at $z \approx 3 H$.

For this set of parameters, the concentration of cosmogenic Ne after 2 Myr is $2.4 \times 10^{-11} \, {\rm cm}^{3} \, {\rm STP} \, {\rm g}^{-1}$, implying $\Phi = 3800$ to produce the observed abundance in {\it PLAC-21}; and on average ${}^{21}{\rm Ne}/{}^{22}{\rm Ne} = 0.68$.
This value of $\Phi$ is in line with observations of protostars, and the predicted value of $\neratio$ is an excellent match to the value in {\it PLAC-21}, which is $0.69 \pm 0.02$ \citep{KoopEtal2018}.

These results depend on the rigidity. 
For $R_0 = 70, 100$, and 130 MV, the values of $\Phi$ needed to match {\it PLAC-21} are $2.2 \times 10^4$, $8000$, and $3800$.
The resultant values of ${}^{21}{\rm Ne}/{}^{22}{\rm Ne}$ are 0.55, 0.63, and 0.69.
These results suggest that high rigidity consistent with impulsive flares is needed to match the ${}^{21}{\rm Ne}/{}^{22}{\rm Ne}$ ratio of cosmogenic Ne in {\it PLAC-21}, and that relatively low values of $\Phi < 10^4$ are required.
The value of $\Phi$ also depends on the surface density of gas, and is proportional to $\Sigma$.}

{Improved models are needed to better predict the production of cosmogenic Ne in hibonite grains diffusing through the disk, but it is clear that the mechanism is plausible.
Hibonite grains can acquire their observed amounts of cosmogenic Ne while diffusing through the disk at radii where the stellar wind impinges on the disk, roughly out to 1 or 2 AU, as depicted in {\bf Figure~\ref{fig:neon}}.
An SEP flux $\Phi < 10^4$ times the present-day flux, which is within the range expected from observations of protostars, and a rigidity $\approx 130$ MV, would reproduce the amount and the ${}^{21}{\rm Ne}/{}^{22}{\rm Ne}$ ratio of cosmogenic Ne in grains like {\it PLAC-21}.
The low midplane temperatures ($< 500$ K) in the disk at these distances \citep{DeschEtal2018} allows the Ne to be retained as it accumulates over 2 Myr until the hibonite grain diffuses into regions of the disk unable to be irradiated by SEPs.
Eventually these hibonite grains would accrete into chondrites.}


\bigskip

\subsection{Hypothesis \#2: Grains were directly irradiated while above the disk}

We now consider the alternative hypothesis, in which hibonite grains are irradiated only after being launched in a disk wind. 
For the reasons described in \S 2.3, such a scenario for a particle as large as {\it PLAC-21} ($a = 60 \, \mu{\rm m}$) is unlikely.
Using Equation~\ref{eq:launch} above from \citet{GiacaloneEtal2019}, and even assuming very favorable conditions ($r = 0.1 \, {\rm AU}$ and $T = 1400 \, {\rm K}$), launching of a grain like {\it PLAC-21} would require $\dot{M}_{\rm out} > 4 \times 10^{-7} \, M_{\odot} \, {\rm yr}^{-1}$. 
The mass accretion rate onto the star, $\dot{M}_{\rm in}$, would have to be about an order of magnitude higher, $\approx 4 \times 10^{-6} \, {\rm yr}^{-1}$.
{Such protostellar mass accretion rates can not persist long enough to be relevant.}
On average, $\dot{M}_{\rm in}$ varies with protostellar age $t$ as $\dot{M}_{\rm in} \approx 10^{-7} \, \left( t / {\rm Myr} \right)^{-1.2} \, M_{\odot} \, {\rm yr}^{-1}$
\citep{BaraffeEtal2002}, so the mass fluxes needed to launch a particle like {\it PLAC-21} are only sustained for $< 0.05$ Myr.
If launching in a disk wind were necessary for hibonite grains to have Ne, it is difficult to reconcile the short window for launching with the observation that 78\% of hibonite grains carry significant cosmogenic Ne.

An additional difficulty with the direct irradiation of hibonite grains is that they must be irradiated for a cumulative time $\approx 6.4 \, {\rm Myr}$ assuming the present-day flux at 1 AU. 
Yet the average fluence received by launched grains that are irradiated is 0.76 yr times the flux at 1 AU \citep{YangEtal2024}.
To receive the necessary fluence to produce the amount of cosmogenic Ne observed in {\it PLAC-21}, the SEP flux from the early Sun would have to be $6 \times 10^6$ times the present-day flux \citep{YangEtal2024}.
This is two orders of magnitude higher than even the highest X-ray fluxes among protostars would suggest and is implausible.

{As mentioned above, it is not even clear how SEPs---which gyrate around magnetic field lines threading the stellar wind---can penetrate into the disk wind, to irradiate grains there.}

Most significantly, irradiation of grains above the disk will not reproduce the $\neratio$ observed in grains like {\it PLAC-21}.
Because grains would be exposed to an unattenuated SEP flux, the $\neratio$ ratio of their cosmogenic Ne would be close to the unshielded value, $\approx 0.49$. The value $\neratio \approx 0.69$ observed in {\it PLAC-21} is inconsistent with irradiation by SEPs above the disk.

\subsection{Summary}

If hibonite grains could only acquire their cosmogenic Ne while launched above the disk in a magnetocentrifugal outflow, this would demand that they were launched above the disk for a ten-year period during which the average SEP flux was $> 10^6$ times the present-day flux, two orders of magnitude higher than even the most active protostars.
However, launching of particles as large as {\it PLAC-21} ($a = 60 \, \mu{\rm m}$) is not even possible, except perhaps very close to the Sun ($< 0.1$ AU), in the first 0.05 Myr of the disk, and even then such particles would not escape to the chondrite-forming region.
The ${}^{21}{\rm Ne}/{}^{22}{\rm Ne}$ ratio of the particles would also remain $\approx 0.49$, incompatible with the observed value in {\it PLAC-21}, $\approx 0.69$.
{It bears repeating that the SEPs would also have little opportunity to leave the magnetic field lines threading the stellar wind, to irradiate the material being launched by a disk wind anchored in the magnetic field of the disk.}
In contrast, hibonite grains in the disk will naturally spend enough time near the surface to receive an SEP fluence sufficient to produce the amounts in particles like {\it PLAC-21}, if the early Sun's SEP flux was enhanced by a factor of only $\Phi < 1 \times 10^4$ times the present-day flux, for the first few Myr of the disk lifetime. 
We calculate a final ${}^{21}{\rm Ne}/{}^{22}{\rm Ne}$ in the cosmogenic Ne $\, \approx \, 0.68$, exactly consistent with the value in {\it PLAC-21}.

\section{Heterogeneity of $\beratio$?}
\label{sec:beten}

Additional limits on the SEP flux impinging the disk could come from determinations of initial $(\beratio)_0$ ratios in meteoritic inclusions, especially CAIs.
The short-lived radionuclide $\beten$ decays with a half-life of 1.39 Myr \citep{ChmeleffEtal2010,KorschinekEtal2010}, so it must have been produced in the Sun's molecular cloud shortly before the Sun's formation, or within the protoplanetary disk.
There are basically three hypotheses for the existence of live $\beten$ incorporated into CAIs: \#1.\ all the $\beten$ was inherited from the molecular cloud; \#2.\ all the $\beten$ was created in the disk by SEP irradiation; or, \#3.\ roughly comparable amounts were contributed from each source.
These various hypotheses make many similar, but sometimes distinct, predictions, and it is easy for data to appear consistent with more than one hypothesis.
According to the scientific method, {\it support} or {\it evidence} for one hypothesis is developed by ruling out the others as the source for $\beten$ in the majority of CAIs, from both astrophysical and meteoritical perspectives.

Each of these hypotheses is falsifiable.
The inheritance model \citep{DeschEtal2004,DunhamEtal2022,DeschEtal2023pp7} demands production of $\beten$ by GCRs spalling CNO nuclei through reactions like ${}^{16}{\rm O}(p,x){}^{10}{\rm Be}$.
One issue is whether sufficient GCR fluxes can be found in the Sun's molecular cloud to create the observed amounts of $\beten$.
This model also predicts uniform $(\beratio)_0$ among all CAIs.
Finding convincing evidence of significantly higher $(\beratio)_0$ ratios in even a small fraction of CAIs would rule out inheritance as the sole source of $\beten$.

The irradiation model \citep{McKeeganEtal2000,GounelleEtal2001,MacPhersonEtal2003,ChaussidonEtal2006,GounelleEtal2006,LiuEtal2010,GounelleEtal2013,SossiEtal2017,Jacquet2019,FukudaEtal2019,FukudaEtal2021}
demands production of $\beten$ by SEPs spalling CNO nuclei.
An important issue is whether sufficient quantities of $\beten$ can be generated in the disk by plausible SEP fluxes.
Presuming this is the case, models like that of \citet{Jacquet2019} predict that a wide range of $(\beratio)_0$ ratios should be found in CAIs formed at even slightly different times and places in the disk.
If the $(\beratio)_0$ ratios among CAIs are uniform within measurement errors, this would place severe restrictions on how much $\beten$ could be produced in this way.

It is ambiguous from the literature whether $(\beratio)_0$ ratios are considered uniform or heterogeneous.
The vast majority ($> 80\%$) of CAIs appear to record a uniform value $(\beratio)_0 \approx 7 \times 10^{-4}$ \citep{DunhamEtal2022}, favoring the inheritance model.
On the other hand, eight reported isochrons out of the $\approx 58$ in the literature 
appear to record $\beratio \approx 20-100 \times 10^{-4}$ \citep{GounelleEtal2013,SossiEtal2017,FukudaEtal2019,FukudaEtal2021,MishraMarhas2019,DunhamEtal2020,BekaertEtal2021}.
As well, a subset of CAIs including FUN (Fractionation and Unknown Nuclear effects) CAIs, PLACs (PLAty Crystals of hibonite) and SHIBs (Spinel-HIBonite inclusions) also tend to record lower values $(\beratio)_0 \approx 3-5 \times 10^{-4}$.
In the inheritance model, these lower values, at least, are interpreted as the CAIs having formed or been reset for Be-B systematics after $\approx 1$ Myr of disk evolution.
Proponents of the irradiation model reject this on the basis that these inclusions generally lack evidence for live $\altwosix$ and are presumed to have formed before other CAIs.
As a result, a hybrid hypothesis has developed in which these inclusions record the $\beten$ inherited from the molecular cloud, while all other CAIs record an additional amount of $\beten$ created in the disk by SEP irradiation \citep{WielandtEtal2012,TatischeffEtal2014,Jacquet2019}.
This hybrid hypothesis is falsified if the $(\beratio)_0$ ratios among the FUN CAIs or PLACs and SHIBs are not uniform.

Below we discuss the astrophysical models and how well they support production of $\beten$ in the molecular cloud or in the protoplanetary disk.
We then present all the literature data for $(\beratio)_0$ ratios in CAIs, paying special attention to whether different populations record uniform $(\beratio)_0$ values or not.
Finally, we examine the three hypotheses above in light of these data.

\subsection{Astrophysical Models}

\subsubsection{Production of $\beten$ in the molecular cloud}

Soon after the discovery of $\beten$, it was regarded as a ``smoking gun" for SEP irradiation in the disk, because it was calculated that GCR spallation of CNO nuclei in the molecular cloud could not yield $\beratio$ greater than about $1 \times 10^{-4}$ \citep{GounelleEtal2001}.
\citet{DeschEtal2004} suggested that the production of $\beten$ in the molecular cloud by GCR spallation could be augmented by trapping of low-energy GCRs that were themselves $\beten$ nuclei, potentially providing the bulk of $\beten$ in the solar nebula. 
Subsequent measurements of the fluxes of low-energy GCRs show that they are much lower than extrapolations would suggest, and this mechanism is not viable \citep{TatischeffEtal2014}.
Although some $\beten$ can be produced by neutrino spallation in supernovas \citep{BanerjeeEtal2016}, it does not appear possible to inject significant quantities of such $\beten$ into the early solar system \citep{DeschEtal2023pp7}.
These recognitions surely have hindered acceptance of the inheritance model.

More recently, though, it has been recognized that the calculation of $\beten$ production by GCR spallation has been greatly underestimated because it is based on the present-day GCR flux in the Sun's neighborhood.
GCRs are accelerated by supernovae in the nearest $\sim 1$ kpc, in the last 15 Myr.
Not only were the rates of star formation and supernovae a factor of 1.5 to 2 times higher 4.5 Gyr ago, they are much higher in spiral arms (where the Sun's molecular cloud would have been located) than in inter-arm regions like the Sun is in today.
Accounting for these factors, $\beratio \approx 7 \times 10^{-4}$ is actually a very probable value for the Sun's molecular cloud \citep{DunhamEtal2022,DeschEtal2023pp7}.

\subsubsection{Production of $\beten$ in the protoplanetary disk}

It is inevitable that some $\beten$ also is produced in protoplanetary disks by SEP irradiation.
The main target nucleus (${}^{16}{\rm O}$) is abundant, and ${}^{9}{\rm Be}$ is rare, ensuring a high $\beratio$ in irradiated materials.
At issue is whether the $\beratio$ approached $7 \times 10^{-4}$ in the region where CAIs formed ($< 1$ AU), in the first $\sim 10^{5}$ yr of disk evolution.

We note that in some models like the ``X-wind" model \citep{ShuEtal2001,GounelleEtal2001}, CAIs form at $< 0.1$ AU in a region devoid of hydrogen gas.
As SEPs lose most of their energy to ionizing the gas, its absence leads to higher efficiencies of $\beten$ production; and the proximity to the Sun leads to higher production rates.
However, this model faces many challenges to match measurements of CAIs, not least of which is that the oxygen fugacity of this hydrogen-poor environment is incompatible with the reducing conditions recorded by CAIs, and it is nearly impossible to launch cm-sized CAIs even from this region
\citep[\S 2.3;][]{DeschEtal2010}.
We consider only models in which $\beten$ is produced by irradiation of disk gas, at about 1 AU, like the model of \citet{Jacquet2019}, to be viable.

\citet{Jacquet2019} determined that steady-state abundances $\beratio \approx 6 \times 10^{-4}$ were possible if the luminosity of $E > 10$ MeV SEPs was $0.005 \, L_{\odot}$, i.e., $9 \times 10^{5}$ greater than the present-day flux from the Sun.
Given these conditions, he also determined that conditions become more favorable for $\beten$ production over time, such that $\beratio \propto t$.
As would be expected, the production of $\beten$ also falls off sharply with distance from the Sun, $r$, falling off roughly as $r^{-3/2}$. 
Unless all CAIs formed at the exact same time and place in the nebula, variations would be expected.

We use the formulism of \S 3.2 to test whether sufficient quantities of $\beten$ could be made.
Most $\beten$ is produced from SEPs through the reaction ${}^{16}{\rm O}(p,x){}^{10}{\rm Be}$, for which 
{we approximate the cross section as 
\begin{equation}
\sigma_{10}(E) = 
\begin{cases}
0 \, {\rm mbarn}, & 
E < 50 \, {\rm MeV} \\
2 \, {\rm mbarn} \, \left( E  / 500 \, {\rm MeV} \right) & 
50 \, {\rm MeV} \leq E < 500 \, {\rm MeV} \\
2 \, {\rm mbarn} & 500 \, {\rm MeV} \leq E 
\end{cases}
\end{equation}
based on \citet{MoskalenkoMashnik2003}.

Following the same treatment as for cosmogenic Ne, we calculate 
\begin{equation}
\left. \frac{{}^{10}{\rm Be}}{{}^{9}{\rm Be}} \right|_{z} = (0.65) \, \left( \frac{{}^{16}{\rm O}}{{}^{9}{\rm Be}} \right) \, \int_{0}^{\infty} \frac{dF(E,z)}{dE} \, \sigma_{10}(E) \, dE \, \left( \Delta t \right).
\end{equation}
We set ${}^{16}{\rm O}/{}^{9}{\rm Be} = 1.9 \times 10^7$ \citep{Lodders2003}.

For the irradiation model described in \S 2, with $F(> 10 \, {\rm MeV} = 100 \, {\rm cm}^{-2} \, {\rm s}^{-1}$ and $R_0 = 130$ MV and $\Delta t = 2$ Myr, 
we calculate ${}^{10}{\rm Be}/{}^{9}{\rm Be} = 1.3 \times 10^{-8}$.
This is an upper limit, because $\Delta t$ cannot exceed the mean life of ${}^{10}{\rm Be}$, 2.00 Myr.
But in fact $\Delta t$ is more aptly given by the timescale for gas to empty from the inner disk and onto the star.
Repeating the calculation for various values of $\Delta t$ and $\Sigma$, we find the ${}^{10}{\rm Be}/{}^{9}{\rm Be}$ ratio is proportional to $\Sigma / (\Delta t)$, and therefore the mass accretion rate $\dot{M}$:
\begin{equation}
\frac{{}^{10}{\rm Be}}{{}^{9}{\rm Be}} = 2 \times 10^{-10} \, \Phi \, \left( \frac{ \dot{M} }{ 10^{-8} \, M_{\odot} \, {\rm yr}^{-1} } \right).
\end{equation}
\citet{Jacquet2019} found very similar results, calculating $\beratio = 6 \times 10^{-4}$ for $\dot{M} = 10^{-7} \, M_{\odot} \, {\rm yr}^{-1}$ and $\Phi = 1.6 \times 10^{5}$, only a factor of 2 greater than what we would predict. 

It is important to note that production of ${}^{10}{\rm Be}$ differs from the production of cosmogenic Ne in that the $\beratio$ ratio must pertain to the whole disk, as CAI minerals condense from that gas. 
Most of the $\beten$ so produced is lost onto the star. 
Most of the cosmogenic Ne produced by irradiation is also lost onto the star, but all that matters is the irradiation to which a few hibonite grains diffusing upstream were exposed.
}

To produce $\beratio = 7 \times 10^{-4}$ {in a disk with typical mass accretion rates} therefore demands $\Phi \sim 1 \times 10^6$.
This is approximate but probably a lower limit: while production of $\beten$ from spallation of C and N nuclei will increase the production rate by tens of percent, the disk will also take $> 10^6$ yr to build up to the steady-state production rate, suggesting the $\beratio$ ratio is an order of magnitude smaller at only $t = 10^5$ yr, when most CAIs are forming.
We conclude that for SEP fluxes in the range inferred for other protostars, irradiation in the disk cannot contribute more than about 1\% of the observed $\beten$.


But if the SEP fluxes were high enough ($\sim 10^6$ times present-day flux at 1 AU) to provide sufficient $\beten$, we concur with \citet{Jacquet2019} that $\beratio$ ratios would be quite variable, tending to increase with time $\propto t^{+1}$, and decreasing with heliocentric distance as $r^{-3/2}$. 
In disk models, temperatures are high enough to form CAIs across a broad range of radii $r \approx 0.5 - 2$ AU \citep{YangCiesla2012} or $r \approx 0.1 - 0.3$ AU \citep{DeschEtal2018}, suggesting variations in $r$ on the order of factors of $> 3$. 
These conditions last from $t < 0.1$ Myr to $t > 0.5$ Myr, suggesting variations in $t$ of a factor $> 5$. 
Models like these suggest some fraction of CAIs should exhibit $(\beratio)_0$ ratios factors of $\approx 4$ above and below the average value. 
Models are not adequate to quantify that fraction, but tens of percent might be expected.

[Similar considerations apply to the hypothesis that $\beten$ atoms were implanted from solar wind into CAIs \citep{BrickerCaffee2010}.
This would require CAIs to pass near (0.06 AU) the Sun before being sent out to the outer solar nebula, and it is likely that they would be quite variable in their $(\beratio)_0$ ratios.]

Large variability in $(\beratio)_0$ ratios would be a signpost for production of $\beten$ by irradiation, which would in turn demand $\Phi \sim 10^6$ in the early isk.

\subsection{Heterogeneity of CAI $(\beratio)_0$ ratios?}

Before quantifying the observed heterogeneity in $(\beratio)_0$ among CAIs, we first categorize CAIs into three main classes, with different attributes pertinent to $(\beratio)_0$ ratios.

We classify most CAIs as ``normal". 
These tend to be large (mm- to cm-sized), multiminerallic, and often melted (types A, B or C).
While CAIs differ mineralogically across different chondrites \citep{MacPhersonEtal2005}, possibly because they are preserved differently, their oxygen isotopes and $\epsilon^{50}{\rm Ti}$ anomalies are very similar (at least, among ordinary, CV, CO and CK chondrites), strongly suggesting they formed together in a common source region \citep{EbertEtal2018}.

Carbonaceous chondrites of type CH or CB (or Isheyevo, which contains lithologies of both) appear to have formed by reassembly after an impact between two asteroids \citep{KrotEtal2005} at about $t = 5.75$ Myr \citep{DeschEtal2023anchors2}.
These CAIs are mineralogically distinct but appear to predate the impact \citep{KrotEtal2008} and almost certainly source from the same region as other CAIs.
However, we consider them separately because they may have been altered by residence in the impact plume caused by the collision, e.g., reacting with plume gas containing spallogenic B from the asteroidal regolith.

Finally, there are the examples of CAIs with ``low" initial $(\alratio)_0$, including FUN  CAIs (Fractionation and Unknown Nuclear Effects CAIs), PLACs (PLAty Crystals of hibonite), and SHIBs (Spinel-HIBonite inclusions).
These are often lumped together, although it should be noted that many FUN CAIs (including Axtell {\it 2771}, {\it CMS-1} and {\it KT1}) are like normal CAIs in that they are large and multiminerallic; and they have initial $(\alratio)_0$ lower than the canonical $5.23 \times 10^{-5}$, but only by a factor of 2 or 3. 
In contrast, PLACs and other similar inclusions are often smaller ($< 100 \, \mu{\rm m}$), are dominated by hibonite, and have initial $(\alratio)_0$ too low to be explained by late resetting or formation in the solar nebula. 
These examples of Low-$\alratio$ Corundum/Hibonite Inclusions (LAACHIs) appear to form a distinct population with a different astrophysical origin \citep{DeschEtal2023laachi}.

Below we review the evidence for or against heterogeneity of $(\beratio)_0$ ratios within these three populations (CV/CO/CR, CH, and FUN CAIs/PLACs).

\subsubsection{Normal CAIs from CV, CO, and CR chondrites}

In {\bf Table~\ref{table:normalcais}}, we list the initial $(\beratio)_0$ ratios inferred from 54 isochrons for 52 CAIs from CV, CO, and CR chondrites.
We draw from the Supplementary Material of \citet{DunhamEtal2022}, who re-analyzed literature data to ensure that all isochrons assumed similar Relative Sensitivity Factors and were regressed using the same algorithm \citep{YorkEtal2004}.
We list the number of spots used in each regression ($N$) and the goodness-of-fit (MSWD, or Mean Squares Weighted Deviation).
Data can only be considered to form a linear relationship if MSWD is sufficiently close to 1 \citep{WendtCarl1991}.
If the data do not form a line, then the excesses in ${}^{10}{\rm B}$ must be attributed to possible radiogenic ${}^{10}{\rm B}$ from decay of $\beten$, {\it plus} some other mechanism (e.g., aqueous alteration, GCR exposure, etc.). 
Unless this other process is identified and carefully modeled, no inferences can be made about the initial $(\beratio)_0$. 
For the most part, the isochrons built in each of the 54 analyses in Table~\ref{table:normalcais} are acceptable, but 12 are not.
Before proceeding, we discuss these individual cases.

\begin{deluxetable}{lcccccc}
\tabletypesize{\scriptsize}
\tablewidth{0pt}
\tablecaption{$\beratio$ ratios in normal CAIs in CV,CO,CM,CR chondrites}
\label{table:normalcais}
\tablehead{
\colhead{Chondrite} & 
\colhead{CAI} & 
\colhead{$(\beratio)_0$ ($10^{-4}$)} &
\colhead{$N$} &
\colhead{MSWD} & 
\colhead{valid?} & 
\colhead{Ref.} 
}
\startdata
CV3ox Allende & 3529-41 & \textit{9.5 $\pm$ 1.9} & 10 & 0.4 & yes? & {a}$^{*}$ \\
CV3ox Allende & 3529-41 & \textit{10.0 $\pm$ 0.9} & 51 & 12.0 & no & {b}$^{*}$ \\
CV3ox Allende & 001 & $5.5 \pm 2.6$ & 7 & 1.0 & yes & {c}$^{*}$ \\
CV3ox Allende & 002 & $7.0 \pm 1.1$ & 12 & 1.5 & yes & {c}$^{*}$ \\
CV3ox Allende & 3898 & $4.8 \pm 1.7$ & 7 & 0.3 & yes & {d} \\
CV3ox Allende & 3 fg & \textit{54.5 $\pm$ 9.4} & 19 & 4.4 & no & {e}${}^{*}$ \\
CV3ox Allende & 2 cg & \textit{9.3 $\pm$ 0.8} & 20 & 8.2 & no & {e}${}^{*}$ \\
CV3ox Allende & T23A & $6.1 \pm 2.2$ & 9 & 0.9 & yes & {f} \\
CV3ox Allende & T68 & $5.7 \pm 2.2$ & 7 & 2.0 & yes & {f} \\
CV3ox Allende & 08 & \textit{7.26 $\pm$ 2.84} & 6 & 4.54 & no & {g} \\
CV3ox Allende & 65 & $10.4 \pm 5.6$ & 9 & 0.54 & yes & {g} \\
CV3ox Allende & 3529-42 & \textit{90.6 $\pm$ 36.2} & 6 & 16.9 & no & {g} \\
CV3ox Allende & 75 & $6.30 \pm 1.46$ & 5 & 0.93 & yes & {g} \\
CV3ox NWA 7891 & ZT4 & $7.3 \pm 1.0$ & 9 & 0.9 & yes & {f} \\
CV3ox NWA 3118 & ZT7 & $5.3 \pm 2.6$ & 6 & 1.3 & yes & {f} \\
CV3ox NWA 6991 & B4 & $6.7 \pm 2.2$ & 8 & 0.6 & yes & {f} \\
CV3ox NWA 6991 & Lisa & \textit{22 $\pm$ 7} & 8 & 1.1 & yes? & {h} \\
CV3red Efremovka & E31 & $6.9 \pm 1.2$ & 7 & 1.0 & yes & {i}$^{*}$ \\
CV3red Efremovka & E36 & $7.0 \pm 1.1$ & 20 & 1.4 & yes & {j}$^{*}$ \\
CV3red Efremovka & E38 & $5.7 \pm 1.2$ & 17 & 0.8 & yes & {c}$^{*}$ \\
CV3red Efremovka & E38 & $6.5 \pm 1.0$ & 8 & 0.9 & yes & {i}$^{*}$ \\
CV3red Efremovka & E40 & \textit{16.0 $\pm$ 5.2} & 8 & 3.0 & no & {k}$^{*}$ \\
CV3red Efremovka & E44 & $7.6 \pm 2.4$ & 13 & 0.7 & yes & {c}$^{*}$ \\
CV3red Efremovka & E48 & $8.2 \pm 3.4$ & 5 & 0.8 & yes & {c}$^{*}$ \\
CV3red Efremovka & E48 & $7.1 \pm 0.4$ & 9 & 0.9 & yes & {i}$^{*}$ \\
CV3red Efremovka & E65 & $7.0 \pm 1.8$ & 17 & 0.8 & yes & {j}$^{*}$ \\
CV3red Efremovka & E66 & \textit{7.6 $\pm$ 1.8} & 20 & 2.3 & no & {j}$^{*}$ \\
CV3red Efremovka & E69 & $6.2 \pm 2.0$ & 12 & 1.3 & yes & {c}$^{*}$ \\
CV3red Efremovka & E104 & \textit{8.0 $\pm$ 0.9} & 13 & 2.5 & no & {i}$^{*}$ \\
CV3red Efremovka & 6456-1 & $7.6 \pm 1.6$ & 6 & 1.9 & yes & {d} \\
CV3red Leoville & 3535-3b & $6.7 \pm 2.4$ & 9 & 1.7 & yes & {d} \\
CV3red Leoville & ZT1 & $9.0 \pm 4.3$ & 9 & 0.6 & yes & {f} \\
CV3red Vigarano & 477-4b & $5.3 \pm 1.7$ & 13 & 1.0 & yes & {d} \\
CV3red Vigarano & 477-5 & $7.4 \pm 1.9$ & 13 & 1.6 & yes & {d} \\
CV3red Vigarano & 1623-9 & $5.8 \pm 1.9$ & 8 & 1.2 & yes & {d} \\
CV3red Vigarano & 3137-2 & $7.2 \pm 3.0$ & 10 & 0.8 & yes & {f} \\
CV3red NWA 5028 & Agave & $11.6 \pm 4.6$ & 6 & 1.0 & yes & {f} \\
CV3red NWA 5028 & Cholla & $7.8 \pm 6.6$ & 4 & 0.2 & yes & {f} \\
CV3red NWA 5028 & Saguaro & $8.1 \pm 1.5$ & 9 & 0.8 & yes & {f} \\
CO3 Yamato 81020 & Y20-9-1 & \textit{21.7 $\pm$ 5.0} & 8 & 2.5 & no & {l} \\
CO3 Yamato 81020 & Y20-1X1 & \textit{29.7 $\pm$ 5.6} & 9 & 1.1 & yes? & {l} \\
CO3 DaG 027 & Anning & $8.5 \pm 3.0$ & 6 & 11 & yes & {f} \\
CO3 DaG 027 & Goeppert & $6.9 \pm 5.2$ & 3 & 0.1 & yes & {f} \\
CO3 DaG 027 & Jemison & $7.9 \pm 3.0$ & 5 & 1.3 & yes & {f} \\
CO3 DaG 027 & Krafft & $10.9 \pm 5.6$ & 6 & 2.0 & yes & {f} \\
CO3 DaG 027 & Mesquite & $9.8 \pm 6.6$ & 6 & 0.8 & yes & {f} \\
CO3 DaG 005 & Bascom & $8.8 \pm 3.4$ & 4 & 0.3 & yes & {f} \\
CO3 DaG 005 & Mitchell & $8.7 \pm 7.6$ & 3 & 0.01 & yes & {f} \\
CO3 DOM 08006 & 31-2 & $14.6 \pm 6.6$ & 9 & 0.9 & yes & {f} \\
CR2 NWA 801 & Cereus & $9.5 \pm 7.6$ & 5 & 0.9 & yes & {f} \\
CR2 NWA 801 & Palo Verde & $10.7 \pm 6.6$ & 5 & 2.1 & yes & {f} \\
CR2 MIL 090657 & Creosote & $2.1 \pm 6.4$ & 8 & 0.4 & yes & {f} \\
CR2 MIL 090657 & Yucca & $8.7 \pm 3.4$ & 5 & 0.4 & yes & {f} \\
CR2 Shisr 033 & Dalea & $7.3 \pm 1.4$ & 5 & 0.2 & yes & {f} 
\enddata
\tablecomments{Entries in italics considered to represent invalid isochrons. Uncertainties are $2\sigma$. * Denotes isochron recalculated by \citet{DunhamEtal2022}. 
References: 
{\it a}. \citet{McKeeganEtal2000};
{\it b.} \citet{ChaussidonEtal2006};
{\it c.} \citet{SugiuraEtal2001};
{\it d.} \citet{MacPhersonEtal2003};
{\it e.} \citet{SossiEtal2017};
{\it f.} \citet{DunhamEtal2022};
{\it g.} \citet{BekaertEtal2021};
{\it h.} \citet{DunhamEtal2020}.
{\it i.} \citet{WielandtEtal2012};
{\it j.} \citet{SrinivasanChaussidon2013};
{\it k.} \citet{MishraMarhas2019};
{\it l.} \citet{FukudaEtal2021}.
}
\end{deluxetable}

Allende CAI {\it 3529-41} has been analyzed twice. 
The original analysis of \citet{McKeeganEtal2000}, after reanalysis by \citet{DunhamEtal2022}, has acceptable MSWD. 
However, the analysis by \citet{ChaussidonEtal2006} has exceptionally high and disqualifying MSWD. 
This CAI evidently suffered extensive aqueous alteration that affected its B isotopes  \citep{DeschOuellette2006}.
For these reasons we do not include either analysis of {\it 3529-41}.
CAI {\it Lisa} has acceptable MSWD, but has clearly been irradiated on the parent body, as evidenced by the high neutron fluence it received \citep{ShollenbergerEtal2018}.
This could have created spallogenic B after its formation that could have steepened the isochron without worsening the MSWD \citep{DunhamEtal2020}.
We therefore exclude this CAI.
Several other isochrons have high and disqualifying MSWD: the two regressions, of fine-grained and of coarse-grained CAIs, built by \citet{SossiEtal2017}; the regressions for Efremovka CAIs {\it E40} \citep{MishraMarhas2019} and {\it E104} \citep{WielandtEtal2012}; the regressions for Allende CAIs {\it 08} and {\it 3529-42} \citep{BekaertEtal2021}; 
and the regression for Yamato 81020 CAI {\it Y20-9-1} \citep{FukudaEtal2021}.
We interpret these CAIs to have had their B isotopes disturbed enough to affect their isochrons, e.g., by irradiation after the CAI formed, or by aqueous alteration on the parent body.
We note that the ${}^{10}{\rm B}/{}^{11}{\rm B}$ ratios of every analysis spot of all of these CAIs except {\it Y20-9-1} lie between the chondritic value 0.2475 \citep{ZhaiEtal1996} and the spallogenic value $\approx 0.44$ \citep{YiouEtal1968}; this indicates that all of these points could be mixtures of chondritic B, some radiogenic ${}^{10}{\rm B}$, and spallogenic B.

All but two of the 54 analyses from CV, CO, and CR chondrites either have linear correlations between ${}^{10}{\rm B}/{}^{11}{\rm B}$ and ${}^{9}{\rm Be}/{}^{11}{\rm B}$ with acceptable MSWD, consistent with a well-determined initial $(\beratio)_0$; or they have ${}^{10}{\rm B}/{}^{11}{\rm B}$ ratios between 0.25 and 0.44 that are uncorrelated (poor MSWD) with ${}^{9}{\rm Be}/{}^{11}{\rm B}$, consistent with mixtures with spallogenic B.
The only two exceptions are those of Yamato 81020 CAIs {\it Y20-9-1} and {\it Y20-1X1} \citep{FukudaEtal2021}.
We exclude these from our analysis, for reasons we discuss below (\S~\ref{sec:fukuda}). 

The 42 valid isochrons (including duplicate measurements of Efremovka CAIs {\it E38} and {\it E40}) of individual CAIs (indexed by $i$) yield initial ratios $x_i = (\beratio)_0$ with $1\sigma$ uncertainties $\delta x_i$ that have weighted mean $x_{\rm avg} = 6.95 \times 10^{-4}$ (and $2\sigma$ uncertainty in the weighted mean of $0.24 \times 10^{-4}$).
The individual $x_i$ are distributed normally about this mean with standard deviation equal to the typical measurement uncertainty.
Defining the $z$-score $z_i = (x_i - x_{\rm avg}) / \delta x_i$, a normal distribution would have: 68.3\%, or 29/42 CAIs with $\left|z_i\right| < 1$; 27.2\%, or 11/42, with $1 < \left|z_i\right| < 2$; and 4.6\%, or 2/42, with $\left|z_i\right| > 2$.
The actual distribution is 25/42, 13/42, and 4/42, very close to a Gaussian distribution, and only slightly broader.
The goodness-of-fit of this distribution about the mean value $6.95 \times 10^{-4}$ is $\chi_{\nu}^{2} = 1.19$.
As $P(\chi_{\nu}^2 > 1.19) = 19\%$, these data are consistent with a single population scattered around their mean according to their measurement errors.
This is to say, the variations in the 42 valid isochrons (with acceptable MSWD) can be entirely attributed to measurement uncertainties, and there is no evidence for intrinsic variations in their initial $(\beratio)_0$ ratios.
The high initial $(\beratio)_0$ ratios that have been used to argue for heterogeneity of $\beten$ \citep{SossiEtal2017,MishraMarhas2019,FukudaEtal2021} are not valid isochrons and are inconsistent with having ${}^{10}{\rm B}$ excesses solely attributable to $\beten$ decay, but instead are consistent with mixtures with spallogenic B.
Clearly CAIs {\it Y20-9-1} and {\it Y20-1X1}, if included, would negate these conclusions; we discuss them separately below (\S~\ref{sec:fukuda}).
 
\subsubsection{CAIs from CH Chondrites and CH/CB Isheyevo}

In {\bf Table~\ref{table:isheyevocais}}, we list the same data as in Table~\ref{table:normalcais}, but for CAIs from CH chondrites and the CH/CB chondrite Isheyevo.
These CAIs are considered separately because of the possibility that their Be-B systematics may have been altered by the impact associated with their parent body; for example, the impact plume may have contained spallogenic B created by irradiation of the asteroid surface. 
We note that for CH/CB/Isheyevo CAIs, it is common in the literature to build isochrons by regressing together data from multiple CAIs.

\begin{deluxetable}{lcccccc}
\tabletypesize{\scriptsize}
\tablewidth{0pt}
\tablecaption{$\beratio$ ratios in CAIs in CH chondrites and CH/CB Isheyevo.}
\label{table:isheyevocais}
\tablehead{
\colhead{Chondrite} & 
\colhead{CAI} & 
\colhead{$(\beratio)_0$ ($10^{-4}$)} &
\colhead{$N$} &
\colhead{MSWD} & 
\colhead{valid?} & 
\colhead{Ref.} 
}
\startdata
Isheyevo & 411 & \textit{104 $\pm$ 26} & 4 & 4.7 & no & {a}$^{*}$ \\
Isheyevo & 501+503+2006+2012 & \textit{4.9 $\pm$ 8.2} & 13 & 3.3 & no & {a}$^{*}$ \\
Isheyevo & 501+508+2006+2012 & \textit{13.45 $\pm$ 7.25} & 13 & 0.7 & yes? & {a}$^{*}$ \\
Isheyevo & Ten CAIs & $5.88 \pm 4.52$ & 30 & 1.03 & yes & {a}$^{*}$ \\
CH SaU & SA301 & \textit{61 $\pm$ 24} & 3 & 0.5 & yes? & {b}$^{*}$ \\
Isheyevo & Five CAIs & $8.6 \pm 2.8$ & 16 & 1.4 & yes & {c} 
\enddata
\tablecomments{Entries in italics considered to represent invalid isochrons. Uncertainties are $2\sigma$. *Denotes isochron recalculated here. 
References: 
{\it a}. \citet{GounelleEtal2013};
{\it b.} \citet{FukudaEtal2019};
{\it c.} \citet{DunhamEtal2022}.
}
\end{deluxetable}

\citet{GounelleEtal2013} presented data for 21 CAIs from Isheyevo, which showed excesses in ${}^{10}{\rm B}/{}^{11}{\rm B}$ above chondritic values.
They claimed that in the majority of these, there was no correlation with ${}^{9}{\rm Be}/{}^{11}{\rm B}$ and therefore no evidence for incorporation of live $\beten$.
\citet{GounelleEtal2013} claimed one CAI ({\it 411}) yielded an isochron with $(\beratio)_0 = (104 \pm 16) \times 10^{-4}$, again not reporting MSWD.
We have performed the regression and find $(\beratio)_0 = (104 \pm 26) \times 10^{-4}$, MSWD $= 4.7$, which is very high and disqualifying.
They also claimed that four CAIs ({\it 501}, {\it 503}, {\it 2006}, and {\it  2012}) could be regressed together, yielding a single isochron with slope $(\beratio)_0 = (13.1 \pm 4.3) \times 10^{-4}$, but did not report MSWD.
We have repeated this calculation and find $(\beratio)_0 = (4.90 \pm 8.16) \times 10^{-4}$, with MSWD $= 3.25$.
We believe \citet{GounelleEtal2013} actually regressed CAIs {\it 501}, {\it 508}, {\it 2006}, and {\it 2012}, which we find yields $(\beratio)_0 = (13.5 \pm 7.3) \times 10^{-4}$, with MSWD $= 0.7$, which is a valid isochron.
However, \citet{GounelleEtal2013} gave no justification why only those four CAIs (out of 21) were picked to be regressed together. 
We find that fully ten CAIs ({\it 409, 413, 501, 502, 508, 2004, 2006, 2009, 2010, 2012}) can be regressed together, yielding $(\beratio)_0 = (5.88 \pm 4.52) \times 10^{-4}$, with MSWD $= 1.03$.
Therefore, consistent with the premise of \citet{GounelleEtal2013}, half (10/21) of CAIs (not 4/21) retain a common, primordial $(\beratio)_0 = (5.88 \pm 4.52) \times 10^{-4}$ [not $(13.1 \pm 4.3) \times 10^{-4}$].
In none of the other CAIs, including {\it 411}, has a value ${}^{11}{\rm B}/{}^{10}{\rm B} > 0.44$ been measured.
All of these CAIs that fail to fit any isochron, including {\it 411}, are consistent with their B isotopes having been disturbed.
Indeed, many Isheyevo CAIs appear to have mixed with a reservoir of spallogenic B with ${}^{10}{\rm B}/{}^{11}{\rm B} = 0.44$ and [B] $\approx 4$ ppb \citep{GounelleEtal2013}.

\citet{FukudaEtal2019} measured eight CAIs in the CH chondrite Sayh al Uhaymir (SaU). 
Unfortunately, in all the seven CAIs except {\it SA301}, only two measurements were made. 
It is therefore impossible to calculate the MSWD of the line to test the possibility the ${}^{10}{\rm B}$ is radiogenic.
Especially since all of the ${}^{11}{\rm B}/{}^{10}{\rm B}$ ratios in these CAIs are $< 0.44$, it is impossible to rule out the possibility that their ${}^{11}{\rm B}/{}^{10}{\rm B}$ ratios were randomly disturbed by addition of spallogenic B. 
The one CAI with three measured points is {\it SA301}, with inferred $(\beratio)_0 = (61 \pm 24) \times 10^{-4}$ and MSWD $= 0.5$.
The low MSWD means this CAI is consistent with having incorporated live $\beten$. 
However, \citet{FukudaEtal2019} also showed a strong correlation between ${}^{10}{\rm B}/{}^{11}{\rm B}$ and $1/[{\rm B}]$ (the inverse of the B concentration), with even lower MSWD. 
Therefore this CAIs is more consistent with its B being a mixture of chondritic B and some other reservoir. 
Because this reservoir has ${}^{10}{\rm B}/{}^{11}{\rm B} > 0.44$, it is not spallogenic.
Nevertheless, as we discuss in \S~\ref{sec:fukuda}, we interpret this as a mixture of two reservoirs and not a valid isochron.

Finally, \citet{DunhamEtal2022} measured five CAIs ({\it Bell}, {\it Meitner}, {\it Rubin}, {\it Tereshkova}, {\it Tharp}) and regressed them together, finding $(\beratio)_0 = (8.6 \pm 2.8) \times 10^{-4}$.
The low value of MSWD $= 1.4$ shows that they comprise a single population, and it is justified to regress these together.

We therefore conclude that 10 out of 21 CAIs analyzed by \citet{GounelleEtal2013} represent a population of CAIs with a common $(\beratio)_0 = (5.88 \pm 4.52) \times 10^{-4}$, and that all 5 CAIs analyzed by \citet{DunhamEtal2022} form a population with common $(\beratio)_0 = (8.6 \pm 2.8) \times 10^{-4}$.
In fact, all 15 CAIs appear to form a single population with weighted mean $(\beratio)_0 = 7.85 \times 10^{-4}$ (with $2\sigma$ uncertaintiy of $2.38 \times 10^{-4}$). 
The CAIs from \citet{GounelleEtal2013} and \citet{DunhamEtal2022} agree with this weighted mean at the $0.5\sigma$ and $0.9\sigma$ levels, respectively.

Although CAIs in Isheyevo were subjected to an impact that could have altered their B isotopes (e.g., by mixing in spallogenic B from the asteroid surface), this does not appear to have happened.
The mean of the 15 Isheyevo CAIs, $(\beratio)_0 = (7.85 \pm 2.38) \times 10^{-4}$, is only $0.8\sigma$ from the mean of the 42  CV/CO/CR CAI analyses, $(6.94 \pm 0.25) \times 10^{-4}$.
This suggests they form a common population, and indeed the all 57 CAIs with valid isochrons cluster around their weighted mean $(6.95 \pm 0.24) \times 10^{-4}$ with MSWD of 1.03, an excellent fit. 
Only the three CAIs {\it SA301} \citep{FukudaEtal2019} and {\it Y20-9-1} and {\it Y20-1X1} \citep{FukudaEtal2021} appear to have valid isochrons that do not match this value.
Excepting these three CAIs (which we discuss next), the 57 CAIs from CV, CO, CR and Isheyevo chondrites all are consistent with having formed in a reservoir with $\beratio = 6.95 \times 10^{-4}$.
These 57 CAIs easily could have falsified the hypothesis that $\beten$ was inherited from the molecular cloud, but they did not.

\subsubsection{The three CAIs with good MSWD and high $(\beratio)_0$}
\label{sec:fukuda}

Of all the CAIs from CV, CO, CR and CH or Isheyevo chondrites with valid MSWD, only three have $(\beratio)_0$ significantly greater than $7 \times 10^{-4}$.
These are: {\it SA301} from the CH chondrite SaU \citep{FukudaEtal2019}; {\it Y20-1X1} from the CO chondrite Yamato 81020; and {\it Y20-9-1} from Yamato 81020, which does not have valid MSWD but which we include here
\citep{FukudaEtal2021}.
In addition, all have some analysis spots with ${}^{10}{\rm B}/{}^{11}{\rm B} > 0.44$, so that a mixing with a reservoir of spallogenic B can be excluded.
At face value, the data are consistent with forming isochrons, and with a radiogenic origin for the ${}^{10}{\rm B}$ excess.
On the other hand, the data may also be consistent with other explanations.
In fact, we show that these are more likely and that a radiogenic origin can effectively be dismissed.

Before discussing the isochrons themselves, we first consider how likely is is that these three CAIs all represent valid isochrons with high $(\beratio)_0 > 7 \times 10^{-4}$, considering all were measured by the same group.
Considering CAIs across all chondrite types come from the same region \citep[e.g.,][]{EbertEtal2018}, the proportion of high-$(\beratio)_0$ CAIs must be similar ($p \approx 3/69 \approx 4.3\%$) across all chondrites, and other examples should have been discovered by other groups.
The probability that {\bf zero} such CAIs were discovered by any other groups in any of the $\approx 66$ other analyzed CAIs is $(1 - p)^{66} \approx 5\%$.
Meanwhile, \citet{FukudaEtal2019} effectively analyzed only one CAI, {\it SA301}, and \citet{FukudaEtal2021} analyzed only two CAIs, {\it Y20-1X1} and {\it Y20-9-1}, yet claimed high $(\beratio)_0$ for all three.
The probability of finding three such examples in all three CAIs for which this could be tested is $p^3 \approx 8 \times 10^{-5}$.
The joint probability that the high $(\beratio)_0$ ratios and that \citet{FukudaEtal2019,FukudaEtal2021} found all three in three attempts, while nobody else found any, is $\approx 4 \times 10^{-6}$ for $p = 3/69 = 4.3\%$, and is lower for all other values of $p$.
It is {\bf extremely} unlikely that this distribution of $(\beratio)_0$ values is real and occurred by chance, and is much more likely to be the result of how \citet{FukudaEtal2019,FukudaEtal2021} analyzed their data.

\begin{figure}[ht]
\centering
\includegraphics[width=\linewidth]{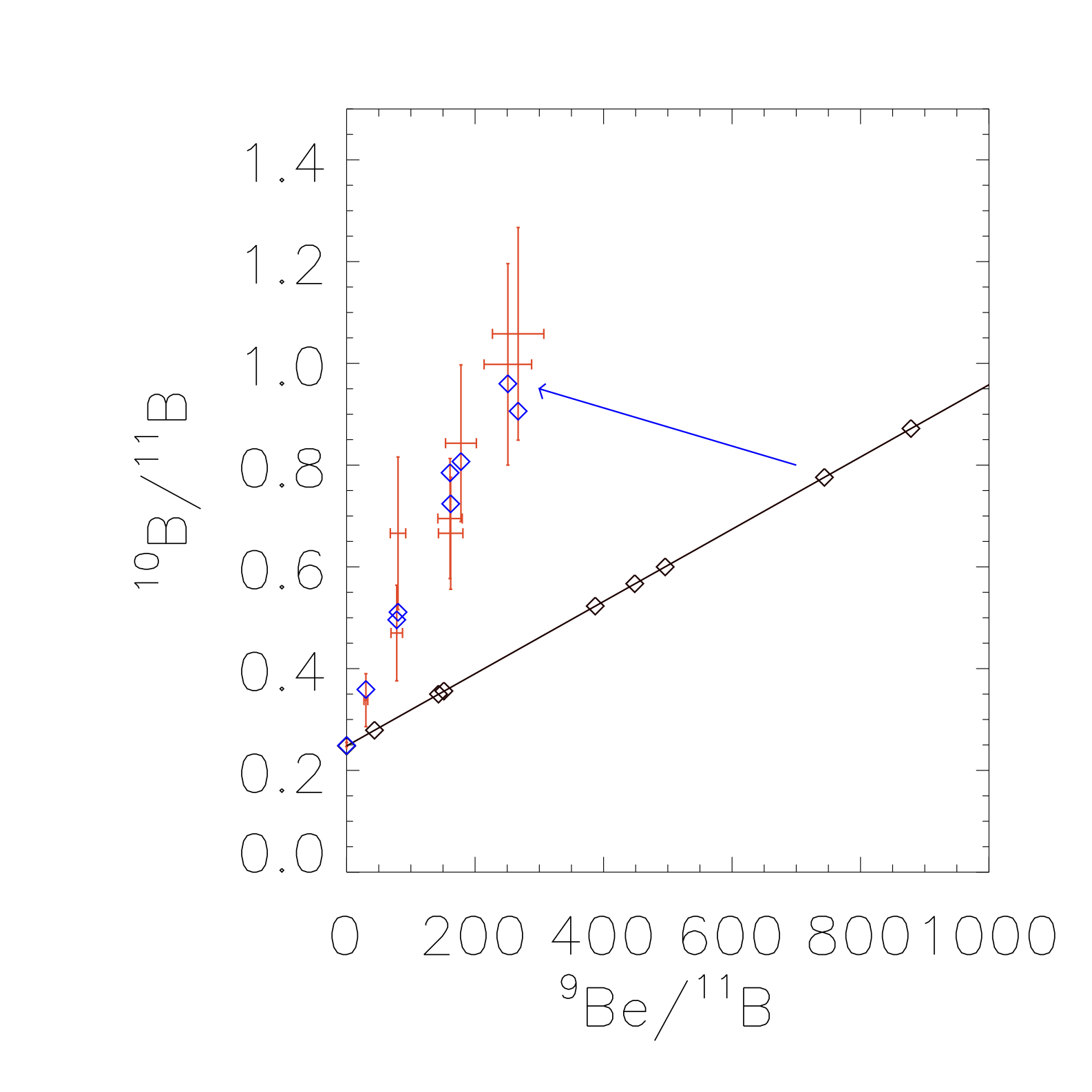}
\caption{Effect on inferred isochron slope of under-subtracting background noise. The data (including $2\sigma$ error bars) for CAI {\it Y20-1X1} reported by \citet{FukudaEtal2021} are plotted in red,
consistent with $(\beratio)_0 \approx 30 \times 10^{-4}$.
If the ratios actually were arrayed along an isochron (black line) with slope $({}^{10}{\rm Be}/{}^{9}{\rm Be})_0 = 7 \times 10^{-4}$ (black diamonds), producing counts $0.2 \, ({\rm [B]} / 50 \, {\rm ppb})$ cps, but the reported measurements inadvertently included unsubtracted noise (0.05 cps), this would result in the reported data having lower ${}^{9}{\rm Be}/{}^{11}{\rm B}$ and higher ${}^{10}{\rm B}/{}^{11}{\rm B}$ values (blue diamonds), i.e, shifting them in the direction of the arrow.
We have chosen the ${}^{9}{\rm Be}/{}^{11}{\rm B}$ values of the `real' data to match the reported ${}^{9}{\rm Be}/{}^{11}{\rm B}$ values, but the ${}^{10}{\rm B}/{}^{11}{\rm B}$ values are an unexpectedly good match.
We conclude that the inferences of high $({}^{10}{\rm Be}/{}^{9}{\rm Be})_0$ values for CAIs {\it SA301} \citep{FukudaEtal2019}, and {\it Y20-1X1} and {\it Y20-9-1} \citep{FukudaEtal2021} are likely artefacts of under-subtracting electronic noise.
}
\label{fig:fukudanoise}
\end{figure}

We speculate that \citet{FukudaEtal2019,FukudaEtal2021} undercorrected during subtraction of the background signal when counting ${}^{10}{\rm B}^{+}$ and ${}^{11}{\rm B}^{+}$ ions.
\citet{FukudaEtal2019,FukudaEtal2021} did not report whether this background was subtracted during Be-B analyses, but did report that the background was not subtracted during Al-Mg analyses (where it would indeed be negligible).
The noise in the electron multipliers was reported to yield a count rate up to 0.05 cps \citep{FukudaEtal2019}.
The count rate of ${}^{10}{\rm B}^{+}$ ions was reported as being `typically' 0.2 cps, but it was not reported what boron concentration this referred to.
We speculate that this is the count rate for [B] $\approx 50$ ppb.
If so, then spots with 5 ppb boron (like melilite spot \#1 in {\it SA301}) would be dominated by noise, which would introduce equal counts of ${}^{10}{\rm B}^{+}$ and ${}^{11}{\rm B}^{+}$.
If not subtracted properly, this would effectively lower the ${}^{9}{\rm Be}/{}^{11}{\rm B}$ ratios and raise the ${}^{10}{\rm B}/{}^{11}{\rm B}$ ratios, increasing the slope of the isochron, as illustrated for the case of CAI {\it Y20-1X1} in {\bf Figure~\ref{fig:fukudanoise}}.
It would also be consistent with the findings that {\it SA301} \citep{FukudaEtal2019} and {\it Y20-1X1} \citep{FukudaEtal2021} show the expected linear correlation between ${}^{10}{\rm B}/{}^{11}{\rm B}$ and $1/{\rm [B]}$, indicating a mixing with a reservoir with ${}^{10}{\rm B}/{}^{11}{\rm B} \approx 1$.
(CAI {\it Y20-9-1} mostly does as well, but is also clearly disturbed.)

\bigskip

We cannot confirm that \citet{FukudaEtal2019,FukudaEtal2021} under-subtracted background noise when analyzing their data.
But given the plausibility of this scenario, and the fact that all three unusual CAIs were analyzed only by this one group, who found this result in all three CAIs they analyzed that could yield this result, we interpret them as artefacts of the data acquisition that arise at low [B] concentrations ($< 10$ ppb).
We do not consider the three CAIs {\it SA301}, {\it Y20-9-1} or {\it Y20-1X1} \citep{FukudaEtal2019,FukudaEtal2021} to record isochrons with supracanonical $({}^{10}{\rm Be}/{}^{9}{\rm Be})_0$.
As such, there are {\bf zero} CAIs that unambiguously record radiogenic $\beten$ with $(\beratio)_0 > 7 \times 10^{-4}$. 
Such a finding would have falsified the inheritance hypothesis, but did not.

\subsubsection{FUN CAIs, PLACs and SHIBs}

The hypothesis that all $\beten$ derives from the molecular cloud also could be falsified if FUN CAIs, PLACs and SHIBs could be proved to have formed before other CAIs, so that their lower values $(\beratio)_0 \approx 3-5 \times 10^{-4}$ could not be attributed to late formation or resetting.
In {\bf Table~\ref{table:funcais}} we list the initial $(\beratio)_0$ ratios of various FUN CAIs, PLACs and SHIBs, again following \citet{DunhamEtal2022}. 
As in Tables~\ref{table:normalcais} and~\ref{table:isheyevocais}, we again report MSWD.
We find that only one of the isochrons in Table~\ref{table:funcais} is disqualified.

We find the weighted mean of the $(\beratio)_0$ ratios of the eight objects in Table~\ref{table:funcais} is $4.39 \times 10^{-4}$, with $2\sigma$ uncertainty of $0.25 \times 10^{-4}$.
While the PLACs and SHIBs and {\it HAL} are reasonably close to this value, the $(\beratio)_0$ ratios of the similar objects {\it SHAL} and {\it Hidalgo} are $4.4\sigma$ below and $5.9\sigma$ above this mean, respectively.
The FUN CAIs range from $2.0\sigma$ and $1.6\sigma$ below this mean, to $2.4\sigma$ above it.
The goodness-of-fit of these eight data to their average is $\chi_{\nu}^{2} = 10.0$, which is an exceedingly high value. 
Although these objects all have low and similar $(\beratio)_0$, they decidedly do not form a single population with uniform $(\beratio)_0$.

The same conclusions are reached if only considering the three FUN CAIs: their mean is $4.37 \times 10^{-4}$, but $\chi_{\nu}^{2} = 6.2$.
If only the {\it HAL}-like inclusions {\it HAL}, {\it SHAL} and {\it Hidalgo} are considered, the mean is $4.2 \times 10^{-4}$, but $\chi_{\nu}^{2} = 27$. 
The {\it HAL}-like objects do not form a population with a common $(\beratio)_0$, nor do the FUN CAIs form a population with a common $(\beratio)_0$.

FUN CAIs, and PLACs and SHIBs, each form heterogeneous populations; but this is what is expected if they formed or were reset at random times $t$ after forming from a reservoir with canonical $\beratio = 7 \times 10^{-4}$ at $t\!=\!0$.
This is testable using dating information from the Al-Mg systematics, so we also include in Table~\ref{table:funcais} the initial $({}^{26}{\rm Al}/{}^{27}{\rm Al})_0$ ratios, where known.
If an inclusion formed from the same uniform reservoir of ${}^{26}{\rm Al}$ that most CAIs formed from, then the time of formation inferred from Al-Mg systematics, $\Delta t_{26}$, can be inferred:
\begin{equation}
\Delta t_{26} = 1.03 \, \ln \left[ \frac{ 
({}^{26}{\rm Al}/{}^{27}{\rm Al})_{\rm SS} }{ 
({}^{26}{\rm Al}/{}^{27}{\rm Al})_{0} } \right]
\, {\rm Myr},
\label{eq:AlMg}
\end{equation}
where we take the mean-life of ${}^{26}{\rm Al}$ to be 1.03 Myr \citep{DeschEtal2023anchors1}, and where we will define $({}^{26}{\rm Al}/{}^{27}{\rm Al})_{\rm SS} = 5.23 \times 10^{-5}$ at $t\!=\!0$.

This standard dating technique probably cannot be applied to PLACs, {\it HAL}, {\it SHAL}, or {\it Hidalgo}. 
These hibonite-dominated inclusions are examples of Low-${}^{26}{\rm Al}/{}^{27}{\rm Al}$ Corundum-Hibonite Inclusions (LAACHIs) that almost uniquely do not appear to have formed from the same uniform reservoir of ${}^{26}{\rm Al}$ as other CAIs \citep{DeschEtal2023laachi}.
For example, if {\it HAL} had formed from the same reservoir, its low initial ($\alratio)_0$ would imply a time of formation $\Delta t_{26} > 7 \, {\rm Myr}$, long after it had been incorporated into Allende.
The $(\alratio)_0$ ratios of inclusions dominated by hibonite and corundum can be low because of late resetting, or they could be low because they didn't form with the canonical amount of ${}^{26}{\rm Al}$.
Because of the ambiguity, we cannot use Al-Mg systematics in these objects to rule out a late resetting or formation.

However, Al-Mg dating probably {\it can} be applied to the large, multiminerallic FUN CAIs {\it CMS-1} and {\it KT1}, which should have formed from the canonical reservoir \citep{DeschEtal2023laachi}.
If found to form early, these could be used to rule out a late resetting; but in fact their Al-Mg and Be-B systematics are  consistent with a late time of formation of these two FUN CAIs.
As discussed by \citet{DeschEtal2023anchors2}, Be-B systematics in FUN CAI {\it CMS-1} imply a formation time $\Delta t_{10} > 0.70 \, {\rm Myr}$, while its Al-Mg systematics imply $\Delta t_{26} = 0.63 \pm 0.21$ Myr; a formation time of 0.8 Myr is consistent with both systems.
Likewise, Be-B systematics in FUN CAI {\it KT1} imply $\Delta t_{10} = 0.75 \pm 0.15$ Myr, while its Al-Mg systematics imply $\Delta t_{26} > 0.76$ Myr; a formation time of 0.8 Myr is consistent with both systems for this CAI as well. 

\begin{deluxetable}{lccccccc}
\tabletypesize{\scriptsize}
\tablewidth{0pt}
\tablecaption{$\beratio$ ratios in FUN CAIs, PLACs and SHIBs}
\label{table:funcais}
\tablehead{
\colhead{Sample}  & \colhead{$(\beratio)_0$ ($10^{-4}$)} & 
\colhead{N} & 
\colhead{MSWD} & 
\colhead{valid?} & 
\colhead{Ref.} & \colhead{$(\alratio)_0$} &
\colhead{Ref.} 
}
\startdata
Type A FUN CAI Axtell {\it 2771} & \textit{3.0} $\pm$ \textit{1.2} & 6 & 5.2 & no & {a} & - & - \\
 " \hspace{0.5in} " \hspace{0.5in} " & $4.0 \pm 0.4$ & 10 & 1.1 & yes & {b}$^{*}$ & - & - \\
Type B FUN CAI {\it CMS-1} & $1.8 \pm 3.2$ & 18 & 0.7 & yes & {c} & $(2.85 \pm 0.57) \times 10^{-5}$ & {d} \\
Type A FUN CAI {\it KT1} & $5.0 \pm 0.5$ & 12 & 0.9 & yes & {b}$^{*}$ & $(-2.2 \pm 4.7) \times 10^{-5}$ & {e} \\
Hibonite FUN CAI {\it HAL} & $4.4 \pm 1.5$ & 6 & ? & yes & {f} & $(5.2 \pm 1.7) \times 10^{-8}$ & {g} \\
Hibonite FUN CAI {\it SHAL} & $3.04 \pm 0.62$ & 6 & 0.85 & yes & {h}$^{*}$ & n/a & - \\
Hibonite FUN CAI {\it Hidalgo} & $7.56 \pm 1.07$ & 5 & 1.2 & yes & {i} & $(1.51 \pm 0.02) \times 10^{-5}$ & {i} \\
20 Murchison SHIBs and PLACs & $5.3 \pm 1.0$ & 23 & 1.1 & yes & {j} & n/a & - \\
3 Murchison PLACs & $5.2 \pm 2.6$ & 9 & 1.4 & yes & {k}$^{*}$ & $< 5 \times 10^{-6}$ & {k} 
\enddata
\tablecomments{Entries in italics considered to represent invalid isochrons. Uncertainties are $2\sigma$. *Denotes isochron recalculated by \citet{DunhamEtal2022}. 
References: 
{\it a.} \citet{MacPhersonEtal2003}; 
{\it b.} \citet{WielandtEtal2012};
{\it c.} \citet{DunhamEtal2022}; 
{\it d.} \citet{WilliamsEtal2017};
{\it e.} \citet{LarsenEtal2011};
{\it f.} \citet{MarhasGoswami2003lpsc};
{\it g.} \citet{FaheyEtal1987};
{\it h.} \citet{LiuKeller2017lpsc};
{\it i.} \citet{LiuEtal2024};
{\it j.} \citet{LiuEtal2010};
{\it k.} \citet{MarhasEtal2002}.
}
\end{deluxetable}

\subsection{Hypothesis Testing}

\subsubsection{Hypothesis \#1: All $\beten$ was inherited from the molecular cloud}

The main prediction of this hypothesis is that all CAIs should have formed from a well-mixed reservoir with $\beratio = (\beratio)_{\rm SS} = 7 \times 10^{-4}$ at $t\!=\!0$.
There should be no CAIs with initial $(\beratio)_0$ exceeding this canonical value outside of measurement errors, and CAIs with lower $(\beratio)_0$ must be interpretable as having formed later.

While several CAIs appear to have $(\beratio)_0 > 7 \times 10^{-4}$, none of these presents a clear case for radiogenic ${}^{10}{\rm B}$ from decay of $\beten$.
This would require a linear correlation between ${}^{10}{\rm B}/{}^{11}{\rm B}$ vs. ${}^{9}{\rm Be}/{}^{11}{\rm B}$ with valid MSWD.
In cases with disqualifying MSWD, it is more likely that the Be-B system has been disturbed by introduction of spallogenic B, unless analysis spots have ${}^{10}{\rm B}/{}^{11}{\rm B} > 0.44$. 
Of the $\approx 69$ CAIs in CV, CO, CR and CH/CB chondrites, only the three CAIs {\it SA301}, {\it Y20-1X1} and {\it Y20-9-1} may present a contradiction to the inheritance model, and we argue that it is much more likely that these inclusions record such high $(\beratio)_0$ and ${}^{10}{\rm B}/{}^{11}{\rm B}$ as an artefact of how \citet{FukudaEtal2019,FukudaEtal2021} acquired or analyzed their data.

The inclusions with $(\beratio)_0 < 7 \times 10^{-4}$ we interpret as having been formed or reset late. 
If the FUN CAIs {\it CMS-1}, {\it KT1} and Axtell {\it 2771} formed at $t \approx 0.8$ Myr, this would reconcile their Be-B and Al-Mg systematics.
The hibonite-dominated PLACs and SHIBs may have formed at $t\!=\!0$ and been reset at $t \approx 1$ Myr for their Be-B systematics in transient thermal events in the nebula. 
This would not be obvious from their Al-Mg systematics, because the initial $(\alratio)_0$ ratios of these ``LAACHI" inclusions are already low because of their unique origins \citep{DeschEtal2023laachi}.
A late formation or resetting for CAIs with $(\beratio)_0 < 7 \times 10^{-4}$ cannot be ruled out. 

Although only $80-90\%$ of CAIs actually record $(\beratio)_0 \approx 7 \times 10^{-4}$, we find that essentially {\bf all} CAIs are consistent with formation in a nebula with uniform $(\beratio)_{\rm SS} = 7 \times 10^{-4}$. 
This is a probable starting value for molecular clouds. 
Inheritance remains a viable hypothesis.

\subsubsection{Hypothesis \#2: All $\beten$ was created by SEP irradiation in the disk}

This hypothesis is viable only if SEP irradiation can produce reservoirs in the disk with $\beratio \approx 7 \times 10^{-4}$. 
This would require substantial enhancements in the SEP flux above the present-day value. 
{
According to \citet{Jacquet2019} and the calculations we have performed here, enhancements by factors $\sim 10^6$ are required.} 

In principle, this hypothesis can  be falsified by finding uniform $(\beratio)_0$ ratios among CAIs, because irradiation models predict initial $(\beratio)_0$ values that vary with time $t$ and location $r$ of a CAI's formation, as $\propto t^{+1} \, r^{-3/2}$.
Although the times and locations of CAIs' formation are not known, it does not seem likely that all CAIs in such different carbonaceous chondrites (CV, CO, CR and CH/CB) could form at identical times and locations.
\citet{DunhamEtal2022}, considering models like those of \citet{YangCiesla2012} and \citet{DeschEtal2018} [in which CAIs form over several $\times 10^5$ yr and from $\approx 0.1 - 1$ AU] estimated that roughly 25\% of CAIs (roughly 15 out of 60) should have $(\beratio)_0$ measurably different from the average value. 
Considering the analyses with valid MSWD, at most 2 or 3 ($< 5\%$), and probably 0 ($0\%$) of the 60+ analyses record higher values.
This represents a severe challenge to irradiation models, even if flux enhancements $\Phi > 10^6$ were justified. 

\subsubsection{Hypothesis \#3: FUN CAIs contain only $\beten$ inherited from the molecular cloud}

Although a hybrid model might seem to offer the best chance to match observations, it is perhaps the most robustly refuted.
In this model, it is assumed that FUN CAIs and PLACs and SHIBs acquired their $(\beratio)_0 \approx 4 \times 10^{-4}$ from the molecular cloud, and all other CAIs received additional $\beten$ raising their $(\beratio)_0$ by $3 \times 10^{-4}$.
Thus irradiation need provide only half of the $\beten$ as in Hypothesis \#2.
But the demands on the radiation field are practically the same: instead of needing a SEP flux $> 10^6$ times the present-day flux, an enhancement factor $\Phi > 5 \times 10^5$ is demanded.
This still far exceeds the {\Steve $< 3 \times 10^3$} enhancement supported by observations of protostars.
Also, even though half of the $\beten$ would be uniform in such a model, significant variations should still exist in the portion derived from irradiation, and tens of percent of CAIs probably still would have $(\beratio)_0$ measurably higher than the average, still exceeding the $0-5\%$ of CAIs that show variations.

The strongest argument against this model is that it predicts that the $(\beratio)_0$ ratios among FUN CAIs, PLACs and SHIBs should be uniform.
Although all of these inclusions record similar $(\beratio)_0 \approx (3-5) \times 10^{-4}$, they are quite distinct.
The three FUN CAIs differ from each other in their initial $(\beratio)_0$ far outside measurement errors, and the three {\it HAL}-like inclusions ({\it HAL}, {\it SHAL}, {\it Hidalgo}) differ from each other even more significantly.

\subsubsection{Summary}

From an astrophysical perspective, SEP irradiation could form a significant fraction of the $\beten$ in the solar nebula only if the SEP flux of the early Sun exceeded the present-day flux by a factor $\Phi > 10^6$.
As almost all protostars are observed to have enhancements {\Steve $\Phi \approx 3 \times 10^2 - 3 \times 10^3$}, only a few percent at most of $\beten$ can be expected to form by irradiation.
In contrast, GCR spallation of molecular cloud gas will produce $\beratio = 7 \times 10^{-4}$ with high probability.

From a meteoritic perspective, properly interpreted data yield no evidence for the variations in $(\beratio)_0$ expected if much of the $\beten$ were produced by SEP irradiation in the disk.
Proper interpretation means rejection of regressions with high MSWD; they are not valid as isochrons.
Such CAIs probably record evidence of disturbance, e.g., spallogenic B due to direct irradiation on the parent body, or aqueous alteration mixing with a spallogenic reservoir.
Even these possibilities could be ruled out if the ${}^{10}{\rm B}/{}^{11}{\rm B}$ ratios of analysis spots exceeded 0.44; but this is not the case for any of the regressions with high MSWD.
Only the three CAIs {\it SA301}, {\it Y20-9-1} and {\it Y20-1X1} analyzed by \citet{FukudaEtal2019} and \citet{FukudaEtal2021} could provide contradictory evidence, but we argue their high $(\beratio)_0$ are artefacts of the analysis done by that group.
We conclude that no CAIs have unambiguous $(\beratio)_0 > 7 \times 10^{-4}$.
FUN CAIs, PLACs and SHIBs have $(\beratio)_0 < 7 \times 10^{-4}$, but these are interpretable as arising from late formation or resetting, and do not demand a reservoir with different $\beratio$.
The low values of these CAIs are too variable to be attributed to inheritance.

The evidence from $\beten$ is that insignificant quantities of $\beten$ were created in the disk. 
Assuming there would be noticeable variations in $(\beratio)_0$ ratios if $> 10\%$ of $\beten$ were produced by SEP irradiation, an upper limit $\Phi < 10^5$ can be assumed.

\section{Live $\beseven$ in the early Solar System?}
\label{sec:beseven}

The ultra-short-lived radionuclide $\beseven$ decays to ${}^{7}{\rm Li}$ with a half-life of only 53 days \citep{JaegerEtal1996}.
If its one-time presence were confirmed in meteoritic samples, it would demand production within the solar nebula, almost certainly by intense SEP irradiation at levels correpsonding to $\Phi > 10^6$.
Evidence for live $\beseven$ in the early Solar System has been claimed several times, based on reported linear correlations between inferred excesses in initial $\liratio$ and ${}^{9}{\rm Be}/{}^{6}{\rm Li}$ ratios in different analysis spots in three different CAIs: 
Allende {\it 3529-41} \citep{ChaussidonEtal2006}, Efremovka {\it E40} \citep{MishraMarhas2019}, and the Dar al Gani 027  inclusion known as {\it Hidalgo} \citep{LiuEtal2024}.
However, an exceedingly important point is that in all three CAIs, the actual {\it measured} values of $\liratio$ in each spot are consistent with chondritic ratio of 12.06 \citep{SeitzEtal2007} or lower; all excesses above this level are {\it inferred} based on the {\it model-dependent} corrections for cosmogenic Li.

Corrections for cosmogenic Li are necessary: during the transit of a meteoroid through space before landing on Earth, GCRs and secondary particles will spall nuclei (especially O) and create Li nuclei, with $\liratio < 2$.
Whatever $\liratio$ a spot starts with---values $\approx 12$ if chondritic, or $> 12$ if there has been Rayleigh distillation or addition of radiogenic ${}^{7}{\rm Li}$---the ratio will be lowered by the addition of cosmogenic Li.
Because spots with smaller [Li] and larger ${}^{9}{\rm Be}/{}^{6}{\rm Li}$ will see their $\liratio$ values decreased more than spots with larger [Li], addition of cosmogenic Li necessarily shifts all points downward but arrayed along a line with negative slope in ${}^{7}{\rm Li}/{}^{6}{\rm Li}$ vs.\ ${}^{9}{\rm Be}/{}^{6}{\rm Li}$ space.
This effect must be corrected for.

However, this correction must be done properly, to avoid non-physical results.
During correction, relatively more cosmogenic Li is subtracted from points with low [Li] and high ${}^{9}{\rm Be}/{}^{6}{\rm Li}$, and the correction therefore introduces a {\it positive} slope in the $\liratio$ vs.\ ${}^{9}{\rm Be}/{}^{6}{\rm Li}$.
If the amount of cosmogenic Li to be subtracted is overestimated, a positive slope mimicking an isochron would be the result, whether or not the inclusion contained live $\beseven$.
In addition, it must be recognized that measurement errors are magnified during the correction, leading to very large uncertainties in the inferred pre-exposure values.

While some correction for cosmogenic Li is necessary, we show that all three CAIs above have been overcorrected, introducing an overly large positive slope that has been interpreted as evidence for live $\beseven$.
In addition, the uncertainties in the correction have been vastly underestimated, making the $\liratio$ values seem inconsistent with single values (zero slopes). 
In what follows, we correct the data in the manner described by \citet{DeschOuellette2006}, but updated to account for the high oxygen wt\% and very low [Li] of {\it Hidalgo}, and to propagate the uncertainties in the data.
We find that in all three cases, the inferred $\liratio$ values (before irradiation) are consistent with uniform values uncorrelated with ${}^{9}{\rm Be}/{}^{6}{\rm Li}$, removing any evidence for live $\beseven$.

\subsection{Allende CAI 3529-41}

\citet{ChaussidonEtal2006} claimed evidence for live $\beseven$ on the basis of $\liratio$ and ${}^{9}{\rm Be}/{}^{6}{\rm Li}$ measurements they made in 68 spots in melilite, fassaite and anorthite in Allende CAI {\it 3529-41}.
They restricted further analysis to 37 spots with [Be] concentrations they considered consistent with crystallization from a melt.
These measured spots had $\liratio$ ratios ranging from 
$10.41 \pm 0.23$ [spot II.7(1)]
to $11.99 \pm 0.82$ [spot III.6]. i.e., there were no observed excesses above the chondritic value.
\citet{ChaussidonEtal2006} inferred excesses after making corrections for cosmogenic Li, after which the $\liratio$ values of spots ranged from $10.46 \pm 0.27$ to $13.46 \pm 0.83$. 
They claimed the data formed an isochron with slope $({}^{7}{\rm Be}/{}^{9}{\rm Be})_0 = (6.1 \pm 1.3) \times 10^{-3}$.

\citet{DeschOuellette2006} identified several problems with this analysis (but see response by \citet{ChaussidonEtal2006reply}).
First, the regression was incorrectly performed without using measurement errors for each point ({\it Isoplot} Model 2).
The actual slope should be
${}^{7}{\rm Be}/{}^{9}{\rm Be} = (9.8 \pm 0.6) \times 10^{-3}$.
\citet{ChaussidonEtal2006} did not report MSWD.
For this best-fit line found by \citet{DeschOuellette2006}, MSWD $= 4.26$, which is an exceptionally poor fit; the probability of MSWD at least this large for $N = 37$ points is $< 0.1\%$.
Most of this bad fit is due to the many points (mostly in melilite) with ${}^{9}{\rm Be}/{}^{6}{\rm Li} < 30$, which \citet{ChaussidonEtal2006} considered disturbed by late-stage aqueous alteration but included in the regression anyway. 
Most importantly, \citet{ChaussidonEtal2006} almost certainly overcorrected for cosmogenic Li.

Corrections for cosmogenic Li are needed.
During passage of a meteoroid to Earth, nuclei of Li (${}^{7}{\rm Li}$ and ${}^{6}{\rm Li}$), B (${}^{11}{\rm B}$ and ${}^{10}{\rm B}$), and Be (including ${}^{9}{\rm Be}$, cosmogenic $\beten$, and cosmogenic $\beseven$ that quickly decays to ${}^{7}{\rm Li}$) are produced when primary GCR protons and $\alpha$ particles across a range of energies spall nuclei of O, Al, Ca, Mg and Si;
additional nuclei are produced when these same nuclei are spalled by secondary protons and neutrons, created at rates that depend on the composition.
Due to the complexity of this process, it would be ideal to experimentally determine Li production rates in carbonaceous chondrites by irradiation by a spectrum of energetic particles.

Instead, what has actually been experimentally determined are production rates of $\beten$ after irradiation of ordinary chondrite material by a single energy of protons \citep{LeyaEtal2000stony}, or production rates of $\beten$ and $\beseven$ by irradiation of stony material \citep{LeyaEtal2000gabbro}.
The corrections for cosmogenic Li made by both \citet{ChaussidonEtal2006} and \citet{DeschOuellette2006} are extrapolations from these experiments, anchored to either the $\beten$ or $\beseven$ production rates. 
\citet{ChaussidonEtal2006} anchored the production rate of Li nuclei to the $\beten$ production rate \citep{LeyaEtal2000stony} by comparing the the cross sections for ${}^{16}{\rm O}(p,x){}^{6}{\rm Li}$ and 
${}^{16}{\rm O}(p,x){}^{7}{\rm Li}$ plus 
${}^{16}{\rm O}(p,x){}^{7}{\rm Be}$ (because it decays to ${}^{7}{\rm Li}$) to the cross section for 
${}^{16}{\rm O}(p,x){}^{10}{\rm Be}$.
They used the cross sections compiled by \citet{ReadViola1984} and \citet{MichelEtal1989}, but at a single proton energy of 600 MeV,
Ultimately, they computed that the 5.2 Myr of cosmic ray exposure experienced by Allende \citep{SchererSchultz2000} would lead to increases of $3.34 \times 10^{14}$ atoms/kg of ${}^{6}{\rm Li}$ and $6.10 \times 10^{14}$ atoms/kg of ${}^{7}{\rm Li}$.
\citet{DeschOuellette2006} anchored the production rates of Li nuclei to the production rate of ${}^{7}{\rm Be}$ in stony meteorites \citep{LeyaEtal2000gabbro}, using energy averages of the improved cross sections of \citet{MoskalenkoMashnik2003}.
Ultimately, they computed that 5.2 Myr of cosmic ray exposure would lead to increases of $2.12 \times 10^{14}$ atoms/kg of ${}^{6}{\rm Li}$ and $3.42 \times 10^{14}$ atoms/kg of ${}^{7}{\rm Li}$.
The estimated $\liratio$ ratios of spallogenic Li are similar (1.82 vs.\ 1.62), but \citet{ChaussidonEtal2006} predict a factor of 1.7 more cosmogenic Li than \citet{DeschOuellette2006} for the same cosmic ray exposure age. 

The approach of \citet{DeschOuellette2006} is to be favored. 
From a modeling perspective, \citet{ChaussidonEtal2006} used cross sections for the reactions at a specific proton energy, 600 MeV, at which many of the reactions have resonances, instead of the energy-averaged cross sections used by \citet{DeschOuellette2006}.
From an experimental perspective, \citet{DeschOuellette2006} predict 5.0 cosmogenic ${}^{7}{\rm Be}$ nuclei produced per cosmogenic $\beten$ nucleus, closer to the experimentally determined proportion of 3.5 \citep{LeyaEtal2000gabbro} than the value of 7.4 predicted by the approach of \citet{ChaussidonEtal2006}, consistent with \citet{ChaussidonEtal2006} predicting a factor of $7.4/5.0 \approx 1.5$ times more cosmogenic Li than \citet{DeschOuellette2006}.



\citet{Leya2011} confirmed many of the conclusions of \citet{DeschOuellette2006}.
He analyzed the data of \citet{ChaussidonEtal2006} using updated calculations for the Li isotope production cross sections, and concluded that they had indeed overestimated the correction for cosmogenic Li.
For example, in one anorthite spot with the largest correction (III.1-6), \citet{ChaussidonEtal2006} had corrected from $\liratio = 12.0$ to $\liratio = 13.5$, but \citet{Leya2011} recommended a correction to only $\liratio \approx 12.9$, suggesting that this one point had been overcorrected by a factor of 1.7, just as \citet{DeschOuellette2006} suggested.
\citet{Leya2011} noted that the melilite data appear to show a normal distribution about a single (near-chondritic) value in their $\liratio$ ratios, indicating that Li in melilite is most likely disturbed. 
Whether Li isotopes were disturbed or not in fassaite and anorthite is less clear, but \citet{Leya2011} found that even in these data, a linear relationship between ${}^{7}{\rm Li}/{}^{6}{\rm Li}$ and ${}^{9}{\rm Be}/{}^{6}{\rm Li}$, demanded by the hypothesis of live ${}^{7}{\rm Be}$, must be rejected.

%

After making their (smaller) correction for cosmogenic Li, \citet{DeschOuellette2006} found that the maximum $\liratio$ of any spots in CAI {\it 3529-41} were $12.89 \pm 0.89$ (III.6) and $12.73 \pm 0.61$ (III.1-6), both consistent with a chondritic value. 
The linear regression yielded slope $({}^{7}{\rm Be}/{}^{9}{\rm Be})_0 = (9.2 \pm 0.6)\times 10^{-4}$, with disqualifyingly high MSWD $=4.14$, almost identical to the previous fit.
That is because most of the slope was driven by points with ${}^{9}{\rm Be}/{}^{6}{\rm Li} < 30$ that \citep{ChaussidonEtal2006} considered late-stage alteration products but included anyway. 
Removing these points from the regression, \citet{DeschOuellette2006} found a slope $({}^{7}{\rm Be}/{}^{9}{\rm Be})_0 = (1.0 \pm 1.2) \times 10^{-4}$, with acceptable MSWD $= 0.72$.
These data that are probably not late-stage alteration products now do conform to a linear trend, but one with slope unresolved from zero.

There definitely is extensive alteration of Li isotopes in Allende CAI {\it 3529-41}, including significant contributions of spallogenic Li to many points, especially those with small ${}^{9}{\rm Be}/{}^{6}{\rm Li} < 30$ \citep{DeschOuellette2006}.
This is also evidenced by the extremely poor MSWD of the Be-B isochron, which suggests alteration of the B isotopes (\S 5.2.1).
But there is no evidence for live $\beseven$ incorporated into this CAI.

\subsection{Efremovka CAI E40}

\citet{MishraMarhas2019} claimed evidence for incorporation of live $\beseven$ in the CV3 Efremovka CAI {\it E40}, based on a correlation between ${}^{7}{\rm Li}/{}^{6}{\rm Li}$ and ${}^{9}{\rm Be}/{}^{6}{\rm Li}$ as measured in 22 spots within the CAI.
After correcting for cosmogenic Li using the identical approach of \citet{ChaussidonEtal2006} and assuming an 11.4 Myr cosmic ray exposure age, and regressing the inferred pre-exposure ${}^{7}{\rm Li}/{}^{6}{\rm Li}$ values, they found a slope $({}^{7}{\rm Be}/{}^{9}{\rm Be})_0 = (1.2 \pm 1.0) \times 10^{-3}$ and intercept $(\liratio)_0 = 12.02 \pm 0.16$, with MSWD of 1.9.
We have regressed the same data using the algorithm of \citet{StephanTrappitsch2023} and find very similar results: $({}^{7}{\rm Be}/{}^{9}{\rm Be})_0 = (1.23 \pm 0.71) \times 10^{-3}$ and intercept $({}^{7}{\rm Li}/{}^{6}{\rm Li})_0 = 12.02 \pm 0.11$, and MSWD $= 1.88$.
The slope is resolved from zero; however, the MSWD is disqualifyingly high. 
For $N = 22$ points, the probability of an MSWD at least this high is $< 1\%$.
This is a strong indication that an additional process has affected the data.
Although we do not identify the source of the disturbance, it may be justifiable to treat one spot, 14Dec-E40-2, as an outlier.
Removing it from the regression, we find slope and intercept $({}^{7}{\rm Be}/{}^{9}{\rm Be})_0 = (1.00 \pm 0.72) \times 10^{-3}$ and $({}^{7}{\rm Li}/{}^{6}{\rm Li})_0 = 12.02 \pm 0.11$, with MSWD $= 1.45$, which is acceptable.
Therefore, if it is justified to remove 14Dec-E40-2 from the regression, and if the correction for cosmogenic Li has been done properly, then there would be evidence for live $\beseven$, although at a level barely resolved from zero.


We have repeated their calculation using the corrections for cosmogenic Li based on the approach of \citet{DeschOuellette2006}, but updated to allow Li isotopic compositions not necessarily near $\liratio \approx 12$ (appropriate for Allende {\it 3529-41} and Efremovka {\it E40}, but not {\it Hidalgo}).
Formulas for the values of both $({}^{9}{\rm Be}/{}^{6}{\rm Li})_{\rm corr}$ and $({}^{7}{\rm Li}/{}^{6}{\rm Li})_{\rm corr}$, and their uncertainties, appear in the Appendix.
As described above (\S 6.1), this correction yields a factor of 1.7 less cosmogenic Li than the approach of \citet{ChaussidonEtal2006} would, although accounting for the fact that CAI {\it E40} is made of melilite, \citet{MishraMarhas2019} probably overcorrected by a factor of only 1.4.

In addition, the cosmic ray exposure age is probably less than \citet{MishraMarhas2019} assumed. 
Although the conference abstract of \citet{MurtyEtal1996lpsc} reported $t_{\rm CRE} = 11.4 \pm 1.7 \, {\rm Myr}$ based on cosmogenic Ne, the more comprehensive (and peer-reviewed) work of \citet{SchererSchultz2000} reported $t_{\rm CRE} = 9.4 \, {\rm Myr}$,
citing \citet{MazorEtal1970}, who actually reported 8.8 Myr.
\citet{MishraMarhas2019} overestimated the cosmic ray exposure age by a factor of at least 1.2.

We have made the corrections for cosmogenic Li using the approach of \citet{DeschOuellette2006}, including adjustments to the ${}^{9}{\rm Be}/{}^{6}{\rm Li}$ ratios, and $t_{\rm CRE} = 9.4 \, {\rm Myr}$, and assuming {\it E40} is made of melilite.
We find the corrected data have weighted mean $\liratio = 12.07 \pm 0.09$, and almost all points are consistent with this value, but with disqualifyingly high MSWD $= 2.3$, driven by two outliers (14Dec-E40-2 and 19Dec-E40-4), off by $3.8\sigma$. 
Removing these, the remaining 20 spots are consistent within $< 1.5\sigma$ with a uniform value $\liratio = 12.02 \pm 0.10$.
The MSWD of the data about this weighted mean is 0.99.

Alternatively, we can linearly regress the data to find an isochron with slope $({}^{7}{\rm Be}/{}^{9}{\rm Be})_0 = (8.68 \pm 6.11) \times 10^{-4}$, intercept $(\liratio)_0 = 12.02 \pm 0.10$, but disqualifyingly high MSWD $= 2.06$ due to the same two $3.5\sigma$ outliers (14Dec-E40-2 and 19Dec-E40-4). 
Removing them and redoing the regression, we find slope $({}^{7}{\rm Be}/{}^{9}{\rm Be})_0 = (6.38 \pm 6.52) \times 10^{-4}$, intercept $(\liratio)_0 = 11.98 \pm 0.10$, and acceptable MSWD $= 0.85$; for 20 points, the probability of MSWD $< 0.85$ is 36\%.
Using our correction and $t_{\rm CRE} = 9.4$ Myr, we find the slope is (just) unresolved from zero. 
While larger values of $t_{\rm CRE}$ might yield slopes resolved from zero, there is an equal probability of smaller values of $t_{\rm CRE}$, which would yields slopes even closer to zero. 

Just as was the case for CAI {\it 3529-41}, high slopes are driven by making a larger correction for cosmogenic Li than are warranted and (to a lesser extent) on including some spots that appear disturbed. 
Our linear regression shows a slope not resolved from zero, and data that are entirely consistent with a uniform value in CAI {\it E40}.


\subsection{Dar al Gani 027 CAI Hidalgo}

Liu et al. (2024) claimed evidence for live $\beseven$ incorporated into the CO3.0 Dar al Gani 027 (DaG 027) CAI {\it Hidalgo}, based on a seeming correlation between the $\liratio$ and ${}^{9}{\rm Be}/{}^{6}{\rm Li}$ values among four spots.
They corrected for cosmogenic Li using the approach of \citet{ChaussidonEtal2006} and \citet{DeschOuellette2006}, but used a slightly higher production rate of ${}^{7}{\rm Be}$ based on CI chondrites \citep{Leya2011} and assuming a cosmic ray exposure age $t_{\rm CRE} = 6.2$ Myr, the mean age of five tentatively paired meteorites from the same strewn field.
(Adjustment for spallogenic Be is not necessary because [Be] $\sim 300$ ppb, while [Li] $\sim 0.05$ ppb.)
Their measured values of ${}^{9}{\rm Be}/{}^{6}{\rm Li}$ and $\liratio$ for each spot, and the values they used after correcting for cosmogenic Li, are listed in {\bf Table~\ref{table:hidalgo}}.
They then regressed the corrected data and found a slope $({}^{7}{\rm Be}/{}^{9}{\rm Be})_0 = (2.0 \pm 1.5) \times 10^{-4}$ and intercept $(\liratio)_0 = -4.7 \pm 33.0$, with MSWD $=0.49$.
We have repeated the regression with the algorithm of \citet{StephanTrappitsch2023} and find essentially identical slope 
$({}^{7}{\rm Be}/{}^{9}{\rm Be})_0 = (2.0 \pm 1.5) \times 10^{-4}$ and intercept $(\liratio)_0 = -4.7 \pm 33.1$, with MSWD $=0.48$.
The MSWD is acceptable for $N = 4$ points (the probability of MSWD being at least this low is 38\%), so the data do appear to array along a line with positive slope. 
This would provide evidence for incorporation of live $\beseven$ if the correction for cosmogenic Li were done properly.

We have calculated the corrections for cosmogenic Li using the equations found in the Appendix.
For this CAI, it is especially worth noting that after subtracting cosmogenic Li, $x = ({}^{9}{\rm Be}/{}^{6}{\rm Li})_{\rm meas}$ is corrected to $X_{\rm corr} = x / (1 - \Delta_{\rm cr})$, and $y = ({}^{7}{\rm Li}/{}^{6}{\rm Li})_{\rm meas}$ is corrected to $Y_{\rm corr} = (y - {\cal R} \, \Delta_{\rm cr}) / (1 -\Delta_{\rm cr})$, where ${\cal R}$ is the $\liratio$ of cosmogenic Li and $\Delta_{\rm cr} = \Delta^{6}{\rm Li} / {}^{6}{\rm Li}$ is the fraction of measured ${}^{6}{\rm Li}$ that is cosmogenic.
For this CAI, for which much of the ${}^{6}{\rm Li}$ is cosmogenic ($\Delta_{\rm cr}$ close to 1), these corrections converge to $Y_{\rm corr} / X_{\rm corr} \approx (y - {\cal R}) / x$, regardless of the exact value of $\Delta_{\rm cr}$.
The spots reported by \citet{LiuEtal2024} have different [Li] but all have $\Delta_{\rm cr} \approx 1$, and therefore the corrected values must array along a line with similar slope $(y - {\cal R}) / x$ and intercept near the origin.
For the four spots they reported, the average value of $(y - {\cal R}) / x$ is $1.9 \times 10^{-4}$.
Therefore, if \citet{LiuEtal2024} overcorrected their data, they would obtain exactly the result they reported: the corrected values arrayed along a well-defined line corresponding to slope $({}^{7}{\rm Be}/{}^{9}{\rm Be}) = (2.0 \pm 1.5) \times 10^{-4}$ and intercept near the origin.
This is very suggestive that their inferred slope is a mathematical artefact, due to overcorrection of the data.

Nevertheless, most other processes altering $\liratio$ ratios (e.g., evaporation of Li) would yield a uniform $\liratio$ within the inclusion, so radiogenic contributions might still be inferred if a linear regression yielded a slope resolved from zero, or if the spots had values of $Y_{\rm corr}$ that were not uniform within uncertainties.
To test this, a good estimate must be made of the uncertainties in $X_{\rm corr}$ and $Y_{\rm corr}$.
These are discussed in the Appendix, and we report these in Table~\ref{table:hidalgo} as well.
As it happens, \citet{LiuEtal2024} have greatly underestimated the uncertainties in $Y_{\rm corr}$, which makes the slope of the linear fit appear resolved from zero when it is not, and the values of $Y_{\rm corr}$ appear not consistent with a uniform value when in fact they are.

\begin{deluxetable}{lcccccccc}
\tabletypesize{\scriptsize}
\tablewidth{0pt}
\tablecaption{Data for DaG 027 CAI {\it Hidalgo}}
\tablehead{
\colhead{$\,$} &
\multicolumn4c{Measured$^{a}$} &
\multicolumn2c{Corrected$^{a}$} &
\multicolumn2c{Our correction} \\
\colhead{Spot} & 
\colhead{[Be]$^{b}$} & 
\colhead{[Li]$^{b}$} & 
\colhead{${}^{9}{\rm Be}/{}^{6}{\rm Li}$$^{c}$} & 
\colhead{${}^{7}{\rm Li}/{}^{6}{\rm Li}$} &
\colhead{${}^{9}{\rm Be}/{}^{6}{\rm Li}$$^{c}$} & 
\colhead{${}^{7}{\rm Li}/{}^{6}{\rm Li}$} &
\colhead{${}^{9}{\rm Be}/{}^{6}{\rm Li}$$^{c}$} & 
\colhead{${}^{7}{\rm Li}/{}^{6}{\rm Li}$} 
}
\label{table:hidalgo}
\startdata
spot 1 & 216 & 0.051 &
$3.12 \pm 0.363$ & 
$8.76 \pm 1.49$ & 
$27.2 \pm 10.2$ & 
$62 \pm 37$ & 
$9.18 \pm 2.96$ & 
$22.64 \pm 10.32$ \\
spot 2 & 183 & 0.042 &
$2.74 \pm 0.335$ & 
$7.47 \pm 1.41$ & 
$42.1 \pm 16.2$ & 
$89 \pm 53$ & 
$8.95 \pm 3.60$ & 
$20.76 \pm 11.48$ \\
spot 3 & 270 & 0.043 & 
$3.96 \pm 0.517$ & 
$7.55 \pm 1.47$ & 
$48.0 \pm 18.7$ & 
$71 \pm 43$ & 
$12.54 \pm 5.03$ & 
$20.43 \pm 11.31$ \\
spot 4 & 388 & 0.061 & 
$4.32 \pm 0.402$ & 
$8.10 \pm 1.28$ & 
$13.9 \pm 0.503$ & 
$22 \pm 13$ & 
$8.89 \pm 1.58$ & 
$14.96 \pm 4.40$
\enddata
\tablecomments{{\it a}. \citet{LiuEtal2024}. {\it b}. Both [Be] and [Li] concentrations are in ppb. {\it c}. ${}^{9}{\rm Be}/{}^{6}{\rm Li}$ ratios have been divided by $10^4$. All uncertainties are $2\sigma$.
}
\end{deluxetable}

Because our corrected ${}^{9}{\rm Be}/{}^{6}{\rm Li}$ and $\liratio$ values, listed in Table~\ref{table:hidalgo}, differ significantly from those of \citet{LiuEtal2024}, we explicitly go through the example of spot 1 to justify them.
\citet{LiuEtal2024} measured [Li] $= 0.051$ ppb and $\liratio = 8.76 \pm 1.49$ in spot 1.
The number of ${}^{6}{\rm Li}$ atoms in 1 kg of sample is therefore 
\[
\#^{6}{\rm Li} = (0.051 \times 10^{-9}) \, \left[ 6.015 + 7.016 \, (8.76 \pm 1.49) \right]^{-1} \, (1.661 \times 10^{-27})^{-1} = (4.55 \pm 0.68) \times 10^{14}
\]
and the number of ${}^{7}{\rm Li}$ atoms per kg is 
\[
\#^{7}{\rm Li} = (8.76 \pm 1.49) \times \#^{6}{\rm Li} = (39.86 \pm 4.67) \times 10^{14}.
\]
\citet{LiuEtal2024} assumed $4.01 \times 10^{14}$ atoms/kg of ${}^{6}{\rm Li}$, and $4.66 \times 10^{14}$ atoms/kg of ${}^{7}{\rm Li}$ were cosmogenic, and therefore inferred abundances before GCR irradiation of roughly $0.54 \times 10^{14}$ atoms/kg of ${}^{6}{\rm Li}$ and $35.20 \times 10^{14}$ atoms/kg of ${}^{7}{\rm Li}$, yielding $\Delta_{\rm cr} = 0.88$ and resulting in a pre-exposure ratio $Y_{\rm corr} \approx 65$.
\citet{LiuEtal2024} reported $Y_{\rm corr} = 62$, but also an unjustifiably low uncertainty of $\pm 37$ ($\pm 60\%$).
Because the uncertainty in $\#^{6}{\rm Li}_{\rm meas}$ is $\pm 0.68 \times 10^{14}$ atoms/kg, and the corrected $\#^{6}{\rm Li}_{\rm corr} = 0.54 \times 10^{14}$ atoms/kg, the uncertainty in $\#^{6}{\rm Li}_{\rm corr}$ should be $\pm 126\%$, and the relative error in $Y_{\rm corr}$ should be similar.
Reproducing their correction, we indeed find $Y_{\rm corr} = 63 \pm 85$, about $130\%$ uncertainty ($2\sigma$).
Thus, \citet{LiuEtal2024} have underestimated the uncertainties by a factor of two.

We also infer a smaller correction.
As discussed in the Appendix, we calculate that $2.99 \times 10^{14}$ atoms/kg of ${}^{6}{\rm Li}$ and $4.83 \times 10^{14}$ atoms/kg of ${}^{7}{\rm Li}$ were added in 6.2 Myr.
We therefore infer that before irradiation, there were $(1.56 \pm 0.68) \times 10^{14}$ atoms/kg of ${}^{6}{\rm Li}$, and $(35.03 \pm 0.47) \times 10^{14}$ atoms/kg of ${}^{7}{\rm Li}$, resulting in $\Delta_{\rm cr} = 0.66$ and corresponding to a pre-exposure ratio $Y_{\rm corr} \approx 22.5 \pm 9.8$ 
(or more precisely, $Y_{\rm corr} = 22.64 \pm 10.32$).
Similar results apply to the other analysis spots.

The effects of these differences are as follows.
Accepting the correction made by \citet{LiuEtal2024} but propagating the uncertainties correctly, we would find slope roughly $(2.0 \pm 2.5) \times 10^{-4}$, i.e., not resolved from zero. 
Indeed, the weighted mean of the $Y_{\rm corr}$ would be 23.4, and all the corrected ${}^{7}{\rm Li}/{}^{6}{\rm Li}$ ratios would match this value within $1\sigma$.
Using our recommended correction, the slope would be $(-5.7 \pm 10.4) \times 10^{-4}$, definitely not resolved from zero. 
The weighted mean of the $Y_{\rm corr}$ would be 17.0, and all the corrected ${}^{7}{\rm Li}/{}^{6}{\rm Li}$ ratios would match this value within $1\sigma$.

\begin{figure}[ht]
\centering
\includegraphics[width=0.49\linewidth]{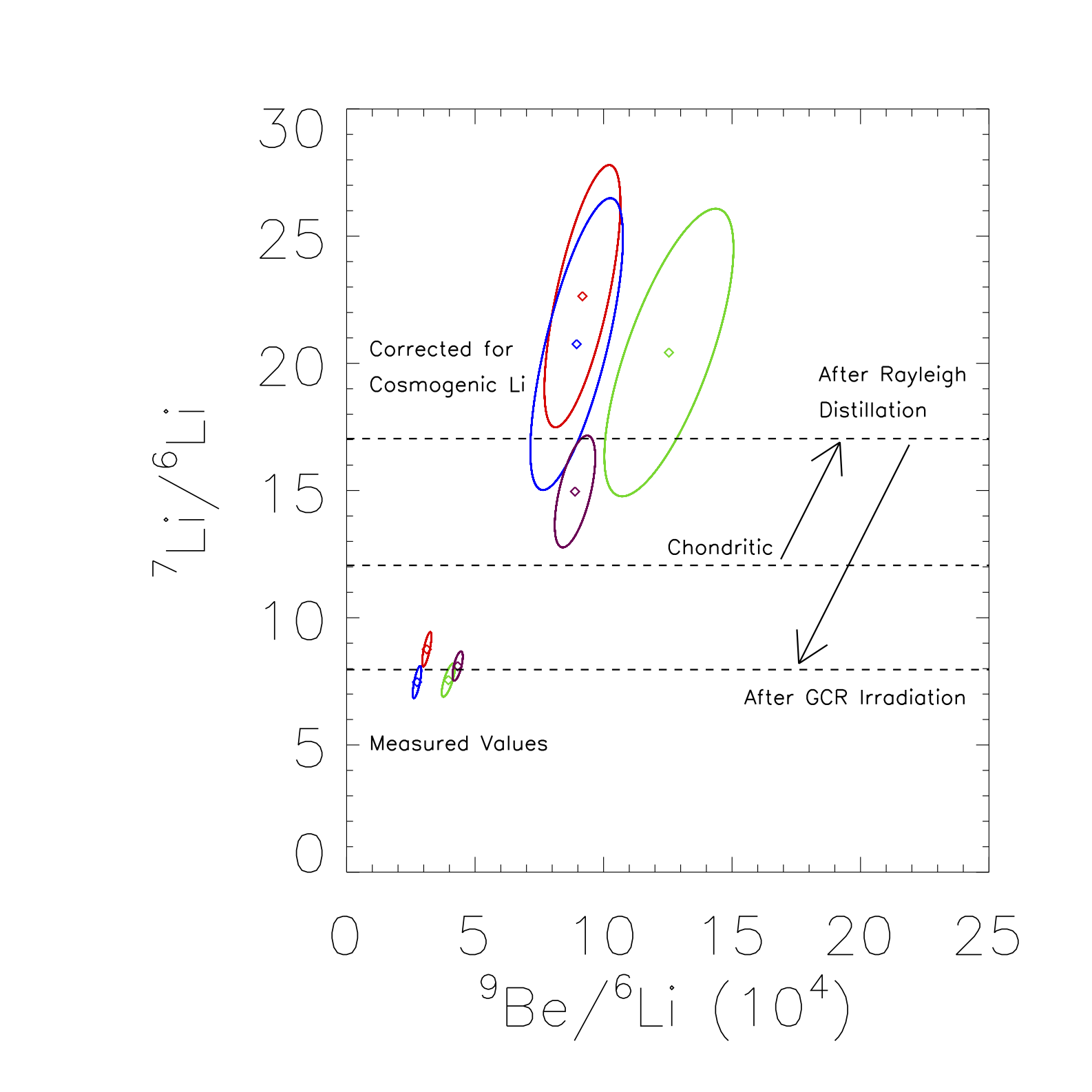}
\includegraphics[width=0.49\linewidth]{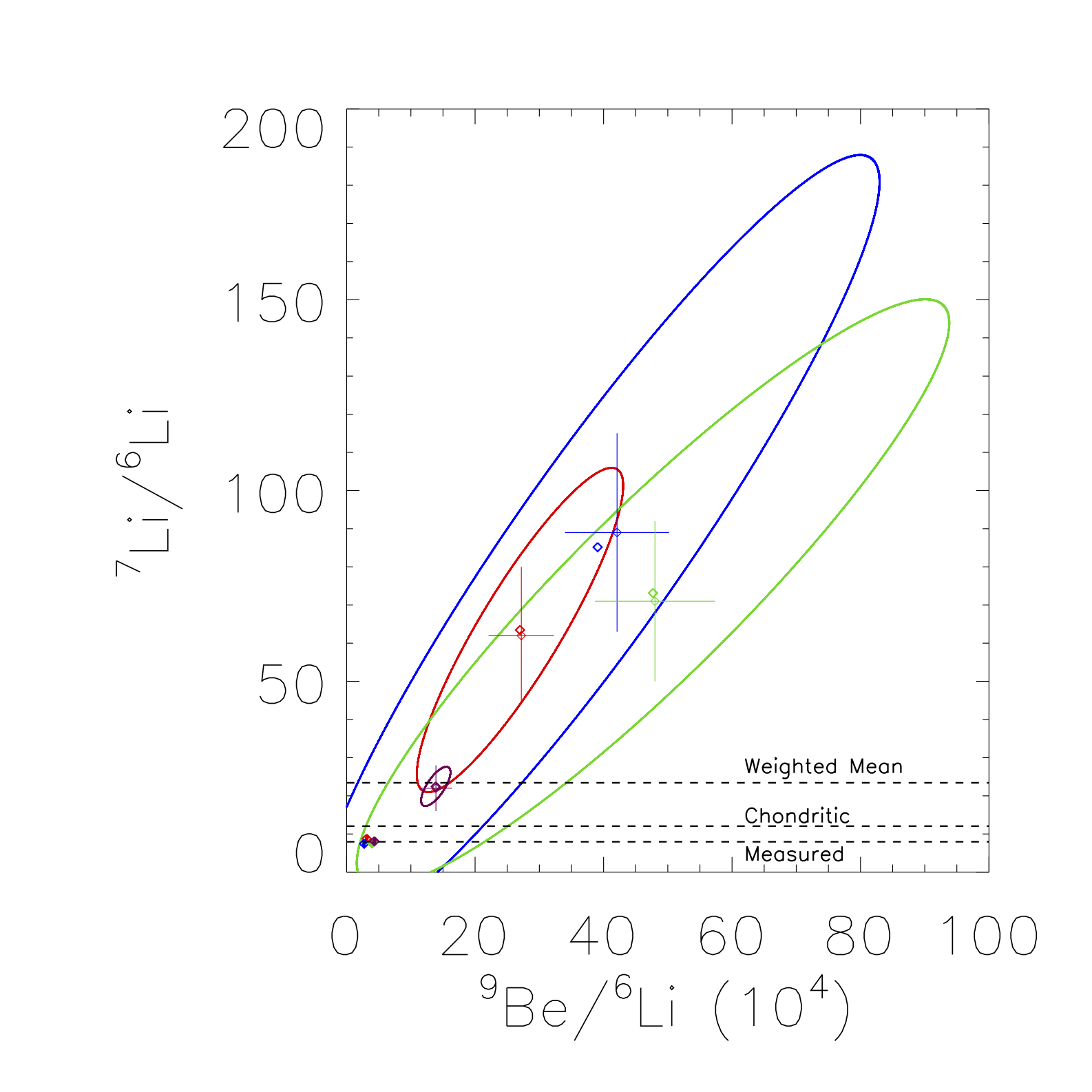}
\caption{({\it Left}):
${}^{7}{\rm Li}/{}^{6}{\rm Li}$ vs. ${}^{9}{\rm Be}/{}^{6}{\rm Li}$ data for CAI {\it Hidalgo}, as measured (lower left corner) by \citet{LiuEtal2024}, with weighted mean $\liratio = 8.0$, and after our corrections for cosmogenic Li.
The spots before GCR irradiation were consistent with a single value, the weighted mean $\liratio = 17.0$.
{\it Hidalgo} is consistent with an initially chondritic $\liratio = 12.0$, raised to a uniform $\liratio = 17.0$ by evaporation and Rayleigh distillation in the nebula, with subsequent addition of cosmogenic Li on the meteoroid (totalling 50 - 70\% of Li), lowering ${}^{9}{\rm Be}/{}^{6}{\rm Li}$ and $\liratio$ to 8.0.
({\it Right}):
${}^{7}{\rm Li}/{}^{6}{\rm Li}$ vs. ${}^{9}{\rm Be}/{}^{6}{\rm Li}$ data for CAI {\it Hidalgo}, as measured (lower left corner) by \citet{LiuEtal2024}, and after their larger correction  (small diamonds with horizontal and vertical error bars) for cosmogenic Li (totalling up to $> 90\%$ of Li).
These would not be consistent with a single value, which led \citet{LiuEtal2024} to infer an isochron with non-zero slope ${}^{7}{\rm Be}/{}^{9}{\rm Be} = (2.0 \pm 1.5) \times 10^{-4}$.
This slope is exactly what would result from overcorrecting the data. 
We reproduce (diamonds) their corrections (large crosses) for cosmogenic Li but find much large uncertainties (large error ellipses).
Even with their greater correction, the $\liratio$ ratios before irradiation would be consistent with a single value (here the weighted mean $\liratio \approx 23.4$).
Data are taken from Table~\ref{table:hidalgo}; see text for details. 
Error bars and error ellipses have been made $1\sigma$. 
Note the difference in scale between the two plots.
}
\label{fig:beseven}
\end{figure}

In {\bf Figure~\ref{fig:beseven}} we 
illustrate the measured data, the data after the correction made by \citet{LiuEtal2024}, and as corrected here.
We find a smaller correction, with cosmogenic Li comprising up to 70\% of total Li, than did \citet{LiuEtal2024}, who consider cosmogenic Li comprising up to $> 90\%$ of total Li. 
But whichever correction is made, the $\liratio$ ratios of all spots before GCR irradiation are consistent with single values after correctly propagating uncertainties: $\liratio \approx 23.4$ for the correction of \citet{LiuEtal2024}, $\liratio \approx 17.0$ for the correction made here.

\bigskip

These results allow tests of hypotheses.
We propose that CAI {\it Hidalgo} formed as a hibonite-dominated object similar to {\it HAL} or {\it DH-H1} \citep{IrelandEtal1992}.
Similar hibonite inclusions have [Li] concentrations ranging from 6 to 6000 ppb \citep{LiuEtal2010}; we assume for this example a value at the low end, $\approx 6$ ppb.
The object {\it DH-H1} was heated and suffered evaporative loss of Ca and Ti, with accompanying isotopic fractionation of $F_{\rm Ca} = +13\permil$ per amu and $F_{\rm Ti} = +11\permil$ per amu, and final ${\rm TiO}_2$ content 0.15wt\% \citep{IrelandEtal1992}, and unknown Li content.
We assume similar heating for {\it Hidalgo}, which would have led to loss of Ca, Ti, and Li, and isotopic fractionation in each.
The heating would have allowed Li diffusion to homogenize the $\liratio$ ratios within it, and evaporative loss would have raised the uniform $\liratio$ value above a presumably chondritic starting ratio $\liratio = 12.06$ \citep{SeitzEtal2007}.
Finally, GCR spallation during residence in the DaG 027 meteoroid for 6.2 Myr added about 0.01 ppb of spallogenic Li with $\liratio \approx 2$, decreasing the values in measured spots to $\liratio \approx 8$.

The data for {\it Hidalgo} are consistent with this scenario. 
Assuming the pre-exposure $\liratio$ ratios of spots 1-4 cluster around a uniform value, that value is the weighted mean $(\liratio)_0 = 17.0 \pm 3.6$, and all four spots are within about $1\sigma$ of this value, and MSWD $= 0.95$, justifying the assumption of a single value.
There is a 92\% probability that the initial value exceeds the chondritic value 12.06, at least for this value of $t_{\rm CRE}$ and [Li] calibration, strongly suggesting that the $\liratio$ ratios were raised by loss of Li by evaporation, by a factor $f$.
The factor $f$ is found from the formula given by \citet{LiuEtal2024}, presuming the inclusion started with a chondritic Li isotopic ratio:
\begin{equation}
\frac{ (\liratio)_{\rm pre} }{ (\liratio)_{\rm chond} } = f^{(\alpha - 1)},
\end{equation}
where $\alpha = (6/7)^{1/2} = 0.9259$.
For $(\liratio)_{\rm pre} = 17.0$, we infer $f = 0.0094$, i.e., loss of $> 99\%$ of Li.
Given a pre-exposure [Li] $\approx 0.04$ ppb, the pre-heating concentration would be [Li] $\approx 4$ ppb, at the low end of the range of other hibonite inclusions \citep{LiuEtal2010}.
The data appear consistent with a hibonite inclusion with [Li] $= 4$ ppb and $\liratio \approx 12$ that was heated and lost Li, resulting in [Li] $= 0.04$ ppb and $\liratio \approx 17$, to which [Li] $= 0.01$ ppb of cosmogenic Li with $\liratio \approx 1.6$ was added, resulting in [Li] $= 0.05$ ppb and $\liratio \approx 8$.

We have not yet considered the systematic uncertainty in $\Delta_{\rm cr}$ of $\pm 30\%$ or more due to uncertainties in $t_{\rm CRE}$ and [Li]. 
Varying $\Delta_{\rm cr}$ by $\pm 30\%$ yields the same results, with data clustering with statistical significance around single values in the range ($\liratio)_{\rm corr} \approx 13$ to $22$, implying [Li] as large as 140 ppb, well within the range of values in hibonites.

Possible objections to this scenario were raised by \citet{LiuEtal2024}.
They favored a starting composition like chondrites, because evaporation of chondritic material resulted in the observed isotopic fractionations in Ca and Ti, in furnace experiments by \citet{FlossEtal1996}.
Of course, this does not rule out non-chondritic starting compositions, but \citet{LiuEtal2024} disfavored a hibonite-like starting composition based on other experiments by \citet{FlossEtal1998}, which suggest corundum would be present as well.
Due to the cutting effect, it is difficult to assess the presence or absence of corundum in {\it Hidalgo} from a thin section, and anyway it is not clear that corundum is necessarily always produced during evaporation.
There is no doubt that more experiments would help to map out the conditions under which the observed fractionations in Ca and Ti could occur, so that better predictions of Li isotopic fractionation could be made. 
Meanwhile, we regard {\it HAL} and {\it DH-H1}, with similar $F_{\rm Ca}$ and $F_{\rm Ti}$ and ${\rm TiO}_2$ abundances, to be good analogs to {\it Hidalgo}; and we follow \citet{DeschEtal2023laachi} and interpret these inclusions as forming from largely hibonite (not chondritic) material.

Regardless of the details of the evaporation and its unclear effects on Ca and Ti or the mineralogy, the above hypothesis, involving no live $\beseven$, is consistent with the measurements of Be and Li isotopes.
As for the hypothesis that {\it Hidalgo} incorporated live $\beseven$, of course the data {\it could} be consistent with a  slope equivalent to $({}^{7}{\rm Be}/{}^{9}{\rm Be})_0 \leq 5 \times 10^{-4}$; but the interpretation of the data as recording radiogenic ${}^{7}{\rm Li}$ from decay of $\beseven$ is not demanded at all.

\subsection{Summary}

Evidence for live $\beseven$ incorporated into a CAI demands $\liratio$ excesses that show a valid (MSWD $\approx 1$) linear correlation with ${}^{9}{\rm Be}/{}^{6}{\rm Li}$, with slope resolved from zero.
This evidence has been claimed in Allende CAI {\it 3529-41}, Efremovka CAI {\it E40}, and DaG 027 CAI {\it Hidalgo}, but incorrectly. 
With one exception, all 92 analysis spots are chondritic or subchondritic in their $\liratio$ ratios.
Excesses are inferred {\bf only} because of model-dependent corrections for cosmogenic Li.
\citet{DeschOuellette2006} demonstrated that the cosmogenic Li corrections made by \citet{ChaussidonEtal2006} were overestimated by a factor $\approx 1.7$. 
These conclusions were largely confirmed by \citet{Leya2011}.
As a result, \citet{ChaussidonEtal2006} inferred a slope of $\liratio$ vs. ${}^{9}{\rm Be}/{}^{6}{\rm Li}$ in spots before GCR spallation resolved from zero; using their revised correction, \citet{DeschOuellette2006} showed the corrected spots showed a slope not resolved from zero.
This CAI appears very disturbed in its Li isotopes.
Similarly, \citet{MishraMarhas2019}, using the same correction as \citet{ChaussidonEtal2006}, found a slope in pre-exposure spots resolved from zero; here we find the slope is not resolved from zero.
Likewise we have re-examined the corrections for cosmogenic Li made by \citet{LiuEtal2024} and found they were overestimated as well, and the uncertainties greatly underestimated,
and the pre-exposure spots show slope not resolved from zero.
In fact, the slope they inferred, $({}^{7}{\rm Be}/{}^{9}{\rm Be})_0 \approx 2 \times 10^{-4}$, is exactly that which would result from overcorrecting the data for cosmogenic Li.
The CAI {\it Hidalgo} appears to have begun its GCR spallation with a uniform $\liratio > 17$, strongly suggesting isotopic fractionation of Li during an earlier, nebular heating phase. 

In any event, none of these CAIs shows evidence for incorporating live $\beseven$.
Any inferences of what SEP flux they experienced are therefore unjustified.

\section{Conclusions}
\label{sec:conclusions}

\subsection{The likely value of $\Phi$}

\begin{figure}[ht]
\centering
\includegraphics[width=\linewidth]{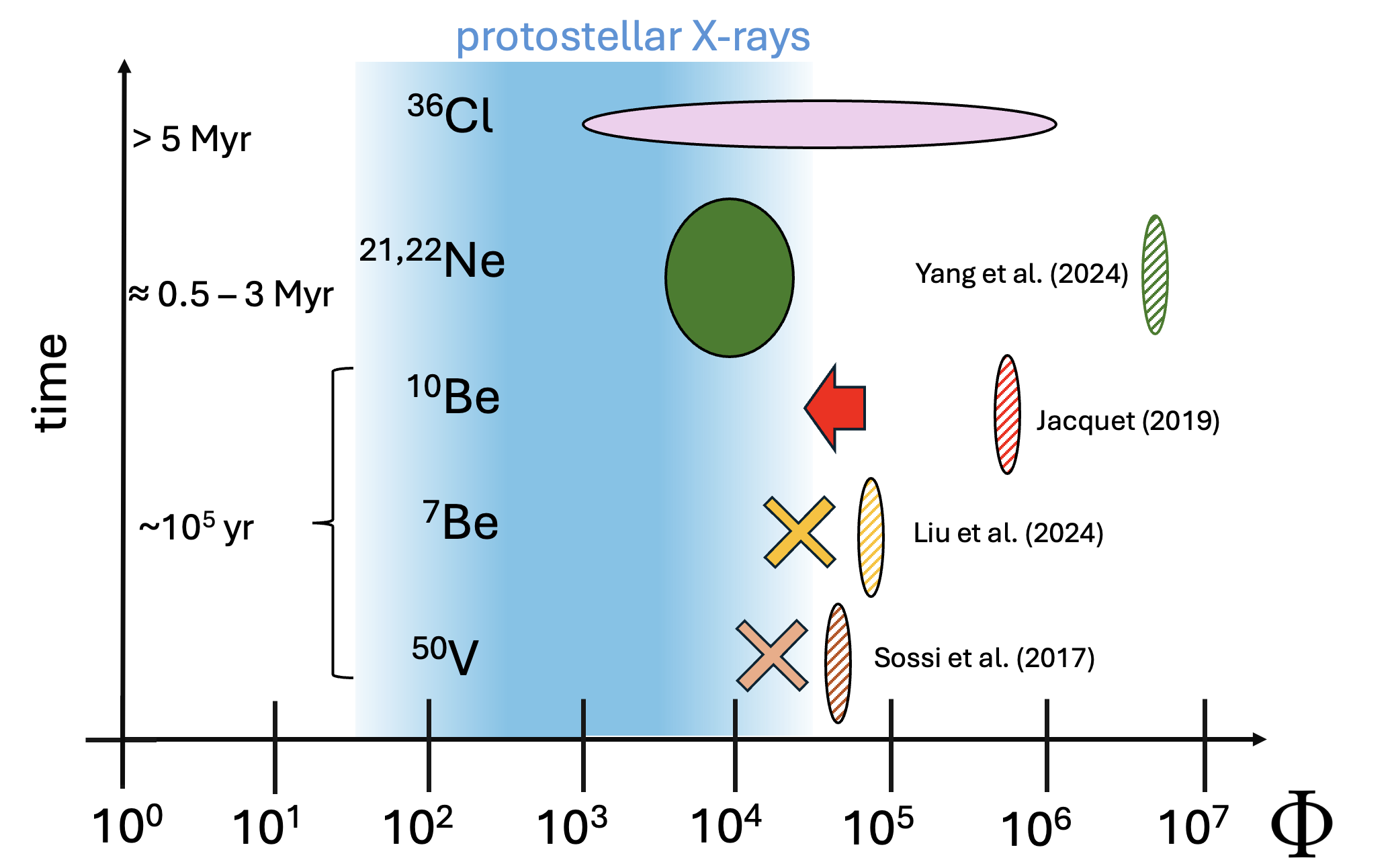}
\caption{
Constraints on $\Phi$ [the factor by which the SEP flux was enhanced above the present-day value, $100 \, {\rm cm}^{-2} \, {\rm s}^{-1}$ particles with $E > 10 \, {\rm MeV}$/nucleon at 1 AU], from various meteoritic data.
\citet{SossiEtal2017} suggested $\Phi \approx 6 \times 10^4$ from V isotope anomalies in CAIs, refuted by \citet{BekaertEtal2021}, who showed these are attributable to evaporation.
\citet{LiuEtal2024} argued for $\Phi \approx 9 \times 10^{4}$ based on a presumed abundance of $\beseven$ in one CAI. 
Our refutation of its existence (\S 6) renders this constraint moot.
\citet{Jacquet2019} argued for $\Phi \approx 9 \times 10^5$ based on perceived heterogeneities in $\beratio$ ratios among CAIs. 
Our demonstration that $\beratio$ ratios were homogeneous at $7 \times 10^{-4}$ (\S 5) puts a limit $\Phi \ll 10^5$ for how much $\beten$ was created by irradiation. 
\citet{YangEtal2024} argued for an exceptionally high value $\Phi = 6 \times 10^6$ based on their interpretaion of how cosmogenic Ne was produced in hibonite grains. 
Our modeling (\S 4) demonstrates that the Ne likely was produced in a disk subject to {\Steve $\Phi \approx 3 \times 10^3$ to $2 \times 10^4$}.
Finally, our modeling of ${}^{36}{\rm Cl}$ production after disk dissipation (\S 3) demands $\Phi < 10^6$, possibly {\Steve $\approx 10^3$}, depending on the chemical composition of ice. 
No meteoritic data {\Steve demand values greater than $\Phi \approx 3 \times 10^3$, which makes the protoSun consistent with the range $\Phi \approx 3 \times 10^2$ to $3 \times 10^3$ }inferred from X-ray observations of protostars \citep{WolkEtal2005}, in the shaded bars (which include some uncertainty about the proton flux per X-ray flux).
The early Sun appears to have been a typical protostar, and data or models suggesting otherwise should be scrutinized.
}
\label{fig:phi}
\end{figure}

In this paper we have considered multiple types of meteoritic data to constrain the factor $\Phi$ by which the SEP flux in the early Solar System was elevated above the present-day value.
We review these in {\bf Figure~\ref{fig:phi}}, for various types of constraints.
These roughly increase with time in the disk: ${}^{50}{\rm V}$, ${}^{7}{\rm Be}$ and ${}^{10}{\rm Be}$ constraints apply to CAIs mostly formed in the first $\sim 10^5$ yr of disk evolution; cosmogenic Ne is produced roughly from $0.5$ to perhaps 3 Myr; 
production of ${}^{36}{\rm Cl}$ requires dissipation of disk gas after $\approx 2.5$ Myr in the inner disk, and probably $> 5$ Myr in the outer disk.
A shaded region depicts the range inferred from observations of X-ray fluxes in protostars (\S 2.1), {\Steve $\Phi \approx 3 \times 10^2$ to $3 \times 10^3$. 
Because the ratio of SEP flux to X-ray flux might be different for protostars compared to the Sun} \citep{FeigelsonEtal2002}, we extend the range to include this uncertainty.
As a reminder, values $\Phi > 10^6$ would imply an extraordinarily large fraction of the early Sun's luminosity was in the form of X rays and SEPs; such values are not supported by observations.

\bigskip
 
The existence of live $\clthreesix$ in CAI minerals that are the products of late-stage alteration provides strong evidence for SEP irradiation.
The high abundances of $\clthreesix$, up to $(\clratio)_0 \approx 2 \times 10^{-5}$, argue against sources (e.g., supernovae) that pre-date the solar nebula, and require late production by SEP irradiation.
During the nebula stage, SEPs lose energy ionizing gas, and any $\clthreesix$ they produced is too diluted by Cl, to match these ratios.
Instead, production within solids is necessary, and the high $(\clratio)_0$ ratios additionally suggest S/Cl ratios in the solids far exceeding chondritic values.
This is possible if Cl remained in the gas phase as HCl, which we argue would be the case.
Enhancements by a factor of {\Steve $\Phi > 1 \times 10^{3}$} appear capable of producing reservoirs with $\clratio$ exceeding the values seen in CAI minerals.
The pink oval next to the ``{$\clthreesix$}" label shows the values of $\Phi$ corresponding to depletions of Cl by factors of $10^{-3}$ to 1. 
It is presumed that gas dissipated from the carbonaceous chondrite-forming region later than about 5 Myr \citep{DeschEtal2018}.

The key constraint on SEP irradiation comes from cosmogenic Ne in hibonite grains. 
We have calculated the production of this Ne in hibonite grains like {\it PLAC-21} as they diffuse through the disk during the {first 2 Myr}. 
For our favored model, {across the range of likely proton rigidities}, the exact amount of cosmogenic Ne is produced if {\Steve $\Phi \approx 3 \times 10^3$ to $2 \times 10^{4}$} depicted by a green oval next to the label ``${}^{21,22}{\rm Ne}$". 
{The exact value is uncertain to within factors of few, because of uncertainties in the rigidity $R_0$, the surface density of gas, $\Sigma$, and also because protostars may eject plasma more evenly with latitude (less concentrated at the equator) than the present-day Sun.
Despite these uncertainties,} values as high as $\Phi \approx 6 \times 10^6$, advocated by \citet{YangEtal2024} (indicated with hatched green oval), can be ruled out, as cosmogenic Ne would be overproduced in all hibonite grains,
{assuming the stellar wind can interact with the disk. If the stellar wind cannot interact with the disk, it is unclear how it would interact with particles in the disk wind.
As well, irradiation in the disk results in the correct value, but irradiation by unattenuated SEPs would result in the wrong ${}^{21}{\rm Ne}/{}^{22}{\rm Ne}$ ratio.}

We have considered the possible production of $\beten$ by SEP irradiation in the disk.
A value $\Phi \sim 9 \times 10^5$ was considered by \citet{Jacquet2019} to be sufficient to produce all the $\beten$ in the disk,
as depicted by the hatched red oval in Figure~\ref{fig:phi} next to the label ``${}^{10}{\rm Be}$".
Nevertheless, models like that of \citet{Jacquet2019} predict considerable variations in the $(\beratio)_0$ ratios recorded by CAIs formed at various times $t$ and heliocentric distances $r$ in the disk, so any variability in $(\beratio)_0$ could be used to constrain $\Phi$.
We have reviewed the literature data and find no credible evidence for deviations (beyond measurement error) in $(\beratio)_0$ from the canonical $7 \times 10^{-4}$ value \citep{DunhamEtal2022} for any CAIs, except for the lower values in FUN CAIs and PLACs, which we interpret as having been formed or reset at $t \approx 1$ Myr.
Assuming that variability in $(\beratio)_0$ at at level $> 10\%$ would be apparent in the data, the uniformity in the data suggests an upper limit $\Phi < 10^5$.
This is depicted as as the red arrow in Figure~\ref{fig:phi}.

Inferences that live $\beseven$ existed in the solar nebula and was incorporated into CAIs as they formed has been used to argue for very high SEP fluxes corresponding to $\Phi \approx 6 \times 10^4$ \citep{LiuEtal2024}, as depicted by the hatched mustard oval next to the label ``${}^{7}{\rm Be}$" in Figure~\ref{fig:phi}.
We have demonstrated that all the evidence for live $\beseven$ comes from overcorrections for cosmogenic Li in the CAIs, plus failure to propagate uncertainties correctly, rendering moot any discussion of what particle fluxes they imply (mustard ``X").

Finally, \citet{SossiEtal2017} had argued for $\Phi \approx 6 \times 10^4$ based on isotopic anomalies in V (brown hatched oval next to the label ``${}^{50}{\rm V}$"). 
This interpretation has been refuted by \citet{BekaertEtal2021}, who showed the data are more consistent with fractionation of V isotopes during evaporation.
This renders moot any interpretations of $\Phi$ from this dataset (brown ``X").

Collectively, the data are consistent with {\Steve $\Phi \approx 3 \times 10^3$ to $2 \times 10^4$ (for a range of SEP spectrum rigidity) throughout disk evolution, 
based on production of cosmogenic Ne in hibonite grains.
Based on the ${}^{21}{\rm Ne}/{}^{22}{\rm Ne}$ ratio in {\it PLAC-21}, we favor high rigidities consistent with $\Phi \approx 3 \times 10^{3}$. 
These values are very consistent with those inferred from X-ray observations of protostars, which suggest $\Phi \approx 3 \times 10^2 - 3 \times 10^3$, no greater than about $4 \times 10^4$.}
The expectation from X-ray observations of protostars is that $\Phi$ should remain roughly constant or decrease slightly over many Myr \citep{WolkEtal2005}.
{\Steve It could be the early Sun was slightly more active than typical protostars, although within the observed range. 
Alternatively, protostars may generate more SEPs per X-ray flux compared to the present-day Sun, as suggested by \citet{FeigelsonEtal2002}.
If $\clthreesix$ is produced by irradiation, the solids must contain $0.1 - 1\%$ of the available Cl (for $\Phi \sim 10^3 - 10^4$), but otherwise irradiation produces the observed abundance of $\clthreesix$.
If $\Phi < 1 \times 10^4$ at early times, we would expect irradiation of the disk to produce very little, $< 1\%$} of the $\beten$ in the solar nebula, and variations of $(\beratio)_0$ among CAIs would be even smaller, in accord with the lack of observed variation in $(\beratio)_0$ ratios.

\subsection{Implications}

The goal of this paper was to use meteoritic data to constrain the SEP flux of the early Solar System.
We have done so using data pertaining to live ${}^{36}{\rm Cl}$ in meteorite parent bodies, cosmogenic Ne in hibonite grains, the evidence from live $\beten$ in CAIs, and the evidence putatively for live $\beseven$ in CAIs.
In order to interpret these data, it has been necessary to model their origins.
Because these astrophysical models are similar in nature, we have considered them together.
The meteoritic data and astrophysical models yield convergent conclusions, from different approaches.

From an astrophysical perspective, values $\Phi > 10^5$ are very unlikely.
Astronomical observations of protostellar X-ray fluxes do not support $\Phi > 10^5$.
For the case of cosmogenic Ne in hibonite grains, {\Steve the model used to argue for $\Phi \sim 10^7$} requires that particles as large as {\it PLAC-21} ($120 \, \mu{\rm m}$) be launched over the disk \citep{YangEtal2024}, which other astrophysical models show does not occur \citep{GiacaloneEtal2019}.
In contrast, astrophysical models of irradiation of hibonite grains while in the disk (\S 3) easily explain the abundance of cosmogenic Ne for reasonable parameters.
If objects like {\it Hidalgo} needed to incorporate live $\beseven$, this would demand their formation near ($\sim 0.03$ AU) the Sun, followed by rapid outward transport, almost certainly by disk winds.
\citet{LiuEtal2024} did not astrophysically model this region, but such an environment as a site of CAI formation has been found to be astrophysically untenable and inconsistent with a host of CAI properties \citep{DeschEtal2010}, and launching in winds even in this environment is unlikely.
It is not impossible to produce the observed abundance of $\beten$ in the disk by irradiation \citep{Jacquet2019}, but this would require atypical conditions.
Equally plausible parameters, based on detailed astrophysical modeling \citep{DeschEtal2018}, make this a less likely option, and favor no production at all of $\beten$ in the disk.

Values $\Phi > 10^5$ can be rejected from a meteoritics perspective as well. 
The demand for $\Phi \sim 10^7$ from cosmogenic Ne requires particles be directly irradiated by SEPs, which would produce cosmogenic Ne with ${}^{21}{\rm Ne}/{}^{22}{\rm Ne} \approx 0.5$, which is inconsistent with the observed ${}^{21}{\rm Ne}/{}^{22}{\rm Ne} \approx 0.7$ in {\it PLAC-21}.
Production of $\beten$ in the protoplanetary disk is demanded only if $(\beratio)_0$ ratios were heterogeneous.
Using the standards of the isotopic geochemistry and cosmochemistry communities \citep{WendtCarl1991,StephanTrappitsch2023}, especially MSWD of isochrons, essentially all samples with reported $(\beratio)_0 > 7 \times 10^{-4}$ have disqualifyingly high MSWD and are not actually isochrons, and are likely simply disturbed;
or, they are likely analytical artefacts. 
Likewise, correct propagation of uncertainties in ${}^{7}{\rm Li}/{}^{6}{\rm Li}$ ratios when making corrections to Li-Be data invalidates all the claims of live $\beseven$ in the solar nebula.
These conclusions require no astrophysical modeling.

The one hypothesis that survives the encounter with all the astrophysical models and meteoritic data is that the early Solar System was characterized by $\Phi \sim 1 \times 10^4$ throughout its first $\approx 5$ Myr, exactly as expected from observations of protostars.
The implication is that the Sun was {\Steve not abnormal}.
Our work stands in contrast to the conclusions of \citet{YangEtal2024}, who claimed evidence for ``an active young Sun beyond expectation".

Another implication is that our work continues the ongoing refutation of the hypothesis that SLRs were created by irradiation.
The discovery of the one-time presence of $\beten$ was once considered the ``smoking gun" for production of SLRs in the solar nebula, and many works \citep{LeeEtal1998,GounelleEtal2001} then and since have explored the ability of SEP irradiation to produce not just $\beten$, but $\altwosix$, ${}^{53}{\rm Mn}$, ${}^{41}{\rm Ca}$, etc.
But today these SLRs are largely understood to have been inherited from the Sun's molecular cloud \citep{DeschEtal2023pp7}, as models like that of 
\citet{Young2014} successfully explain the abundances of $\altwosix$, ${}^{53}{\rm Mn}$, ${}^{41}{\rm Ca}$, ${}^{60}{\rm Fe}$, etc., while formation of the Sun in a spiral arm allows sufficient GCR irradiation to produce $\beratio \approx 7 \times 10^{-4}$ in its molecular cloud \citep{DunhamEtal2022}.
Irradiation as a source of isotopic heterogeneity in V \citep{SossiEtal2017} has been refuted \citep{BekaertEtal2021}, and in this paper we refute the production of meaningful amounts of $\beseven$ and $\beten$ in the disk.
Only some $\clthreesix$ appears to have formed by SEP irradiation, late in disk history, 
{after dissipation of the inner disk.}
Presumably the same fluids carrying ${}^{36}{\rm Cl}$ in the parent bodies would carry other spallogenic nuclei like $\beten$ or high ${}^{10}{\rm B}/{}^{11}{\rm B}$, or have low ${}^{7}{\rm Li}/{}^{6}{\rm Li}$. 
Be may be too insoluble to be carried into CAIs, but aqueous alteration of CAIs could introduce anomalous, spallogenic B or Li. 
Indeed, this seems to be the case for Allende CAI {\it 3529-41} \citep{DeschOuellette2006}.

With little $\beten$ produced in the solar nebula, a canonical amount inherited from the molecular cloud can be defined. 
We find
$(\beratio)_{\rm SS} = (6.95 \pm 0.25) \times 10^{-4}$, slightly lower than (but within uncertainties of) the value $7.1 \times 10^{-4}$ found by \citet{DunhamEtal2022}.
This SLR should be usable for chronometry, although it will be difficult to find later-forming inclusions with the necessary Be/B ratios as high as are seen in CAI melilite.
FUN CAIs are one possibility, and those not dominated by hibonite may also be measured for their Al-Mg systematics.
Exactly two have been measured so far: {\it CMS-1} and {\it KT1}. 
In both, their Al-Mg and Be-B ages are compatible with formation at $t = 0.8$ Myr, i.e., these systems appear concordant.
Although hibonite-dominated FUN inclusions may have formed early (and possibly thermally reset late), large type A/B FUN CAIs probably formed late.

We anticipate that as new measurements are made, new CAIs will be reported to have formed with $(\beratio)_0 > 7 \times 10^{-4}$, or with excesses of $\liratio$ vs. ${}^{9}{\rm Be}/{}^{6}{\rm Li}$ interpreted as the one-time presence of live $\beseven$.
Just because $\beten$ heterogeneities or live $\beseven$ have not been discovered yet does not mean they don't exist or remain to be discovered.
But we urge meteoriticists to apply the standards of their field to the analyses of these data, taking special care with isochrons based on light isotopes that are disturbed especially easily \citep{DunhamEtal2020}, to rule out alternatives.
It is not sufficient for data to resemble an isochron. 
The radiogenic hypothesis demands the relationship (e.g., ${}^{10}{\rm B}/{}^{11}{\rm B}$ vs.\ ${}^{9}{\rm Be}/{}^{11}{\rm B}$) be linear, with valid MSWD $\approx 1$. 
A mixing between two reservoirs should be ruled out, on the basis of invalid MSWD in the ${}^{10}{\rm B}/{}^{11}{\rm B}$ vs.\ 1/[B] correlation.
Potential contributions from various spallogenic reservoirs, including GCR irradiation, should be considered. 
Especially to claim the one-time presence of live $\beseven$, the correction for cosmogenic Li must be done correctly, including correct propagation of uncertainties.
In general, before claiming support for an unexpected hypothesis, just as much attention (or more) should be paid to alternative hypotheses.
According to the scientific method, data do not support a hypothesis; either they are consistent with the predictions of the hypothesis or they rule it out. 
To support a hypothesis, one must falsify all the alternatives.
Here we have ruled out high values $\Phi > 10^5$. 
The hypothesis that withstands scrutiny is that that the early Sun was normal among protostars, with an X-ray flux and SEP flux a factor of $\Phi \sim 1 \times 10^{4}$ times the present day values.

%
%


\vspace{0.1in}
\noindent
We thank two anonymous referees for reviewing the paper. Their critiques helped us significantly improve the paper, especially regarding the heliophysics.
We thank Emilie Dunham, Kohei Fukuda, Ingo Leya, Ming-Chang Liu, Curtis Williams, and Misha Zolotov for useful discussions.
The results reported herein benefitted from collaborations and/or information exchange within NASA's Nexus for Exoplanet System Science (NExSS) research coordination network sponsored by NASA's Science Mission Directorate, {grant 80NSSC23K1356,
PI Steve Desch}. 

%

\vspace{5mm}





\appendix

\section{Cosmogenic Li corrections}

Here we re-derive the corrections for cosmogenic Li produced on the parent-body by spallation on the meteoroid by GCR irradiation. 

Following \citet{DeschOuellette2006}, we anchored production of ${}^{6}{\rm Li}$ and ${}^{7}{\rm Li}$ to production of $\beseven$, assuming a ratio of cross sections ${}^{6}{\rm Li}/{}^{7}{\rm Be} \approx 1.3$ and ${}^{7}{\rm Li}/{}^{7}{\rm Be} \approx 1.1$.
\citet{DeschOuellette2006} used the production rate of $\beseven$ in a gabbro target measured by \citet{LeyaEtal2000stony}, $17 \, {\rm dpm}$ per kg of O, for a primary proton flux $1 \, {\rm cm}^{-2} \, {\rm s}^{-1}$.
An appropriate flux is $4.06 \, {\rm cm}^{-2} \, {\rm s}^{-1}$, and assuming all production of Li was from spallation of O, comprising 42.2wt\% of the gabbro, \citet{DeschOuellette2006} calculated $1.63 \times 10^{14}$ atoms of ${}^{7}{\rm Be}$ produced per kg in 5.2 Myr.
In fact, contributions from Si, Al, and other species contribute, and \citet{DeschOuellette2006} should have found $2.24 \times 10^{14}$ atoms of ${}^{7}{\rm Be}$ per kg of gabbro composition, a value larger by a factor of 1.37.
However, because the recoil distances of spallation-produced Li atoms would be only microns, what matters is the composition of the specific minerals in which the cosmogenic Li is produced.
We assume the same spectrum of secondary particles as in the gabbro; this is appropriate because CV and CO chondrites are similar in composition to the gabbro and would have very similar matrix effects.
(NB: the secondary particle spectrum considered by \citet{Leya2011} was for CI chondrites specifically, which would differ.)
We find almost identical amounts produced in gehlenite ($2.18 \times 10^{14}$), fassaite ($2.31 \times 10^{14}$), anorthite ($2.39 \times 10^{14}$) and hibonite ($2.36 \times 10^{14}$) as in the gabbro ($2.24 \times 10^{14}$). 
We adopt a value $2.3 \times 10^{14}$ atoms of ${}^{7}{\rm Be}$ and $3.0 \times 10^{14}$ atoms of ${}^{6}{\rm Li}$
per kg produced in 5.2 Myr.
\citet{LiuEtal2024} assumed a slightly higher value, $3.4 \times 10^{14}$ atoms of ${}^{6}{\rm Li}$ per kg in 5.2 Myr.

We follow \citet{DeschOuellette2006} in defining $\Delta_{\rm cr} = \Delta^{6}{\rm Li} / {}^{6}{\rm Li}$, where ${}^{6}{\rm Li}$ is the measured number of ${}^{6}{\rm Li}$ atoms per kg, as 
\begin{equation}
\Delta_{\rm cr} = \left( 2.3 \times 10^{14} \, {\rm kg}^{-1} \right) \, \left[ 10^{9} \, \frac{1 \, {\rm ppb}}{ [{\rm Li}]_{\rm meas} } \right] \, \left[ 6.015 + 7.016 \, \left( \frac{{}^{7}{\rm Li}}{{}^{6}{\rm Li}} \right)_{\rm meas} \right] \, \left( 1.66 \times 10^{-27} \, {\rm kg} \right) \, \left( \frac{ t_{\rm CRE} }{ 5.2 \, {\rm Myr} } \right),
\end{equation}
or, more appropriately for DaG 027,
\begin{equation}
\Delta_{\rm cr} = 0.0450 \, \left[ \frac{ 1 \, {\rm ppb} }{ [{\rm Li}]_{\rm meas} } \right] \, \left[ \frac{ 6.015 + 7.016 \, ({}^{7}{\rm Li}/{}^{6}{\rm Li})_{\rm meas} }{ 90.207 } \right] \, \left( \frac{ t_{\rm CRE} }{ 6.2 \, {\rm Myr} } \right).
\label{eq:correction}
\end{equation}
Here 6.015 and 7.016 are the atomic weights of the Li isotopes.
This value of $\Delta_{\rm cr}$ is 41\% higher than the value assumed by \citet{DeschOuellette2006}.

The various authors \citep{ChaussidonEtal2006,MishraMarhas2019,LiuEtal2024} all report for various analysis spots measurements of $x = ({}^{9}{\rm Be}/{}^{6}{\rm Li})_{\rm meas}$, $y = ({}^{7}{\rm Li}/{}^{6}{\rm Li})_{\rm meas}$, and their measurement uncertainties $\delta x$ and $\delta y$.
They also report [Li], the mass fraction of Li, which is uncertain by as much as $\pm 30\%$ due to calibration issues \citep{CarlsonHervig2023}.
Another important input is the cosmic ray exposure time $t_{\rm CRE}$, which also is uncertain.
The values of [Li] and $t_{\rm CRE}$ enter $\Delta_{\rm cr}$, in terms of which it is straightforward to find the corrected (pre-exposure) values $X_{\rm corr} = ({}^{9}{\rm Be}/{}^{6}{\rm Li})_{\rm corr}$ and $Y_{\rm corr} = ({}^{7}{\rm Li}/{}^{6}{\rm Li})_{\rm corr}$: 
\begin{equation}
X_{\rm corr}
= \frac{1}{1 - \Delta_{\rm cr}} \, \left( \frac{{}^{9}{\rm Be}}{{}^{6}{\rm Li}} \right)_{\rm meas}
= \frac{x}{1 - \Delta_{\rm cr}} 
\end{equation}
and
\begin{equation}
Y_{\rm corr}  
= \frac{1}{1-\Delta_{\rm cr} } \, \left[ \left( \frac{ {}^{7}{\rm Li}}{{}^{6}{\rm Li} } \right)_{\rm meas} - {\cal R} \, \Delta_{\rm cr} \right] 
= \frac{ y - {\cal R} \, \Delta_{\rm cr} }{ 1 - \Delta_{\rm cr} },
\end{equation}
where ${\cal R} \approx 1.61$ is the $\liratio$ ratio of cosmogenic Li.
It is assumed that insignificant Be is produced.
The equation for $Y_{\rm corr}$ is identical to Equation 1 of \citet{DeschOuellette2006}.

We find the uncertainties in these quantities by recognizing that random measurement uncertainties enter into $x$ and $y$, but that the uncertainties in $\Delta_{\rm cr}$ (involving [Li] and $t_{\rm CRE}$) are systematic, and should be treated as a model input that is varied, rather than as a random variable.
We therefore calculate the uncertainty first assuming fixed [Li] and $t_{\rm CRE}$.
Uncertainty in $\Delta_{\rm cr}$ still exists because it depends on the amount of ${}^{6}{\rm Li}$ in the sample, and for a given [Li], that is as uncertain as $({}^{7}{\rm Li}/{}^{6}{\rm Li})_{\rm meas}$.
We substitute Equation A2 above for $\Delta_{\rm cr}$, take the partial derivatives with respect to $x$ and $y$, and find:
\begin{equation}
\delta X_{\rm corr} = 
X_{\rm corr} \, \left[ 
\frac{ (\delta x)^2 }{ x^2 } 
+\left( \frac{ 7.016 \, y }{ 6.015 + 7.016 \, y } \right)^2 \, \frac{ \Delta_{\rm cr}^2 }{ ( 1 - \Delta_{\rm cr} )^2 } \, \frac{ (\delta y)^2 }{ y^2 } \right]^{1/2}
\end{equation}
and 
\begin{equation}
\delta Y_{\rm corr} = 
\left[
Y_{\rm corr} 
+\frac{6.015}{7.016}
\right]\, 
\frac{1}{ (1 - \Delta_{\rm cr} ) } \, \left( \frac{ 7.016 \, y }{6.015 + 7.016 \, y} \right) \, \frac{ (\delta y) }{ y }.
\end{equation}
To build an isochron, it is also important to determine the correlation coefficient $\rho_{xy}$.
Because the highly uncertain quantity $({}^{6}{\rm Li})_{\rm corr}$ appears in the denominator of both $X_{\rm corr}$ and $Y_{\rm corr}$, these two quantities will be highly correlated. 
We find this following Equation 20 of \citet{StephanTrappitsch2023}, recognizing that there is very little uncertainty (and no correction) in the measurement of ${}^{9}{\rm Be}$, and therefore $\delta({}^{6}{\rm Li})_{\rm corr} / ({}^{6}{\rm Li})_{\rm corr} \approx \delta X_{\rm corr} / X_{\rm corr}$, so that
\begin{equation}
\rho_{xy} \approx \frac{ (\delta X_{\rm corr} / X_{\rm corr}) }{ (\delta Y_{\rm corr} / Y_{\rm corr}) }.
\end{equation}
This is true regardless of how small the correction is, and all the points in ${}^{7}{\rm Li}/{}^{6}{\rm Li}$ vs. ${}^{9}{\rm Be}/{}^{6}{\rm Li}$ space should be plotted as rotated error ellipses according to Equations 6, 8, and 9 of \citet{StephanTrappitsch2023}.

After following this procedure for all points, assuming a value of $t_{\rm CRE}$ and a calibration for [Li], these latter quantities  should be systematically varied across their plausible ranges. 
This will have the effect of increasing or decreasing $\Delta_{\rm cr}$ for all points systematically.
The uncertainty in quantities such as the slope of the isochron will correspond to a combination of this systematic uncertainty and the measurement uncertainties.

It is important to recognize that $({}^{6}{\rm Li})_{\rm corr}$ is the most uncertain quantity and therefore quantities like $X_{\rm corr}$ and $Y_{\rm corr}$ that rely on it will be highly uncertain as well.
The typical measurement uncertainty in $y$ might be $\pm 15\%$, but as $\Delta_{\rm cr}$ approaches 1, the uncertainties in $X_{\rm corr}$ and $Y_{\rm corr}$ approach $(\pm 15\%) / (1 - \Delta_{\rm cr})$, and $\rho_{xy}$ approaches 1. 
This is understood simply: if there were measured to be $100 \pm 15$ atoms of ${}^{6}{\rm Li}$, but cosmic rays are thought to have created 90 atoms ($\Delta_{\rm cr} \approx 0.9$), then it would be inferred that the pre-exposure abundance would be $10$, but the uncertainty would still be $\pm 15$. 
The uncertainty in the inferred quantity formally would then be $\pm 150\%$.

%
%
%

\bibliography{Irradiation}{}
\bibliographystyle{aasjournal}



\end{document}